\documentclass{article}

\usepackage{amssymb,amsfonts,amsmath}
\usepackage{cite,enumerate,float,indentfirst}
\usepackage{color}

\def\be{\begin{eqnarray}}
\def\ee{\end{eqnarray}}
\def\nn{\nonumber}

\def\Tr{{\rm Tr}\,}

\def\H{\phantom{.}^{\vee}\hspace{-.07cm} H}
\def\lock{\tau}
\def\T24{T}

\definecolor{red}{rgb}{1,0,0}
\definecolor{orange}{rgb}{1,0.5,0}
\definecolor{violet}{rgb}{0.7,0,1}

\hoffset= - 1.0in         

\begin{document}

\title{\vspace{0.1cm}{\LARGE {\bf Tangle blocks in the theory of link invariants}\vspace{.2cm}}
\author{{\bf A. Mironov$^{a,b,c}$}, \
{\bf A. Morozov$^{b,c}$}, \ {\bf An. Morozov$^{b,c,d}$}
}
\date{ }
}

\maketitle

\vspace{-5.5cm}

\begin{center}
\hfill FIAN/TD-04/18\\
\hfill IITP/TH-07/18\\
\hfill ITEP/TH-09/18\\
\end{center}

\vspace{3cm}

\begin{center}
$^a$ {\small {\it Lebedev Physics Institute, Moscow 119991, Russia}}\\
$^b$ {\small {\it ITEP, Moscow 117218, Russia}}\\
$^c$ {\small {\it Institute for Information Transmission Problems, Moscow 127994, Russia}}
$^d$ {\small {\it MIPT, Dolgoprudny, 141701, Russia}}
\end{center}

\vspace{1cm}

\begin{abstract}
The central discovery of $2d$ conformal theory was holomorphic factorization, which expressed correlation functions through bilinear combinations of conformal blocks, which are easily cut and joined without a need to sum over the entire huge Hilbert space of states. Somewhat similar, when a link diagram is glued from tangles, the link polynomial is a multilinear combination of {\it tangle blocks} summed over just a few representations of intermediate states. This turns to be a powerful approach because the same tangles appear as constituents of very different knots so that they can be extracted from simpler cases and used in more complicated ones. So far this method has been technically developed only in the case of arborescent knots, but, in fact, it is much more general. We begin a systematic study of tangle blocks by detailed consideration of some archetypical examples, which actually lead to non-trivial results, far beyond the reach of other techniques. At the next level, the tangle calculus is about gluing of tangles, and functorial mappings from ${\rm Hom}({\rm tangles})$. Its main advantage is an explicit realization of multiplicative composition structure, which is partly obscured in traditional knot theory.
\end{abstract}

\vspace{.5cm}

\tableofcontents

\bigskip

\bigskip

\section{Introduction}

The main task of modern quantum field theory is to find and put under control
relations between non-perturbative correlation functions,
which could provide their Lagrangian-independent description.
This should be an important step in understanding dualities and
developing a background-independent approach to string theory and quantum gravity.
Unfortunately, this is a difficult problem, and it is not yet solved
even in exactly solvable theories, of which the most prominent are
essentially free $2d$ conformal and $3d$ Chern-Simons theory \cite{CS,Wit}.
In the latter case, it is possible to sum up the perturbation theory for
arbitrary observables (Wilson loop averages),
which provides the fascinating quantities known as
link and knot invariants \cite{knotpols}, sometimes also called link/knot "polynomials" where the word "polynomial" refers to one
of their spectacular properties: \ that, in the simply-connected $3d$ spaces
(where there are no instanton-like corrections),
they are rational functions with simple denominators depending only on the number of link components and on the representation and with the numerators being Laurent polynomials with integer coefficients,
of some peculiar quantities $q = \exp \Big(\frac{2\pi i g^2}{1+Ng^2}\Big) $
and $A= \exp \Big(\frac{2\pi i Ng^2}{1+Ng^2}\Big)$.
With mounting number of explicitly calculated link polynomials,
we begin to see that, indeed, they are interconnected by many mysterious relations,
which manifest existence of various "effective theory" formulations,
which have no {\it a priori} connection to the Chern-Simons action,
but instead make transparent particular relations between the
exact observables.
In this paper, we discuss one of such formulations.
It is nearly obvious if one relies on the modern version \cite{modRT1}-\cite{modRT2}
of the Reshetikhin-Turaev (RT)
approach \cite{RT}, which associates the link "polynomials" with a lattice theory
on $2d$ link diagrams (4-valent graphs without boundaries).
Then, if  the graph is cut into parts, one can "glue" the whole correlation function
from those on its parts.
The parts have boundaries, thus the building blocks are not
link polynomials themselves,
but the same blocks can be used to construct different closed graphs,
which provides numerous relations between the corresponding link polynomials.
As we are going to demonstrate, this simple idea can be effectively used in evaluating link polynomials,
which still remains an extremely hard mathematical
problem.
Moreover, it was already proved in at least two contexts:
in arborescent calculus of \cite{arbor} and in the satellite calculus of \cite{sat},
that this is, indeed, a powerful approach.

\bigskip

We begin with reminding the oldest well-known fact relevant to this story.

The link or knot is called {\bf composite} if it can be separated (without disentangling) into two independent parts by cutting
a single line at two places:

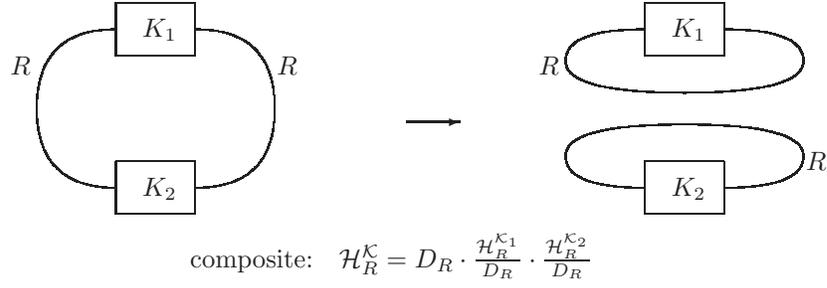
\begin{picture}(300,150)( -125,-75)

\put(0,30){
\put(0,0){\line(1,0){30}}
\put(0,0){\line(0,1){20}}
\put(0,20){\line(1,0){30}}
\put(30,0){\line(0,1){20}}
\put(10,7){\mbox{${ K}_1$}}
}

\put(0,-30){
\put(0,0){\line(1,0){30}}
\put(0,0){\line(0,1){20}}
\put(0,20){\line(1,0){30}}
\put(30,0){\line(0,1){20}}
\put(10,7){\mbox{${ K}_2$}}
}

\qbezier(0,40)(-30,40)(-30,10)
\qbezier(0,-20)(-30,-20)(-30,10)

\qbezier(30,40)(60,40)(60,10)
\qbezier(30,-20)(60,-20)(60,10)

\put(-40,23){\mbox{$R$}}  \put(61,23){\mbox{$R$}}

\put(110,5){\vector(1,0){20}}

\put(28,-50){\mbox{\rm composite:} \ \
${\cal H}_R^{\cal K} = D_R\cdot \frac{{\cal H}_R^{{\cal K}_1}}{D_R} \cdot
\frac{{\cal H}_R^{{\cal K}_2}}{D_R}$}

\put(200,0){
\put(0,30){
\put(0,0){\line(1,0){30}}
\put(0,0){\line(0,1){20}}
\put(0,20){\line(1,0){30}}
\put(30,0){\line(0,1){20}}
\put(10,7){\mbox{${ K}_1$}}
}

\put(0,-30){
\put(0,0){\line(1,0){30}}
\put(0,0){\line(0,1){20}}
\put(0,20){\line(1,0){30}}
\put(30,0){\line(0,1){20}}
\put(10,7){\mbox{${ K}_2$}}
}

\qbezier(0,40)(-30,40)(-30,28)
\qbezier(15,16)(-30,16)(-30,28)
\qbezier(0,-20)(-30,-20)(-30,-8)
\qbezier(15,4)(-30,4)(-30,-8)

\qbezier(30,40)(60,40)(60,28)
\qbezier(15,16)(60,16)(60,28)
\qbezier(30,-20)(60,-20)(60,-8)
\qbezier(15,4)(60,4)(60,-8)

\put(-40,23){\mbox{$R$}}  \put(61,-13){\mbox{$R$}}
}

\end{picture}

\noindent
Then, the {\it reduced} link/knot polynomials decompose into products of
two associated with the closures of the two pieces:
\be
{ H}_R = { H}^{{\cal K}_1}_R\cdot { H}^{{\cal K}_2}_R
\ \ \ \Longleftrightarrow \ \ \
D_R \cdot {H}^{\cal K}_R = {\cal H}^{{\cal K}_1}_R\cdot {\cal H}^{{\cal K}_2}_R
\label{composite}
\ee
Though in ${K}_1$ and ${K}_2$ there are some open lines, i.e. they are tangles, but not links or knots,
the associated quantities are very easy to find:
connecting the two external lines for ${K}_1$ provides a link/knot ${\cal K}_1$,
thus the relevant quantity is just
$\frac{{\cal H}^{{\cal K}_1}_R}{D_R}$.
In more detail, it is a corollary of three relations for {\it unreduced} polynomials:
\be
{\cal H}^{{\cal K}_1}_R = \Tr{\cal B}^{{\cal K}_1}_R = D_R\cdot \underline{ H}^{{\cal K}_1}_R \nn \\
{\cal H}^{{\cal K}_2}_R = \Tr{\cal B}^{{\cal K}_2}_R = D_R\cdot \underline{H}^{{\cal K}_2}_R \nn \\
{\cal H}_R = \Tr( {\cal B}^{{\cal K}_2}_R {\cal B}^{{\cal K}_2}_R) =
D_R\cdot \underline{H}^{{\cal K}_1}_R\cdot \underline{ H}^{{\cal K}_2}_R
\label{compositepieces}
\ee
where underlining denotes the ``partly reduced" polynomial (in the case of knot, it is complete reduction).
What is actually used here, is that, by the arguments of \cite{modRT1}-\cite{modRT2},
the tangles (open blocks) ${\cal B}_R=\underline{H}_R\cdot I_R$  act as {\it unit matrices}
within the representation $R$, the product of two unit matrices is still unit
and its weighted trace  is the quantum dimension $D_R$.
Therefore, the only relevant quantities are the eigenvalues $\hat H_R$ and dimensions $D_R$.

In these pictures and formulas, ${\cal K}_1$ and ${\cal K}_2$ can actually be links,
though cut at two points can be only one of their components.
For example, one can consider the following configuration:

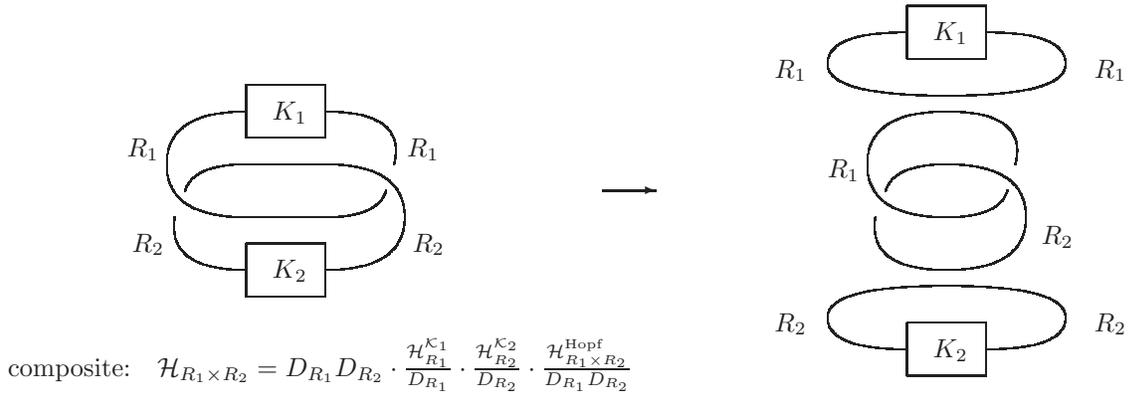
\begin{picture}(300,170)( -100,-80)

\put(0,0){

\put(0,30){
\put(0,0){\line(1,0){30}}
\put(0,0){\line(0,1){20}}
\put(0,20){\line(1,0){30}}
\put(30,0){\line(0,1){20}}
\put(10,7){\mbox{${ K}_1$}}
}

\put(0,-30){
\put(0,0){\line(1,0){30}}
\put(0,0){\line(0,1){20}}
\put(0,20){\line(1,0){30}}
\put(30,0){\line(0,1){20}}
\put(10,7){\mbox{${  K}_2$}}
}

\qbezier(0,40)(-30,40)(-30,20)
\qbezier(0,-0)(-30,-0)(-30,20)

\qbezier(30,20)(60,20)(60,0)
\qbezier(30,-20)(60,-20)(60,0)

\put(0,20){\line(1,0){30}}
\put(0,-0){\line(1,0){30}}

\qbezier(0,-20)(-30,-20)(-27,-0)
\qbezier(0,20)(-20,20)(-23,10)

\qbezier(30,40)(61,40)(56,20)
\qbezier(30,-0)(50,-0)(53,10)

\put(-45,23){\mbox{$R_1$}}  \put(61,23){\mbox{$R_1$}}
\put(-43,-13){\mbox{$R_2$}}  \put(63,-13){\mbox{$R_2$}}

\put(250,0){
\put(0,60){
\put(0,0){\line(1,0){30}}
\put(0,0){\line(0,1){20}}
\put(0,20){\line(1,0){30}}
\put(30,0){\line(0,1){20}}
\put(10,7){\mbox{${ K}_1$}}
}

\put(0,-60){
\put(0,0){\line(1,0){30}}
\put(0,0){\line(0,1){20}}
\put(0,20){\line(1,0){30}}
\put(30,0){\line(0,1){20}}
\put(10,7){\mbox{${ K}_2$}}
}

\put(0,30){
\qbezier(0,40)(-30,40)(-30,28)
\qbezier(15,16)(-30,16)(-30,28)
\qbezier(30,40)(60,40)(60,28)
\qbezier(15,16)(60,16)(60,28)
}
\put(0,-30){
\qbezier(0,-20)(-30,-20)(-30,-8)
\qbezier(15,4)(-30,4)(-30,-8)
\qbezier(30,-20)(60,-20)(60,-8)
\qbezier(15,4)(60,4)(60,-8)
}

\put(-50,53){\mbox{$R_1$}}  \put(71,53){\mbox{$R_1$}}
\put(-30,15){\mbox{$R_1$}}  \put(51,-10){\mbox{$R_2$}}
\put(-50,-43){\mbox{$R_2$}}  \put(71,-43){\mbox{$R_2$}}

\put(15,0){
\qbezier(0,40)(-30,40)(-30,20)
\qbezier(0,-0)(-30,-0)(-30,20)
\qbezier(0,-20)(-30,-20)(-27,-0)
\qbezier(0,20)(-20,20)(-23,10)
}

\put(-15,0){
\qbezier(30,20)(60,20)(60,0)
\qbezier(30,-20)(60,-20)(60,0)
\qbezier(30,40)(61,40)(56,20)
\qbezier(30,-0)(50,-0)(53,10)
}

}

\put(135,10){\vector(1,0){20}}

\put(-90,-60){\mbox{\rm composite:} \ \
${\cal H}_{R_1\times R_2} = D_{R_1}D_{R_2}
\cdot \frac{ {\cal H}^{{\cal K}_1}_{R_1} }{D_{R_1}}
\cdot \frac{{\cal H}^{{\cal K}_2}_{R_2} }{D_{R_2}}
\cdot \frac{{\cal H}^{\rm Hopf}_{R_1\times R_2}}{D_{R_1}D_{R_2}}$}
}

\end{picture}

\noindent
Applying (\ref{composite}) twice we get:
\be
{ H}_{R_1\times R_2} =  { H}^{{\cal K}_1}_{R_1}
\cdot   H^{{\cal K}_2}_{R_2}
\cdot  H^{\rm Hopf}_{R_1\times R_2}
\ \ \ \ \Longrightarrow \ \ \ \
D_{R_1}D_{R_2}\cdot {\cal H}_{R_1\times R_2} =
 { {\cal H}^{{\cal K}_1}_{R_1} }
\cdot  {{\cal H}^{{\cal K}_2}_{R_2} }
\cdot  {{\cal H}^{\rm Hopf}_{R_1\times R_2}}
\ee
More general,

\begin{picture}(300,170)( -100,-80)

\put(0,0){

\put(0,30){
\put(0,0){\line(1,0){30}}
\put(0,0){\line(0,1){20}}
\put(0,20){\line(1,0){30}}
\put(30,0){\line(0,1){20}}
\put(10,7){\mbox{${ K}_1$}}
}

\put(0,-30){
\put(0,0){\line(1,0){30}}
\put(0,0){\line(0,1){20}}
\put(0,20){\line(1,0){30}}
\put(30,0){\line(0,1){20}}
\put(10,7){\mbox{${  K}_3$}}
}

\put(0,0){
\put(0,-5){\line(1,0){30}}
\put(0,-5){\line(0,1){30}}
\put(0,25){\line(1,0){30}}
\put(30,-5){\line(0,1){30}}
\put(10,7){\mbox{${ K}_2$}}
}

\qbezier(0,40)(-30,40)(-30,20)
\qbezier(0,-0)(-30,-0)(-30,20)

\qbezier(30,20)(60,20)(60,0)
\qbezier(30,-20)(60,-20)(60,0)


\qbezier(0,-20)(-30,-20)(-27,-0)
\qbezier(0,20)(-20,20)(-23,10)

\qbezier(30,40)(61,40)(56,20)
\qbezier(30,-0)(50,-0)(53,10)

\put(-45,23){\mbox{$R_1$}}  \put(61,23){\mbox{$R_1$}}
\put(-43,-13){\mbox{$R_2$}}  \put(63,-13){\mbox{$R_2$}}

\put(250,0){
\put(0,60){
\put(0,0){\line(1,0){30}}
\put(0,0){\line(0,1){20}}
\put(0,20){\line(1,0){30}}
\put(30,0){\line(0,1){20}}
\put(10,7){\mbox{${ K}_1$}}
}

\put(0,-60){
\put(0,0){\line(1,0){30}}
\put(0,0){\line(0,1){20}}
\put(0,20){\line(1,0){30}}
\put(30,0){\line(0,1){20}}
\put(10,7){\mbox{${ K}_3$}}
}

\put(0,0){
\put(0,-5){\line(1,0){30}}
\put(0,-5){\line(0,1){30}}
\put(0,25){\line(1,0){30}}
\put(30,-5){\line(0,1){30}}
\put(10,7){\mbox{${ K}_2$}}
}

\put(0,30){
\qbezier(0,40)(-30,40)(-30,28)
\qbezier(15,16)(-30,16)(-30,28)
\qbezier(30,40)(60,40)(60,28)
\qbezier(15,16)(60,16)(60,28)
}
\put(0,-30){
\qbezier(0,-20)(-30,-20)(-30,-8)
\qbezier(15,4)(-30,4)(-30,-8)
\qbezier(30,-20)(60,-20)(60,-8)
\qbezier(15,4)(60,4)(60,-8)
}

\put(-50,53){\mbox{$R_1$}}  \put(71,53){\mbox{$R_1$}}
\put(-30,15){\mbox{$R_1$}}  \put(51,-10){\mbox{$R_2$}}
\put(-50,-43){\mbox{$R_2$}}  \put(71,-43){\mbox{$R_2$}}

\put(15,0){
\qbezier(0,40)(-30,40)(-30,20)
\qbezier(-15,-0)(-30,-0)(-30,20)
\qbezier(0,-20)(-30,-20)(-27,-0)
\qbezier(-15,20)(-20,20)(-23,10)
}

\put(-15,0){
\qbezier(45,20)(60,20)(60,0)
\qbezier(30,-20)(60,-20)(60,0)
\qbezier(30,40)(61,40)(56,20)
\qbezier(45,-0)(50,-0)(53,10)
}

}

\put(135,10){\vector(1,0){20}}

\put(-90,-60){\mbox{\rm composite:} \ \
$\frac{{\cal H}_{R_1\times R_2}}{ D_{R_1}D_{R_2}}
= \frac{ {\cal H}^{{\cal K}_1}_{R_1} }{D_{R_1}}
\cdot \frac{{\cal H}^{{\cal K}_2}_{R_1\times R_2}}{D_{R_1}D_{R_2}}
\cdot \frac{{\cal H}^{{\cal K}_3}_{R_2} }{D_{R_2}}
$}
}

\end{picture}

\noindent
The only {\bf subtlety} with (\ref{composite}) is the need to carefully distinguish
between the knot/link and its mirror.
Normally, the mirror symmetry acts on the knot polynomial in a simple way:
\be
\overline{{\cal H}(A,q) }= {\cal H}(A^{-1},q^{-1})
\label{mirr}
\ee
but if only one of the components of a composite link
is mirror-transformed, then only one of the two factors in (\ref{composite})
is changed by the rule  (\ref{mirr}), and the entire product
gets essentially different.

\bigskip

In this paper, we mostly consider
the {\bf double-line reducible}
links and knots which can be separated
by cutting, say, a {\it pair} of lines
(generalization to multiple lines is straightforward).
Note that this is {\it much more} than just {\bf cabling,
which is a particular case} of the double line going through
${\cal K}_1$ and ${\cal K}_2$ without entangling.

\begin{picture}(300,150)( -25,-75)

\put(0,30){
\put(0,0){\line(1,0){30}}
\put(0,0){\line(0,1){20}}
\put(0,20){\line(1,0){30}}
\put(30,0){\line(0,1){20}}
\put(10,7){\mbox{${\cal K}_1$}}
}

\put(0,-30){
\put(0,0){\line(1,0){30}}
\put(0,0){\line(0,1){20}}
\put(0,20){\line(1,0){30}}
\put(30,0){\line(0,1){20}}
\put(10,7){\mbox{${\cal K}_2$}}
}

\qbezier(0,40)(-30,40)(-30,10)
\qbezier(0,-20)(-30,-20)(-30,10)

\qbezier(30,40)(60,40)(60,10)
\qbezier(30,-20)(60,-20)(60,10)

\put(-40,23){\mbox{$R$}}  \put(61,23){\mbox{$R$}}

\put(-48,-50){\mbox{\rm composite:} \ \
$H_R = H_R^{{\cal K}_1} \cdot H_R^{{\cal K}_2}$}

\put(200,0){

\put(0,30){
\put(0,0){\line(1,0){30}}
\put(0,0){\line(0,1){20}}
\put(0,20){\line(1,0){30}}
\put(30,0){\line(0,1){20}}
\put(10,7){\mbox{${\cal K}_1$}}
}

\put(0,-30){
\put(0,0){\line(1,0){30}}
\put(0,0){\line(0,1){20}}
\put(0,20){\line(1,0){30}}
\put(30,0){\line(0,1){20}}
\put(10,7){\mbox{${\cal K}_2$}}
}

\qbezier(0,42)(-30,42)(-30,10)
\qbezier(0,-22)(-30,-22)(-30,10)

\qbezier(30,42)(60,42)(60,10)
\qbezier(30,-22)(60,-22)(60,10)

\qbezier(0,38)(-26,38)(-26,10)
\qbezier(0,-18)(-26,-18)(-26,10)

\qbezier(30,38)(56,38)(56,10)
\qbezier(30,-18)(56,-18)(56,10)

\put(-58,33){\mbox{$R\otimes R$}}  \put(61,33){\mbox{$R\otimes R$\ {\rm or}}}
\put(-72,23){\mbox{{\rm or}\,$R\otimes \bar R$}}
\put(61,23){\mbox{ $R\otimes \bar R$}}

\put(-30,-55){\mbox{\rm double-line reducible}}
}

\put(295,8){\mbox{$=\ \ \ \sum_Q$}}

\put(380,0){

\put(0,30){
\put(0,0){\line(1,0){30}}
\put(0,0){\line(0,1){20}}
\put(0,20){\line(1,0){30}}
\put(30,0){\line(0,1){20}}
\put(10,7){\mbox{${\cal K}_1$}}
}

\put(0,-30){
\put(0,0){\line(1,0){30}}
\put(0,0){\line(0,1){20}}
\put(0,20){\line(1,0){30}}
\put(30,0){\line(0,1){20}}
\put(10,7){\mbox{${\cal K}_2$}}
}

\put(-40,23){\mbox{$Q$}}  \put(61,23){\mbox{$Q$}}

\linethickness{1mm}

\qbezier(-1,40)(-30,40)(-30,10)
\qbezier(-1,-20)(-30,-20)(-30,10)

\qbezier(31,40)(60,40)(60,10)
\qbezier(31,-20)(60,-20)(60,10)

\put(-40,-55){\mbox{$H_R
= \frac{\sum_Q D_Q H_{R,Q}^{{\cal K}_1} H_{R,Q}^{{\cal K}_2}}{\sum_Q D_Q}$}}
}

\end{picture}

In this picture we put the two representations on the double lines coinciding, though they can generally differ.

In the spirit of RT theory, one can again expect that the
link/knot polynomials will be bilinear compositions of the two \cite{sat}:
\be
\boxed{
H_R = \frac{1}{D_R^2}
\sum_{Q} D_Q \cdot H^{{\cal K}_1}_{R,Q} \cdot H^{{\cal K}_2}_{R,Q}
}
\label{decof}
\ee
where $Q \in R^{\otimes 2}$ or $Q \in R\otimes \bar R$,
depending on whether the two lines are parallel or antiparallel,
and $D_Q$ are the quantum dimensions of the corresponding representations, they
satisfy $\sum_Q D_Q = D_R^2$.
The main difference from (\ref{composite}) is that now we have
the entire sum over all representations $Q$. However, within any given $Q$,
all matrices are still unit, and we can still use exactly the same
formulas (\ref{compositepieces}) merely with $R$ substituted by $Q$.
There are numerous interesting sorts of knots and links,
which belong to this double-line reducible class,
and, hence, can be handled in this way, we study several examples to illustrate how (\ref{decof}) works in this paper.
This extends considerably the set of comprehensible knots,
including some which are not arborescent.
In fact, the arborescent calculus of \cite{arbor}
is also inspired by the same type of argument
and can be considered as a particular big
example of this cut-and-join algorithm.

Of course, generalizations to links/knots separated by cutting arbitrary number of lines
is straightforward, we provide simple illustrations of them in this text.

The simplest tangle is just the single vertex, and, in the RT formalism, it
is described by the quantum ${\cal R}$-matrix ${\cal R}:R_1\otimes R_2\longrightarrow R_2\otimes R_1$ with four indices,
${\cal R}_{jb}^{ai}$, where indices $i,j$ belong to representation $R_1$
and $a,b$, to representation $R_2$.
In any representation $Q\in R_1\otimes R_2$, this ${\cal R}$-matrix
is just a square matrix ${\cal R}_A^B$ with indices $A,B$ belonging to
(labeling the states in) $Q$.
The main property of the ${\cal R}$-matrix is that, in any irreducible representation $Q$,
this ${\cal R}_A^B \sim\delta_A^B$.

Of course, each link diagram can be decomposed in these elementary
${\cal R}$-matrix tangles.
However, it is often useful to construct links and knots from more complicated
building blocks.
For example, in the case of $m$-strand torus diagrams, a useful
tangle is a product of $m-1$ ${\cal R}_{i,i+1}$-matrices acting on the neighbour strands $i$, $i+1$:
${\rm Tor}_m = {\cal R}_{1,2}{\cal R}_{2,3}\ldots {\cal R}_{m-1,m}$.
It is also a unit matrix, but in irreducible representations
from the decomposition of the product $R_1\otimes\ldots\otimes R_m$.

Already in the simplest examples like
the vacuum block, \ref{vacuumblock},
one can see that the tangle blocks depend not only on the
explicit argument $Q$, but also on the "hidden" $R$.
Thus, the $R$-dependence enters in two ways:
through selecting the intermediate representations
and through the "hidden" $R$-dependence of tangle blocks,
which naively depend only on the intermediate representations,
but, in fact, also on what they were made from, i.e. on $R$.
However, there is a sort of blocks, which do {\it not} depend on $R$.
This happens when the external lines can be combined into a {\it cable},
which passes as a whole through the interior like it happens
for the cabled links and knots without external lines \cite{AnoMcabling}.
A natural name for them is {\it cable blocks} or {\it cable tangles}.
The calculational art is to cut the link diagram in such a way
that complicated tangles are $R$-independent cable blocks, while the explicit
$R$-dependence remains only in some simpler tangles (which we sometimes call {\it cut-and-join} blocks).
They can be often handled by the arborescent technique.

The simplest illustration of decomposition into sophisticated {\it cable}
and elementary {\it cut-and-join} blocks is provided
by evaluating the knot polynomials for the Whitehead doubles in \cite{sat}.
It is a straightforward generalization of the standard cabling method
by attaching the cable ends to an additional {\it lock} element $\lock$.
The story is described by the following pictures:
the one at the r.h.s. is for trefoil's twist satellite, the l.h.s. is for that
of arbitrary $m$-strand link diagram.

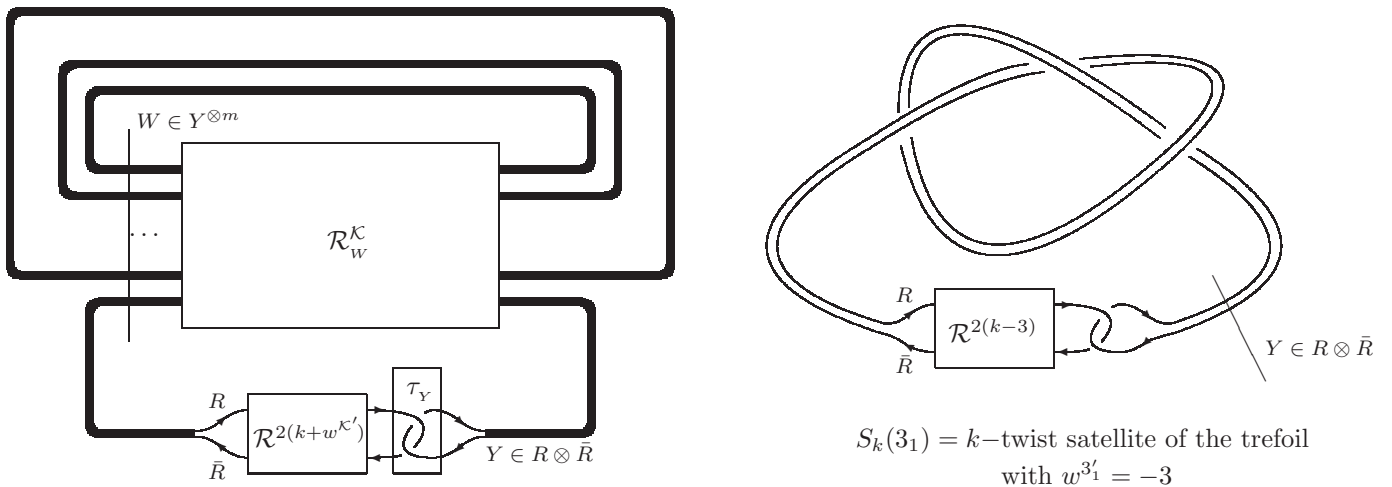
\begin{picture}(300,210)(-120,-130)

\put(250,0){

\qbezier(-44,12)(-30,-65)(50,5)     \qbezier(-48,9)(-30,-69)(50,0)
\qbezier(20,42)(110,51)(50,0)      \qbezier(24,39)(100,48)(50,5)
\qbezier(0,40)(-20,38)(-60,12)      \qbezier(6,37)(-20,34)(-60,8)
\qbezier(-49,22)(-48,90)(52,16)    \qbezier(-45,24)(-45,84)(50,12)
\qbezier(-60,8)(-132,-35)(-55,-58) \qbezier(-60,12)(-140,-35)(-55,-62)
\qbezier(61,5)(125,-35)(55,-58)    \qbezier(64,8)(132,-35)(55,-62)

\put(-65,-105){\mbox{$S_k(3_1)= k-$twist satellite of the trefoil}}
\put(-10,-119){\mbox{with $w^{3_1'} = -3$}}

\put(10,0){

\qbezier(-65,-58)(-60,-59)(-55,-55)\qbezier(-45,-51)(-50,-51)(-55,-55)
\qbezier(-65,-62)(-60,-63)(-55,-67)\qbezier(-45,-69)(-50,-69)(-55,-67)

\qbezier(45,-58)(40,-59)(35,-55)\qbezier(25,-51)(30,-51)(35,-55)
\qbezier(45,-62)(40,-63)(35,-65)\qbezier(25,-69)(30,-69)(35,-65)

\put(-55,-55){\vector(1,1){2}}
\put(-55,-67){\vector(-2,1){2}}
\put(5,-51){\vector(1,0){2}}
\put(3,-69){\vector(-1,0){2}}
\put(35,-55){\vector(1,-1){2}}
\put(35,-65){\vector(-1,-1){2}}

\put(-60,-50){\mbox{\footnotesize $R$}}
\put(-60,-77){\mbox{\footnotesize $\bar R$}}
\put(80,-70){\mbox{\footnotesize $Y\in R\otimes\bar R$}}
\put(60,-40){\line(1,-2){20}}

\qbezier(25,-51)(24,-51)(22,-52)\qbezier(25,-69)(8,-69)(17,-57)
\qbezier(0,-69)(11,-69)(13,-68)\qbezier(0,-51)(30,-51)(18,-65)

\put(-45,-75){\line(0,1){30}}
\put(0,-75){\line(0,1){30}}
\put(-45,-75){\line(1,0){45}}
\put(-45,-45){\line(1,0){45}}

\put(-40,-65){\mbox{${\cal R}^{2(k-3)}$}}


}
}


\put(0,-40){

\put(-70,50){\line(1,0){120}}
\put(-70,-20){\line(1,0){120}}
\put(-70,50){\line(0,-1){70}}
\put(50,-20){\line(0,1){70}}
\put(-15,10){\mbox{${\cal R}^{\cal K}_{_W}$}}
\put(-90,-25){\line(0,1){80}}
\put(-87,56){\mbox{\footnotesize $W \in  Y^{\otimes m}$}}

\qbezier(-65,-59)(-60,-59)(-55,-55)\qbezier(-45,-51)(-50,-51)(-55,-55)
\qbezier(-65,-61)(-60,-61)(-55,-65)\qbezier(-45,-69)(-50,-69)(-55,-65)

\qbezier(45,-59)(40,-59)(35,-55)\qbezier(25,-51)(30,-51)(35,-55)
\qbezier(45,-61)(40,-61)(35,-65)\qbezier(25,-69)(30,-69)(35,-65)

\put(-55,-55){\vector(1,1){2}}
\put(-55,-65){\vector(-1,1){2}}
\put(5,-51){\vector(1,0){2}}
\put(3,-69){\vector(-1,0){2}}
\put(35,-55){\vector(1,-1){2}}
\put(35,-65){\vector(-1,-1){2}}

\put(-60,-50){\mbox{\footnotesize $R$}}
\put(-60,-77){\mbox{\footnotesize $\bar R$}}
\put(45,-70){\mbox{\footnotesize $Y\in R\otimes\bar R$}}

\qbezier(25,-51)(24,-51)(22,-52)\qbezier(25,-69)(8,-69)(17,-57)
\qbezier(0,-69)(11,-69)(13,-68)\qbezier(0,-51)(30,-51)(18,-65)

\put(-45,-75){\line(0,1){30}}
\put(0,-75){\line(0,1){30}}
\put(-45,-75){\line(1,0){45}}
\put(-45,-45){\line(1,0){45}}

\put(-43,-65){\mbox{${\cal R}^{2(k+w^{\cal K'}\!)}$}}

\put(10,-75){\line(0,1){40}}
\put(28,-75){\line(0,1){40}}
\put(10,-75){\line(1,0){18}}
\put(10,-35){\line(1,0){18}}
\put(15,-45){\mbox{$\lock_{_Y}$}}

\linethickness{1mm}
\put(-100,40){\line(1,0){30}}  \put(50,40){\line(1,0){30}}
\put(-110,30){\line(1,0){40}}  \put(50,30){\line(1,0){40}}
\put(-90,15){\mbox{$\ldots$}}
\put(-130,0){\line(1,0){60}}   \put(50,0){\line(1,0){60}}
\put(-100,-10){\line(1,0){30}}   \put(50,-10){\line(1,0){30}}

\put(-100,70){\line(1,0){180}}
\put(-110,80){\line(1,0){200}}
\put(-130,100){\line(1,0){240}}

\put(-100,-60){\line(1,0){35}}  \put(45,-60){\line(1,0){35}}

\qbezier(-100,40)(-105,40)(-105,45)  \qbezier(-100,70)(-105,70)(-105,65)
\put(-105,45){\line(0,1){20}}
\qbezier(-110,30)(-115,30)(-115,35)  \qbezier(-110,80)(-115,80)(-115,75)
\put(-115,35){\line(0,1){40}}
\qbezier(-130,0)(-135,0)(-135,5)  \qbezier(-130,100)(-135,100)(-135,95)
\put(-135,5){\line(0,1){90}}
\qbezier(-100,-10)(-105,-10)(-105,-15)  \qbezier(-100,-60)(-105,-60)(-105,-55)
\put(-105,-55){\line(0,1){40}}

\qbezier(80,40)(85,40)(85,45)  \qbezier(80,70)(85,70)(85,65)
\put(85,45){\line(0,1){20}}
\qbezier(90,30)(95,30)(95,35)  \qbezier(90,80)(95,80)(95,75)
\put(95,35){\line(0,1){40}}
\qbezier(110,0)(115,0)(115,5)  \qbezier(110,100)(115,100)(115,95)
\put(115,5){\line(0,1){90}}
\qbezier(80,-10)(85,-10)(85,-15)  \qbezier(80,-60)(85,-60)(85,-55)
\put(85,-55){\line(0,1){40}}
}

\end{picture}

\noindent
From this picture, one reads:
\be
\boxed{
H^{S_k({\cal K})}_{_R} =
\frac{1}{D_{_R}}\sum_{Y\in R\otimes \bar R}
{\cal H}_{_Y}^{\cal K}\cdot \mu_{_Y}^{2(k+w^{\cal K'})} \cdot\lock^Y_R
}
\label{SatH}
\ee
where $D_R$ is the quantum dimension of representation $R$,
\be\label{mu}
\mu_Y=\pm  q^{C_2(Y)}
 \ee
 are the eigenvalues of the quantum $SL_q(N)$ ${\cal R}$-matrix,  $C_2(Y)$ being the eigenvalue of the second Casimir operator in representation $Y$, and sign is chosen depending on whether $Y$ lies in the symmetric or antisymmetric parts of the product $R\otimes \bar R$.
To this, one has to add the issue of $Q$-independent normalization of the ${\cal R}$-matrix and its eigenvalues. In particular, the normalization determines  the framing (see below).

One of the two building blocks is here given by the unreduced HOMFLY
of the original knot ${\cal K}$
\be
{\cal H}^{\cal K}_Y = \sum_{W\in Y^{\otimes m}}
D_W\cdot {\rm Tr}_{{\rm mult}_{_W}} {\cal R}_W^{\cal K}
= D_Y \cdot H^{\cal K}_Y = {\rm Tr}_Y \hat H^{\cal K}
\ee
and does {\it not} depend on $R$.
The matrix
\be
(\hat H^{\cal K})_{YY'} = \delta_{Y,Y'} \cdot \frac{{\cal H}^{\cal K}_Y}{D_Y}
\ee
is an archetypical example of the cable blocks.

The central role in (\ref{SatH}) is played by the element $\lock^Y_R$,
which is a universal building block,
which can be attached to any ${\cal K}$.
It was actually introduced and calculated long ago in the study of twist knots in \cite{evo}.
Once $\lock$ is known, (\ref{SatH}) reduces the original calculation for the
Whitehead double to that of the original knot,
but in higher representations.

In this paper, we provide more examples of tangles of the both types, simple, universal,
but $R$-dependent {\it cut-and-join}, which have non-trivial matrix structure,
and $R$-independent {\it cable}, which are diagonal matrices and, thus, proportional
to the HOMFLY polynomials, which can be calculated by a variety of already
developed methods, from cabling and skein relations to arborescent calculus
and differential expansions.

We also demonstrate how these building blocs can be combined in different ways
to provide various link/knot invariants ("polynomials").

\subsubsection*{Notation, framing, normalization.}

\paragraph{Notation.} Throughout the text, we use the following notation: the quantum dimension of the representation $R$ is given by the celebrated hook formula
\be
{\cal H}^{\rm unknot}_R = D_R = \prod_{(i,j)\in R}
\frac{\{Aq^{i-j}\}}{\{q^{h_{i,j}}\}}
\ee
where $h_{i,j}$ is the hook length for the element $(i,j)$ of the Young diagram $R$, $A=q^N$ and $\{x\} := x-x^{-1}$, so that the quantum number  $[n]=\frac{\{q^n\}}{\{q\}}$. Since we consider only finite-dimensional representations of $SU(N)$, each of them is associated with a Young diagram. Throughout the text, we do not differ between the representation $R$ and the corresponding Young diagram. We also use the notation $\mu_Y$ for the eigenvalues of the ${\cal R}$-matrix in the channel $R\otimes\bar R=\oplus Y$ and $\lambda_Q$, in the channel $R\otimes R=\oplus Q$.

\paragraph{Normalization.} We denote by calligraphic letters the unreduced knot/link invariants: the HOMFLY-PT invariant is denoted through ${\cal H}$. The reduced HOMFLY-PT invariant, i.e. that divided by the product of quantum dimensions of the representations coloring the link components is denoted through just $H$. Note that the normalization of the link invariants in \cite{katlas} is different: as soon as they consider uncolored links, they normalize the invariants dividing them just by one quantum dimension of the fundamental representation $D_\Box={\{A\}\over\{q\}}$.

We also often need a mirror image of the knot/link, which, in terms of the invariant, is described by the transformation $(A,q)\to (1/A,1/q)$. We denote the corresponding invariants with the bar:
\be
\overline{H(A,q)}:=H\Big({1\over A},{1\over q}\Big)
\ee

\paragraph{Framing.} Typically, all the invariants are given in the topological framing for knots and in the standard (or canonical) framing for links \cite{MarF,Atiah,China1}. This framing contains a trivial $U(1)$-factor
\be\label{U1fac}
U=q^{2\sum_{i>j}|R_i||R_j|L_{ij}/N}
 \ee
 where the sum goes over all components of the link, the $i$-th component being colored with the representation $R_i$, and $L_{ij}$ is the linking number of the $i$-th and $j$-th components. This factor is absent for knots, and we omit it for links unless the inverse is stated\footnote{It is often omitted, see \cite{Ch,Mar}.}. It is essential only when one reduces representations, for the concrete $SU(N)$, removing from the Young diagrams the rows of the length $N$.

We sometimes need to consider various relations between knot/link invariants inspired by group theory identities. Hence, these relations are most naturally realized in the framing consistent with the group theory structure of the invariants, and, in the case of knots, this is nothing but the vertical framing. In the case of links, this is the so called differential framing, which is also consistent with the differential expansion of invariants, hence the name \cite{China1}. We denote the invariants in this framing with a check-superscript:$\phantom{.}^{\vee}\hspace{-.07cm} H$.

\paragraph{Important note.}

In this paper, for the sake of simplicity, we deal only with the cases of trivial multiplicities. Otherwise, there would appear additional matrices in the space of multiplicities, and the formulas become too overload.

\section{Simple building blocks}
\subsection{The vacuum block: the simplest among cut-and-join
\label{vacuumblock}}

Let us demonstrate how our procedure works in the simplest example of the unreduced colored HOMFLY polynomial for the unknot,
which is equal to the quantum dimension of the coloring representation $R$,
\be
{\cal H}^{\rm unknot}_R = D_R
\ee
If the simplest unknot diagram (a circle) is cut into two semicircles,
we get a decomposition over all representations $Y$ from the product of

\begin{picture}(300,90)(-150,-55)

\put(0,0){\circle{40}}
\put(-20,0){\vector(0,1){2}}
\put(-28,-22){\mbox{$R$}}
\put(40,-3){\mbox{$=\ \ \ \ \sum_{Y\in R\otimes \bar R}$}}

\put(20,0){
\put(100,0){\circle{4}} \put(180,0){\circle{4}}
\qbezier(100,0)(100,20)(140,20)
\qbezier(100,0)(100,-20)(140,-20)
\qbezier(180,0)(180,20)(140,20)
\qbezier(180,0)(180,-20)(140,-20)

\put(140,-50){\line(0,1){80}}
\put(142,-50){\mbox{$Y$}}

\put(145,20){\vector(1,0){2}} \put(145,24){\mbox{$R$}}
\put(135,-20){\vector(-1,0){2}} \put(125,-32){\mbox{$\bar R$}}
}

\end{picture}

\noindent
However, actually contributing to  the sum is only the singlet representation $Y=\emptyset$
with dimension $D_\emptyset = 1$.
Thus the formula which we should associate with this picture is
\be
D_R = \sum_{Y\in R\otimes \bar R} D_Y{\cal V}_R^Y \overline{{\cal V}_R^Y}={\cal V}_R^\emptyset \overline{{\cal V}_R^\emptyset}
\ee
where ${\cal V}=\bar {\cal V}$ are the two blocks in the left and right parts
of the second picture, which we can naturally call {\it vacuum}.
It follows that the vacuum block

\begin{picture}(300,80)(-150,-40)

\qbezier(0,0)(0,20)(40,20)
\qbezier(0,0)(0,-20)(40,-20)
\put(35,20){\vector(1,0){2}}
\put(35,-20){\vector(-1,0){2}}
\put(35,25){\mbox{$R$}}
\put(35,-35){\mbox{$\bar R$}}
\put(45,-2){\mbox{$Y$}}

\put(0,0){\circle{4}}
\put(-50,-2){\mbox{${\cal V}_R^Y \ \ \ \ \ =$}}
\put(100,-2){\mbox{$  =\ \ \ \  \ \delta^Y_\emptyset\cdot\sqrt{D_R}$}}

\end{picture}


As mentioned in the Introduction, already in this example
we see that the tangle block depends not only on the
apparent argument $Y$, but also on the "hidden" $R$.
Thus the vacuum block is the simplest example of the cut-and-join blocks.

\subsection{The lock block
\label{lock}}

The central role in (\ref{SatH}) is played by the {\it  lock element} $\lock^Y_R$,
which was actually calculated long ago in the study of twist knots in \cite{evo}.
From this block $\lock$, one can construct a few even simpler objects:

\begin{picture}(300,100)(0,-50)

\put(0,0){\line(1,0){30}}
\put(0,0){\line(0,1){20}}
\put(30,0){\line(0,1){20}}
\put(0,20){\line(1,0){30}}
\put(10,8){\mbox{${\cal K}$}}
\put(-16,15){\mbox{$Y$}}
\put(38,15){\mbox{$Y'$}}
\put(-15,-20){\mbox{$\frac{1}{D_Y}\cdot {\cal H}_Y^{\cal K}\cdot\delta_{Y,Y'}$}}

\put(150,20){\line(1,0){20}}
\put(150,20){\line(0,1){20}}
\put(170,20){\line(0,1){20}}
\put(150,40){\line(1,0){20}}
\put(154,27){\mbox{${\cal R}^k$}}
\put(160,-10){\circle*{6}}
\put(155,-2){\mbox{$\lock_R^Y$}}
\put(95,-30){\mbox{${\cal H}^{{\rm twist}_k}_R
= \sum_{Y\in R\otimes\bar R}D_Y\cdot \lock_R^Y \cdot\mu_Y^{k}$}}

\put(300,10){\circle*{6}}
\put(295,18){\mbox{$\lock_R^Y$}}
\put(260,-15){\mbox{${\cal H}^{{\rm tor}_{[2,2]}}_R
= \left(V^\emptyset_R\right)^2 \cdot\lock_R^\emptyset $}}

\put(290,0){
\put(130,10){\circle*{6}}
\put(190,10){\circle*{6}}
\put(110,8){\mbox{$\tilde\lock_R^Y$}}
\put(197,8){\mbox{$\tilde\lock_R^Y$}}
\put(90,-30){\mbox{${\cal H}^{{\rm tor}_{[2,4]}}_R
= \sum_{Y\in R\otimes\bar R}D_Y\cdot (\tilde\lock_R^Y)^2  $}}
}

\put(272,10){\circle{6}} \put(328,10){\circle{6}}

\linethickness{1mm}

\put(0,10){\line(-1,0){20}}
\put(30,10){\line(1,0){20}}

\qbezier(160,-10)(130,-10)(130,10)
\qbezier(148,30)(130,30)(130,10)
\qbezier(160,-10)(190,-10)(190,10)
\qbezier(172,30)(190,30)(190,10)

\put(275,10){\line(1,0){50}}

\put(-10,0){
\qbezier(460,-10)(430,-10)(430,10)
\qbezier(460,30)(430,30)(430,10)
\qbezier(460,-10)(490,-10)(490,10)
\qbezier(460,30)(490,30)(490,10)
}

\end{picture}

\noindent
Here and below $\lock_R^Y$ is the eigenvalue of the "lock" operator for $Y\in R\otimes \bar R$,
which is diagonal in $Y$, $\lock_R^{YY'}=\lock_R^Y\cdot\delta_{Y,Y'}$,
but depends on  $R$
(which contradicts the idea of effective theory \cite{rama2} in its extreme form).
Also note that there are two slightly different lock operators,
antiparallel $\lock$ and parallel $\tilde\lock$, which differ by the factor $\mu_Y$ (see (\ref{mu})) and the mirror map:
\be
\overline{\tilde \lock^Y_R }= \mu_Y\cdot \lock^Y_R
\ee
e.g. $\tilde\lock_R^\emptyset = \overline{\lock_R^\emptyset}$, while
$\tilde \lock_R^{\rm adj} = -A^{-1} \cdot \overline{\lock_R^{\rm adj}}$:

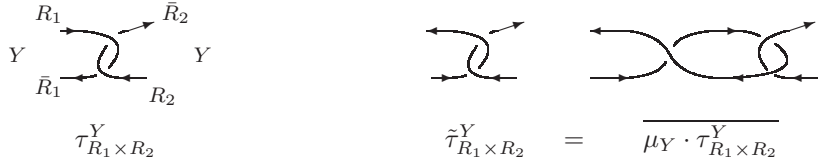
\begin{picture}(300,80)(-290,-50)

\put(-60,60){
\qbezier(25,-51)(24,-51)(22,-52)\qbezier(25,-69)(8,-69)(17,-57)
\qbezier(0,-69)(11,-69)(13,-68)\qbezier(0,-51)(30,-51)(18,-65)
\put(6,-69){\vector(1,0){2}}
\put(32,-69){\vector(-1,0){10}}
\put(2,-51){\vector(-1,0){4}}
\put(25,-51){\vector(3,1){10}}

\put(5,-95){\mbox{$\tilde\lock_{R_1\times R_2}^Y$}}
}

\put(-10,-35){\mbox{$=$}}

\put(20,-35){\mbox{$\overline{\mu_Y\cdot \lock_{R_1\times R_2}^Y}$}}

\put(50,60){
\qbezier(27,-69)(22,-69)(20,-68) \qbezier(25,-51)(8,-55)(17,-66)
\qbezier(0,-69)(35,-68)(20,-56)\qbezier(2,-51)(10,-51)(14,-53)
\put(6,-69){\vector(-1,0){2}}
\put(36,-69){\vector(-1,0){10}}
\put(2,-51){\vector(1,0){4}}
\put(25,-51){\vector(3,1){10}}

\qbezier(0,-69)(-15,-69)(-20,-60)\qbezier(-20,-60)(-25,-51)(-40,-51)
\qbezier(2,-51)(-15,-51)(-18.5,-57)\qbezier(-21.5,-63)(-25,-69)(-40,-69)
\put(-40,-51){\vector(-1,0){10}}
\put(-50,-69){\vector(1,0){15}}
}



\put(-200,60){
\qbezier(25,-51)(24,-51)(22,-52)\qbezier(25,-69)(8,-69)(17,-57)
\qbezier(0,-69)(11,-69)(13,-68)\qbezier(0,-51)(30,-51)(18,-65)
\put(6,-69){\vector(-1,0){2}}
\put(32,-69){\vector(-1,0){10}}
\put(2,-51){\vector(1,0){4}}
\put(25,-51){\vector(3,1){10}}
\put(5,-95){\mbox{$ \lock_{R_1\times R_2}^Y$}}

{\footnotesize
\put(-10,-47){\mbox{$R_1$}}
\put(-10,-75){\mbox{$\bar R_1$}}
\put(-20,-63){\mbox{$Y$}}

\put(38,-47){\mbox{$\bar R_2$}}
\put(33,-78){\mbox{$R_2$}}
\put(50,-63){\mbox{$Y$}}
}
}

\end{picture}

\noindent
where, for the needs of further dealing with links, we consider two generally distinct representations, $R_1$ and $R_2$ and introduce the corresponding index notation $R_1\times R_2$, which is not to be confused with the tensor product $R_1\otimes R_2$. We also often use the subscript $R$ instead of $R\times R$.

In the arborescent calculus of \cite{arbor},
$\lock$ is represented by the simplest {\it fingers}:
\be\label{lockS}
\lock^Y_R ={D_R\over\sqrt{D_Y}} (\bar S \bar T^2\bar S)_{\emptyset Y},
\ \ \ \ \ \ \ \
\tilde\lock^Y_R ={D_R\over\sqrt{D_Y}} (S^\dagger T^2S)_{\emptyset Y}
\ee
Actually there are $4$ such fingers\footnote{If there are non-trivial multiplicities, there are 8 different fingers, since
the reflection w.r.t. to the vertical axis would produce another finger, in this case.}, differing by directions of arrows
(and thus by the choice between ${\cal R}$ and $\tilde{\cal R}$)
and by the choice between ${\cal R}$ (black dot)
and ${\cal R}^{-1}$ (white dot):

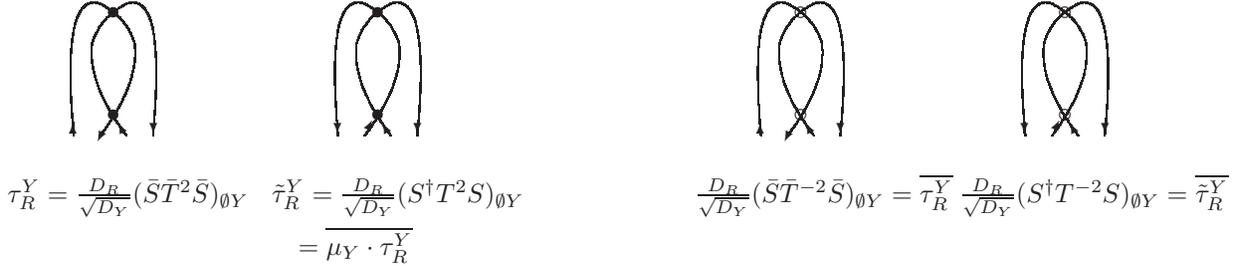
\begin{picture}(300,200)(-150,-100)

\put(-120,0){
\qbezier(5,0)(20,20)(18,35)
\qbezier(18,35)(15,45)(5,50)
\qbezier(5,50)(-10,55)(-5,0)
\qbezier(15,0)(0,20)(2,35)
\qbezier(2,35)(5,45)(15,50)
\qbezier(15,50)(30,55)(25,0)
\put(-5.3,3){\vector(0,1){2}}
\put(25.1,3){\vector(0,-1){2}}
\put(6,2){\vector(-1,-2){2}}
\put(14,1){\vector(-1,2){2}}
\put(10,8){\circle*{4}}
\put(10,47){\circle*{4}}
\put(-30,-25){\mbox{$\lock_R^Y ={D_R\over\sqrt{D_Y}} (\bar S\bar T^2\bar S)_{\emptyset Y}$}}
}

\put(-20,0){
\qbezier(5,0)(20,20)(18,35)
\qbezier(18,35)(15,45)(5,50)
\qbezier(5,50)(-10,55)(-5,0)
\qbezier(15,0)(0,20)(2,35)
\qbezier(2,35)(5,45)(15,50)
\qbezier(15,50)(30,55)(25,0)
\put(-5.3,3){\vector(0,-1){2}}
\put(25.1,3){\vector(0,-1){2}}
\put(6,2){\vector(1,2){2}}
\put(14,1){\vector(-1,2){2}}
\put(10,8){\circle*{4}}
\put(10,47){\circle*{4}}
\put(-30,-25){\mbox{$\tilde\lock_R^Y = {D_R\over\sqrt{D_Y}}(S^\dagger T^2 S)_{\emptyset Y}$}}
\put(-20,-45){\mbox{$=\overline{\mu_Y\cdot \lock_R^Y}$}}
}

\put(140,0){
\qbezier(5,0)(20,20)(18,35)
\qbezier(18,35)(15,45)(5,50)
\qbezier(5,50)(-10,55)(-5,0)
\qbezier(15,0)(0,20)(2,35)
\qbezier(2,35)(5,45)(15,50)
\qbezier(15,50)(30,55)(25,0)
\put(-5.3,3){\vector(0,1){2}}
\put(25.1,3){\vector(0,-1){2}}
\put(6,2){\vector(-1,-2){2}}
\put(14,1){\vector(-1,2){2}}
\put(10,8){\circle{4}}
\put(10,47){\circle{4}}
\put(-30,-25){\mbox{${D_R\over\sqrt{D_Y}}(\bar S\bar T^{-2}\bar S)_{\emptyset Y}=\overline{\lock_R^Y}$}}
}

\put(240,0){
\qbezier(5,0)(20,20)(18,35)
\qbezier(18,35)(15,45)(5,50)
\qbezier(5,50)(-10,55)(-5,0)
\qbezier(15,0)(0,20)(2,35)
\qbezier(2,35)(5,45)(15,50)
\qbezier(15,50)(30,55)(25,0)
\put(-5.3,3){\vector(0,-1){2}}
\put(25.1,3){\vector(0,-1){2}}
\put(6,2){\vector(1,2){2}}
\put(14,1){\vector(-1,2){2}}
\put(10,8){\circle{4}}
\put(10,47){\circle{4}}
\put(-30,-25){\mbox{$ {D_R\over\sqrt{D_Y}}(S^\dagger T^{-2} S)_{\emptyset Y}=\overline{\tilde\lock_R^Y} $}}
}

\end{picture}

\noindent
The Racah matrices $S$ and $\bar S$  depend on $R$,
thus these $\lock^Y_R$ are also $R$-dependent.
For a description of $S$ and $\bar S$ with different $R_1$ and $R_2$ needed for evaluating $\lock_{R_1\times R_2}^Y$,
see \cite{Sdifreps}.

\subsection{Double braids and Hopf links: the cross-switching element ${\cal B}$\label{sB}}

Consider now the following pattern:

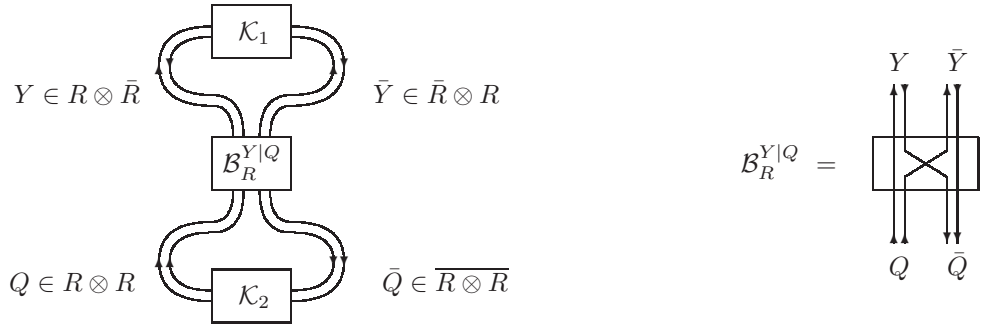
\begin{picture}(300,150)(-100,-70)

\put(0,50){
\put(0,0){\line(1,0){30}}
\put(0,0){\line(0,1){20}}
\put(0,20){\line(1,0){30}}
\put(30,0){\line(0,1){20}}
\put(10,7){\mbox{${\cal K}_1$}}
}

\put(0,0){
\put(0,0){\line(1,0){30}}
\put(0,0){\line(0,1){20}}
\put(0,20){\line(1,0){30}}
\put(30,0){\line(0,1){20}}
\put(4,7){\mbox{${\cal B}_R^{Y|Q}$}}
}

\put(0,-50){
\put(0,0){\line(1,0){30}}
\put(0,0){\line(0,1){20}}
\put(0,20){\line(1,0){30}}
\put(30,0){\line(0,1){20}}
\put(10,7){\mbox{${\cal K}_2$}}
}

\qbezier(0,62)(-20,62)(-20,47)
\qbezier(0,32)(-20,32)(-20,47)
\qbezier(0,32)(8,32)(8,20)
\put(-20,46){\vector(0,1){2}}
\put(-16,48){\vector(0,-1){2}}

\qbezier(0,58)(-16,58)(-16,47)
\qbezier(0,36)(-16,36)(-16,47)
\qbezier(0,36)(12,36)(12,20)

\put(0,20){
\qbezier(0,-62)(-20,-62)(-20,-47)
\qbezier(0,-32)(-20,-32)(-20,-47)
\qbezier(0,-32)(8,-32)(8,-20)
\put(-20,-47){\vector(0,1){2}}
\put(-16,-47){\vector(0,1){2}}

\qbezier(0,-58)(-16,-58)(-16,-47)
\qbezier(0,-36)(-16,-36)(-16,-47)
\qbezier(0,-36)(12,-36)(12,-20)
}

\put(30,0){
\qbezier(0,62)(20,62)(20,47)
\qbezier(0,32)(20,32)(20,47)
\qbezier(0,32)(-8,32)(-8,20)
\put(20,48){\vector(0,-1){2}}
\put(16,46){\vector(0,1){2}}

\qbezier(0,58)(16,58)(16,47)
\qbezier(0,36)(16,36)(16,47)
\qbezier(0,36)(-12,36)(-12,20)
}

\put(30,20){
\qbezier(0,-62)(20,-62)(20,-47)
\qbezier(0,-32)(20,-32)(20,-47)
\qbezier(0,-32)(-8,-32)(-8,-20)
\put(20,-47){\vector(0,-1){2}}
\put(16,-47){\vector(0,-1){2}}

\qbezier(0,-58)(16,-58)(16,-47)
\qbezier(0,-36)(16,-36)(16,-47)
\qbezier(0,-36)(-12,-36)(-12,-20)
}

\put(-75,33){\mbox{$Y\in R\otimes \bar R$}}
\put(61,33){\mbox{$\bar Y\in \bar R\otimes  R$   }}
\put(-80,-38){\mbox{ $Q\in R\otimes   R$}}
\put(61,-38){\mbox{ $\bar Q\in \overline{R\otimes   R}$}}

\put(250,0){
\put(0,0){\line(1,0){40}}
\put(0,0){\line(0,1){20}}
\put(0,20){\line(1,0){40}}
\put(40,0){\line(0,1){20}}

\put(8,-20){\vector(0,1){60}}
\put(32,40){\vector(0,-1){60}}
\put(12,-20){\line(0,1){25}}
\put(28,-20){\line(0,1){25}}
\put(12,40){\line(0,-1){25}}
\put(28,40){\line(0,-1){25}}
\qbezier(12,5)(20,10)(28,15)
\qbezier(28,5)(20,10)(12,15)

\put(8,-18){\vector(0,1){2}}
\put(12,-18){\vector(0,1){2}}
\put(28,-18){\vector(0,-1){2}}

\put(12,38){\vector(0,-1){2}}
\put(28,38){\vector(0,1){2}}
\put(32,38){\vector(0,-1){2}}

\put(6,-32){\mbox{$Q$}}
\put(28,-32){\mbox{$\bar Q$}}

\put(6,45){\mbox{$Y$}}
\put(28,45){\mbox{$\bar Y$}}

\put(-50,7){\mbox{${\cal B}_R^{Y|Q}\ = $}}
}

\end{picture}

\noindent
The left picture was associated with the double braid in \cite{evo} and the switching block ${\cal B}$ is another important tangle, which allows one to evaluate various knots and links, not only double-braids, see examples in s.4. In the arborescent calculus \cite{arbor},
\be\label{sb1}
{\cal B}_R^{Y|Q} = {D_R\over\sqrt{D_YD_Q}}(\bar S \bar T S)_{Y|Q}
\ee
This parallel-antiparallel switching block ${\cal B}$
is also going to play a prominent role in describing the modular properties
of effective topological theory of knots and links in \cite{topfromknots}.

\subsection{Digression: switching functor}

A more sophisticated way of switching between representations $Q$ in the parallel channel and $Y$ in the antiparallel can be provided by a functor:

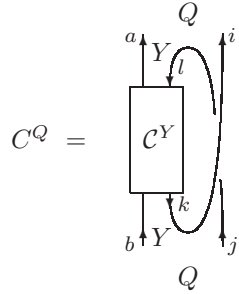
\begin{picture}(300,125)(-150,-40)

\put(0,0){\line(1,0){20}}
\put(0,0){\line(0,1){40}}
\put(0,40){\line(1,0){20}}
\put(20,0){\line(0,1){40}}

\put(5,0){\line(0,-1){20}}
\put(5,40){\vector(0,1){20}}

\qbezier(15,0)(15,-15)(22,-15)
\qbezier(22,-15)(35,-15)(35,60)
\qbezier(15,40)(15,55)(22,55)
\qbezier(22,55)(30,55)(32,30)
\qbezier(34,5)(35,2)(35,-20)

\put(5,-15){\vector(0,1){2}}
\put(35,-15){\vector(0,1){2}}
\put(35,58){\vector(0,1){2}}
\put(15,42){\vector(0,-1){2}}
\put(15,-4){\vector(0,-1){2}}

\put(8,-20){\mbox{$Y$}}
\put(8,50){\mbox{$Y$}}

\put(18,-35){\mbox{$Q$}}
\put(18,65){\mbox{$Q$}}

\put(5,17){\mbox{${\cal C}^ Y$}}

\put(-45,17){\mbox{${C}^Q\ =$}}

{\footnotesize
\put(-2,-22){\mbox{$b$}}
\put(37,-22){\mbox{$j$}}
\put(-2,57){\mbox{$a$}}
\put(37,57){\mbox{$i$}}
\put(18,-7){\mbox{$k$}}
\put(18,45){\mbox{$l$}}
}
\end{picture}

\noindent
It "interrupts" the motion along the right line
by twisting it upside-down around the left one.
In indices,
\be\label{F1}
{\cal D}: \ \ \ \
C_{bj}^{ai} =  \sum_{k,l} {\cal C}_{bl}^{ak}\cdot {\cal R}_{kj}^{li}
\ee
but we rather need it in the Tanaka-Krein representation:
\be
C^Q_{R_1\times R_2} = \sum_Y {\cal D}^Q_Y\cdot {\cal C}^Y_{R_1\times\bar R_2}
\ee
${\cal D}$ here depends on the two incoming representations $R_1$ and $R_2$,
we suppressed them in the notation.
The orientation in the picture appeared relevant in the calculations in \cite{SW}.
There is also an analogue with the arrows in the right part reversed, i.e.
with $R_2$ substituted by conjugate $\bar R_2$.

In this example, ${\cal D}$ is actually a functor connecting two different
realizations of "the same" tangle block ${\cal C}$.
In fact, every tangle block can be embedded into
and interpreted as an element of ${\rm Hom}({\rm tangles})$,
and this is actually the proper way to look at the tangle calculus.
In this paper, we, however, formulate things in a more traditional way.

\subsection{Whitehead block}

Another important block is the Whitehead block:

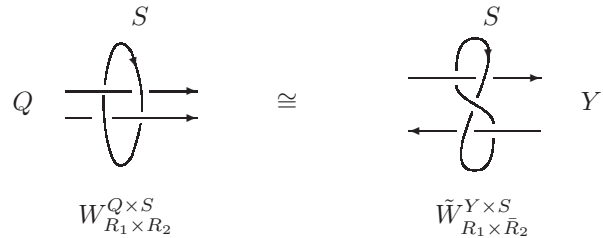
\begin{picture}(300,110)(-150,-60)

\put(0,5){\line(1,0){25}}
\put(32,5){\vector(1,0){18}}
\put(10,-5){\line(-1,0){10}}
\put(18,-5){\vector(1,0){32}}

\qbezier(29,-2.5)(30,6)(25,20)
\qbezier(28.5,-7.5)(27,-15)(25,-20)
\qbezier(25,-20)(18,-30)(15,-8)
\qbezier(25,20)(18,30)(15,8)
\qbezier(15,-8)(14.2,0)(14.5,3)

\put(-20,-2){\mbox{$Q$}}
\put(25,30){\mbox{$S$}}
\put(25,20){\vector(1,-3){2}}

\put(5,-45){\mbox{$W^{Q\times S}_{R_1\times R_2 }$}}

\put(80,-2){\mbox{$\cong$}}

\put(130,0){
\put(0,10){\line(1,0){25}}
\put(32,10){\vector(1,0){18}}
\put(50,-10){\line(-1,0){25}}
\put(18,-10){\vector(-1,0){18}}

\qbezier(18,7)(18,4)(25,0)
\qbezier(32,-7)(32,-4)(25,0)

\qbezier(18,13)(18,25)(25,25)
\qbezier(25,25)(35,25)(26,3)
\qbezier(32,-13)(32,-25)(25,-25)
\qbezier(25,-25)(15,-25)(24,-3)

\put(65,-2){\mbox{$Y$}}
\put(28,30){\mbox{$S$}}
\put(30,19){\vector(0,-1){2}}

\put(10,-45){\mbox{$\tilde W^{Y\times S}_{R_1\times\bar R_2 }$}}

}

\end{picture}

\noindent
Here $Q$ belongs to the decomposition of the product $R_1\otimes R_2$ and $Y$, to the decomposition of the product $R_1\otimes\bar R_2$.

If we contract the horizontal lines in $\tilde W$ with each other, we get the Whitehead link $L_{5a1}$, hence the name.
If we contract the  r.h.s. ends of the lines of $\tilde W$ with their l.h.s. ends,
we get the composite open chain.
The two emerging relations are
\be\label{Wtilde}
D_R\tilde W_{R\times \bar R}^{\emptyset\times S} = {\cal H}_{R\times S}^{L_{5a1}} \nn \\
\sum_Y D_Y\tilde W_{R_1\times \bar R_2}^{Y\times S} =
D_SD_{R_1}D_{R_2}H^{\rm Hopf}_{R_1\times S }\overline{H_{R_2\times S}^{\rm Hopf}}
\ee
Similar relations for $W^Q$ are:
\be\label{W1}
\sum_Q D_Q W^{Q\times S}_{R_1\times R_2} = D_SD_{R_1}D_{R_2} H^{\rm Hopf}_{R_1\times S}
\overline{H^{\rm Hopf}_{R_2\times S}}
\ee
when we contract the  r.h.s. ends of the lines with their l.h.s. ends, producing the composite open chain;
\be\label{W2}
\sum_Q \mu_Q  D_Q W^{Q\times S}_{R\times R} = {\cal H}^{L_{5a1}}_{R\times S}\\
\sum_Q \mu_Q^{-1}  D_Q W^{Q\times S}_{R\times R} = D_SD_R
\label{W3}
\ee
when we contract the  r.h.s. ends of the lines with their permuted l.h.s. ends, producing the same Whitehead or two unknots depending on the way of crossing when permuting; and
\be\label{W4}
D_R W^{\emptyset\times S}_{\bar R\times R} = {\cal \H}^{L_{4a1}}_{R\times S}
\ee
when we contract the horizontal lines with each other, with changing the direction of one of the arrows (i.e. considering $R_2=\bar R_1$), producing ``Solomon's knot'' (link $L4a1$), which is, in fact, the torus link ${\rm tor}_{[2,4]}$.

$W^Q$ and $\tilde W^Y$ are related by the switching block ${\cal B}$,
\be\label{sb2}
\sum_Q D_Q{\cal B}_R^{Y|Q}W_R^{Q\times S}=\tilde W_R^{Y\times S}
\ee

\subsection{Families of cut-and-join blocks\label{cj}}

The $\lock$-block is the simplest in the following family $B(m)$:

\begin{picture}(300,90)(-150,-30)

\put(0,0){\line(1,0){12}}
\put(0,0){\line(0,1){30}}
\put(12,0){\line(0,1){30}}
\put(0,30){\line(1,0){12}}

\put(3,0){\line(-1,-2){10}}
\put(9,0){\line(1,-2){10}}
\put(3,30){\line(-1,2){10}}
\put(9,30){\line(1,2){10}}

\put(2,13){\mbox{$m$}}

\put(-20,13){\mbox{$Y$}}

\put(30,10){
\put(50,35){\mbox{$\lock^Y = B(\bar 2)$}}
\put(50,13){\mbox{$\tilde\lock^Y = B(2)$}}
\put(50,-9){\mbox{$\overline{\lock^Y} = B(-\bar 2)$}}
\put(50,-31){\mbox{$\overline{\tilde\lock^Y} = B(-2)$}}
}

\end{picture}

\noindent
where the vertical box denotes the vertical 2-braid
of length $m$, while $\bar m$ refers to the antiparallel strands.
These blocks are of course very simple
(2-strand torus) cable blocks, if considered
in the vertical channel, but get less trivial in the horizontal one,
and $\lock$ is the simplest of those.

\bigskip

The next important generalization is $B(m,m')$

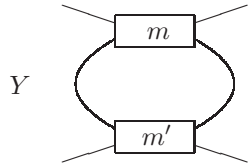
\begin{picture}(300,80)(-150,-50)

\put(0,0){\line(1,0){30}}
\put(0,0){\line(0,1){12}}
\put(30,0){\line(0,1){12}}
\put(0,12){\line(1,0){30}}

\put(0,-40){
\put(0,0){\line(1,0){30}}
\put(0,0){\line(0,1){12}}
\put(30,0){\line(0,1){12}}
\put(0,12){\line(1,0){30}}
}

\put(0,9){\line(-3,1){20}}
\put(0,-37){\line(-3,-1){20}}
\put(30,9){\line(3,1){20}}
\put(30,-37){\line(3,-1){20}}

\qbezier(0,3)(-30,-14)(0,-31)
\qbezier(30,3)(60,-14)(30,-31)

\put(12,3){\mbox{$m$}}
\put(10,-37){\mbox{$m'$}}

\put(-40,-18){\mbox{$Y$}}


\end{picture}

\noindent
with two horizontal 2-strand braids of lengths $m$ and $m'$.
Within this family, the four types of the $\lock$-block correspond
to $B(\pm 1,\mp 1)$ and $B(\pm\bar 1,\mp\bar 1)$.
For $m,m'=2$, one gets another important quadruplet:

\begin{picture}(400,110)(-200,-60)

\put(-200,0){
\put(0,5){\line(1,0){25}}
\put(32,5){\vector(1,0){18}}
\put(25,-5){\vector(-1,0){25}}
\put(32,-5){\line(1,0){18}}

\qbezier(25,-20)(32,0)(25,20)
\qbezier(25,-20)(18,-30)(15,-8)
\qbezier(25,20)(18,30)(15,8)
\qbezier(14.5,-3)(14.2,0)(14.5,3)

\put(-20,-2){\mbox{$Y$}}
\put(25,30){\mbox{$S$}}
\put(25,20){\vector(1,-4){2}}

{\footnotesize
\put(-30,-45){\mbox{$B(2,-2)=H^{\rm Hopf}_{Y\times S}$}}
}

}

\put(-100,0){
\put(0,5){\line(1,0){25}}
\put(32,5){\vector(1,0){18}}
\put(25,-5){\line(-1,0){25}}
\put(32,-5){\vector(1,0){18}}

\qbezier(25,-20)(32,0)(25,20)
\qbezier(25,-20)(18,-30)(15,-8)
\qbezier(25,20)(18,30)(15,8)
\qbezier(14.5,-3)(14.2,0)(14.5,3)

\put(-20,-2){\mbox{$Q$}}
\put(25,30){\mbox{$S$}}
\put(25,20){\vector(1,-4){2}}

{\footnotesize
\put(-20,-45){\mbox{$B(2,-\bar 2)=H^{\rm Hopf}_{Q\times S}$}}
}
}

\put(20,0){
\put(0,5){\line(1,0){25}}
\put(32,5){\vector(1,0){18}}
\put(10,-5){\vector(-1,0){10}}
\put(18,-5){\line(1,0){32}}

\qbezier(28.5,-2.5)(30,6)(25,20)
\qbezier(28,-7.5)(27,-15)(25,-20)
\qbezier(25,-20)(18,-30)(15,-8)
\qbezier(25,20)(18,30)(15,8)
\qbezier(15,-8)(14.2,0)(14.5,3)

\put(-20,-2){\mbox{$Y$}}
\put(25,30){\mbox{$K$}}
\put(25,20){\vector(1,-4){2}}

{\footnotesize
\put(-10,-45){\mbox{$B(2,2)=T^{Y\times S}_{R_1\times\bar R_2 }$}}
}
}

\put(120,0){
\put(0,5){\line(1,0){25}}
\put(32,5){\vector(1,0){18}}
\put(10,-5){\line(-1,0){10}}
\put(18,-5){\vector(1,0){32}}

\qbezier(29,-2.5)(30,6)(25,20)
\qbezier(28.5,-7.5)(27,-15)(25,-20)
\qbezier(25,-20)(18,-30)(15,-8)
\qbezier(25,20)(18,30)(15,8)
\qbezier(15,-8)(14.2,0)(14.5,3)

\put(-20,-2){\mbox{$Q$}}
\put(25,30){\mbox{$S$}}
\put(25,20){\vector(1,-3){2}}

{\footnotesize
\put(-15,-45){\mbox{$B(2,\bar 2)=W^{Q\times S}_{R_1\times R_2}$}}
}

\put(60,-2){\mbox{$\cong$}}

\put(76,0){
\put(0,10){\line(1,0){25}}
\put(32,10){\vector(1,0){18}}
\put(50,-10){\line(-1,0){25}}
\put(18,-10){\vector(-1,0){18}}

\qbezier(18,7)(18,4)(25,0)
\qbezier(32,-7)(32,-4)(25,0)

\qbezier(18,13)(18,25)(25,25)
\qbezier(25,25)(35,25)(26,3)
\qbezier(32,-13)(32,-25)(25,-25)
\qbezier(25,-25)(15,-25)(24,-3)

\put(65,-2){\mbox{$Y$}}
\put(28,30){\mbox{$S$}}
\put(30,19){\vector(0,-1){2}}

{\footnotesize
\put(10,-45){\mbox{$\tilde W^{Y\times S}_{R_1\times\bar R_2 }$}}
}

}
}

\end{picture}

\noindent
In the last case we added another avatar of the same block
related by the switch block.
Of course, such pairs exist for the other three blocks in the picture.
Hopf, as a cable block, does not actually depend on the representations $R_1$, $R_2$ living on lines.

\subsection{Hopf and other torus  blocks: the simplest among the cable blocks\label{cabH}}

By definition the cable blocks are obtained by cutting a cabled knot or a cabled link at one place,
\be
{\cal CB}^{\cal L}_{R,R'} = \frac{{\cal H}^{\cal L}_R}{D_R}\cdot\delta_{R,R'}
\ee
The vacuum block is the simplest example: when ${\cal L}={\rm unknot}$.
If we substitute a single wire by a cable of $p$ wires,
we get a family of new links, which is the $p$-satellite ${\cal S}_k^{(p)}({\cal L})$
of the original ${\cal L}$, where $k$ refers to some twisting of $p$ strands
(when $p=2$ there is no ambiguity in the word "twisting", for $p>2$ one can
consider different types of twists actually labeled by a $p$-strand braid).
The basic formula is then
\be
{\cal H}^{{\cal S}_k^{(p)}({\cal L})}_R
\ \ = \!\!\!\sum_{Q\in R_1\otimes\ldots\otimes R_p} {\cal H}_Q^{\cal L}
\ \ = \!\!\!\sum_{Q\in R_1\otimes\ldots\otimes R_p} D_Q {\cal B}^{\cal L}_{Q,Q}
\ee

The simplest of the cabling blocks is of course the {\it torus} one,
which is exhaustively described by the celebrated
Rosso-Jones formula \cite{RJ,Ch}:
then ${\cal H}_Q^{\rm torus}$ and thus ${\cal B}_Q^{\rm torus}$
can be either extracted  as a component of the sum
\be
{\cal H}^{{\rm torus}_{[m,n]}}_R
= \sum_{Q\in R^{\otimes m}}   D_Q\cdot c_{R,Q}\cdot\lambda_Q^{\frac{2n}{m}}
\ee
or calculated directly:
\be\label{RJ}
{\cal H}^{{\rm torus}_{[m,n]}}_Q
= \sum_{W\in Q^{\otimes m}}  c_{Q,W}\cdot D_W\cdot \lambda_W^{\frac{2n}{m}}
\ee

If some strands are antiparallel, then the representations $Q$ belong to the product
of $R$ and its conjugate $\bar R$, which explicitly depends of $N$ that parameterizes
the gauge group $SL(N)$ (in fact, through Vogel's universality\footnote{
Vogel's universality conjectures that {\it some} formulas from group theory
(like dimensions and Racah matrices in {\it some} representations)
are the same for all simple groups: obtained by substituting particular
group dependent values to a triple of parameters $u,v,w$.
The original conjecture \cite{Vogel} was about the dimensions of representations from the
adjoint tower ($E_8$-sector), but it was soon realized to be only partly true:
some universal formulas for dimensions are at best irrational.
However, the conjecture remains fully true in application to knot theory:
the {\it combinations} of dimensions that appear in knot invariants
are not quite arbitrary, and they {\it are} universal \cite{MMkrM}.
Moreover, the statement is straightforwardly extended to the corresponding
Racah matrices (6j-symbols) and, thus, at least to the entire arborescent calculus
\cite{MMuniv}.
}
of the adjoint sector
\cite{Vogel,MMkrM}, one can relate invariants for other groups to the HOMFLY invariants for $SL(N)$).
Then, to obtain the knot polynomial that depends on $A=q^N$, one needs to make
calculations for a particular integer $N$ and then analytically continue to arbitrary $A$
(continuation is unambiguous because the HOMFLY invariants are rational functions, or just
{\it polynomials}).
This is a somewhat tedious procedure, moreover, it is rarely doable in practice,
because the needed representations are not symmetric, thus very few knot polynomials
are known even in the adjoint representations (see \cite{MMkrM} and \cite{MMuniv} for
adjoint arborescent calculus, which is developed there only for the first representation
from adjoint tower, i.e. for the adjoint representation itself).

In fact, the tangle block calculus provides a breakthrough in this problem \cite{sat}:
a cable block for ${\cal L}$ for a 2-antiparallel wire is not just closed,
but convoluted with the lock element $\lock$, it gives rise to
the twist satellite $S_k({\cal L})$ (instead of the ordinary
${\cal S}_k^{(1,\bar 1)}({\cal L})$), which is the Whitehead double
described by eq.(\ref{SatH}).
This equation can be regarded as an expression for ${\cal H}^{S_k({\cal L})}_R$
for the Whitehead double $R$-colored HOMFLY polynomial through those for the original ${\cal L}$ in $R\times\bar R$,
but it is possible to use it also in the opposite direction:
as expressing the original adjoint polynomials through the fundamental ones (for $R=[1]$)
of the twist satellite.
Since any fundamental polynomial is straightforwardly calculable at least by the
skein-relation method, (\ref{SatH}) is a straightforward method for adjoint calculations.
Higher representations from the adjoint tower (Vogel' universal $E_8$-sector) require
higher symmetric representations $R=[r]$, which are also straightforward for the ordinary cabling method
of \cite{AnoMcabling}.

As the simplest illustration, we consider an example of the Hopf link in the
role of ${\cal L}$: then, (\ref{SatH}) for the fundamental $R=[1]$ becomes
\be
{\cal H}_{[1]\times S}^{S_{2k}({\rm Hopf})}
= \tau^\emptyset_{[1]}D_{[1]} + D_{\rm adj} \cdot\tau^{\rm adj}_{[1]}\cdot A^{2k}\cdot
{\cal H}^{\rm Hopf}_{{\rm adj}\times S}
\label{SatHopf}
\ee
Twisting $2k$ can be only even,
$S=[s]$ here is the representation on the second circle in the Hopf link,
which we do not cut. For $k=0$, the satellite link is nothing but the Whitehead link $L_{5a1}$ from
the table in \cite{katlas}, while, at $k=2$ and $k=-2$, these are links $L_{7n2}$ and $L_{8n2}$ accordingly. We describe them in detail in s.\ref{4.1.1}.

\subsection{The trident block}

The number of lines in the cuts can be arbitrary and not even the same.
As an illustration, let us cut a knot by one and by three lines:

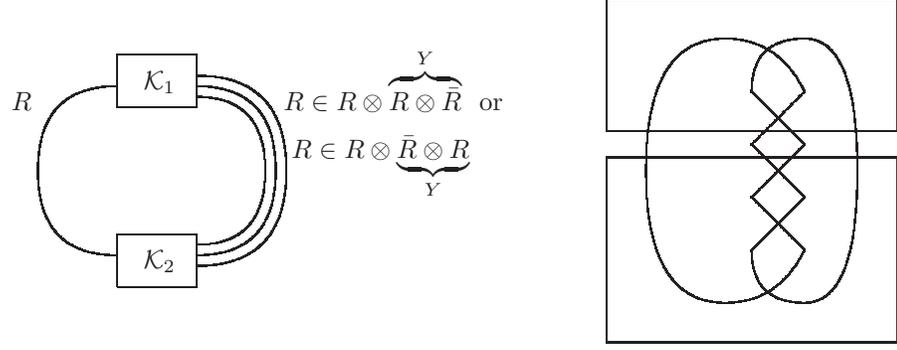
\begin{picture}(300,130)(-150,-65)

\put(0,34){
\put(0,0){\line(1,0){30}}
\put(0,0){\line(0,1){20}}
\put(0,20){\line(1,0){30}}
\put(30,0){\line(0,1){20}}
\put(10,7){\mbox{${\cal K}_1$}}
}

\put(0,-34){
\put(0,0){\line(1,0){30}}
\put(0,0){\line(0,1){20}}
\put(0,20){\line(1,0){30}}
\put(30,0){\line(0,1){20}}
\put(10,7){\mbox{${\cal K}_2$}}
}

\qbezier(0,42)(-30,42)(-30,10)
\qbezier(0,-22)(-30,-22)(-30,10)

\qbezier(30,46)(64,46)(64,10)
\qbezier(30,-26)(64,-26)(64,10)

\qbezier(30,42)(60,42)(60,10)
\qbezier(30,-22)(60,-22)(60,10)


\qbezier(30,38)(56,38)(56,10)
\qbezier(30,-18)(56,-18)(56,10)

\put(-40,33){\mbox{$R $}}
\put(63,33){\mbox{$R\in R\otimes \overbrace{R\otimes \bar R}^Y$\ \ {\rm or} }}

\put(66,15){\mbox{$R\in R\otimes \underbrace{\bar R\otimes   R}_Y$}}

\put(200,10){
\qbezier(0,0)(0,50)(30,50)\qbezier(0,0)(0,-50)(30,-50)
\qbezier(60,30)(50,50)(30,50)\qbezier(60,-30)(50,-50)(30,-50)

\qbezier(40,30)(40,50)(60,50)\qbezier(40,-30)(40,-50)(60,-50)
\qbezier(80,0)(80,50)(60,50)\qbezier(80,0)(80,-50)(60,-50)

\qbezier(40,-30)(50,-20)(60,-10)
\qbezier(40,10)(50,-0)(60,-10)
\qbezier(40,10)(50,20)(60,30)

\qbezier(40,30)(50,20)(60,10)
\qbezier(40,-10)(50,-0)(60,10)
\qbezier(40,-10)(50,-20)(60,-30)

\put(-15,15){\line(1,0){110}}
\put(-15,65){\line(1,0){110}}
\put(-15,15){\line(0,1){50}}
\put(95,15){\line(0,1){50}}

\put(-15,5){\line(1,0){110}}
\put(-15,-65){\line(1,0){110}}
\put(-15,5){\line(0,-1){70}}
\put(95,5){\line(0,-1){70}}
}

\end{picture}

Representation in the three-line cut is still $R\in R\otimes R\otimes \bar R$,
but at the r.h.s. it appears with some multiplicity. Thus, the operator corresponding to such an element, which we will call the trident block, is a rectangular rather than a square matrix acting from a space $R$ to the space $R\otimes R\otimes \bar R$. Its size is $1\times($multiplicity of $R\in R\otimes R\otimes \bar R)$. Therefore, the polynomial of the whole knot is provided by the product of the trident block and a conjugate trident block:
\begin{equation}
\left.{\cal H}_{R}^{\cal K} =\text{Tr}\ {\cal T}^{{\cal K}_1}_{R,R\otimes Y}
\overline{{\cal T}^{{\cal K}_2}_{R,R\otimes Y}}\ \right|_{{Y\in R\otimes \bar R}\atop{R\in R\otimes Y}}
\end{equation}
where the bar over the second component implies transposition of the trident block.

\begin{figure}[h!]
\begin{picture}(300,110)(-120,-60)




\put(0,0){
\put(-15,-10){\vector(0,1){40}}
\put(15,0){
\qbezier(-15,-10)(-15,10)(0,10)
\qbezier(15,-10)(15,10)(0,10)
}
\put(-20,-10){\line(1,0){25}}
\put(-20,-10){\line(0,-1){15}}
\put(-20,-25){\line(1,0){25}}
\put(5,-25){\line(0,1){15}}
\put(-15,-40){\vector(0,1){15}}
\put(0,-40){\vector(0,1){15}}
\put(30,-10){\vector(0,-1){30}}

{\footnotesize
\put(-14,-20){\mbox{${\cal R}^{k}$}}
\put(-8,29){\mbox{$R$}}
\put(-27,-45){\mbox{$R$}}
\put(-8,-45){\mbox{$R$}}
\put(20,-45){\mbox{$\bar{R}$}}
}

}

\put(100,0){
\put(-15,-10){\vector(0,1){40}}
\put(15,0){
\qbezier(-15,-10)(-15,10)(0,10)
\qbezier(15,-10)(15,10)(0,10)
}
\put(-15,-40){\vector(0,1){30}}
\put(0,-40){\vector(0,1){30}}
\put(30,-10){\vector(0,-1){30}}
\multiput(15,10)(0,2){10}
{\line(0,1){1}}
{\footnotesize
\put(-8,29){\mbox{$R$}}
\put(15,29){\mbox{$Y=\emptyset$}}
\put(-27,-45){\mbox{$R$}}
\put(-8,-45){\mbox{$R$}}
\put(20,-45){\mbox{$\bar{R}$}}
}

}


\end{picture}
\caption{\label{tribl} Specific trident block $\mathcal{T}^{(k)}$, where there is just a two strand braid with $k$ crossings. On the right, the trivial trident block $\mathcal{T}^{(0)}\sim\delta_{Y,\emptyset}$ is shown.}
\end{figure}
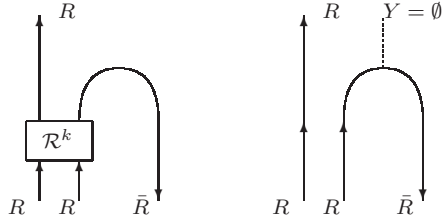

On the r.h.s. of the picture, we show how this applies to the $2$-strand knot $5_1={\rm Torus}_{[2,5]}$. One can cut any $2$-strand link diagram into two pieces of lengths $k_1+k_2=k$. Each part then corresponds to a specific trident block (see Fig.\ref{tribl}) with $k_1$ and $k_2$ crossings correspondingly. Now one can combine the trident pieces ${\cal T}^{(k)}$ in different ways.
Various combinations
\begin{equation}
{\cal H}^{[2,k]} =\text{Tr}\ {\cal T}_{R|Y}^{(k_1)}\cdot \overline{{\cal T}_{R|Y}^{(k_2)}}
\end{equation}
provide a lot of relations. In particular, since $\mathcal{T}^{(0)}_{R|Y}\sim\delta_{Y,\emptyset}$,
$\mathcal{H}^{[2,k]}= {\cal T}^{(k)}_{R|Y}\overline{\mathcal{T}_{R|Y}^{(0)}} =\mathcal{T}^{(k)}_{R|\emptyset}$,
while ${\cal T}_R^{(k)} \overline{{\cal T}_R^{(-k)}}$
is just a product of two unknots.

\section{Calculating simplest building blocks}

\subsection{Simple constructions with $\lock$\label{3lock}}

Let us demonstrate how one can calculate tangle blocks from known knot/link invariants, without manifest usage of formulas through the Racah matrices like (\ref{lockS}). We start with the simplest example. According to our logic,
we expect  that
\be
\lock_{[1]}^\emptyset = \frac{1}{D_{[1]}}\cdot {\cal \H}^{{\rm tor}_{[2,2]}}_{[1]}
= D_{[1]}\cdot{\H}^{{\rm tor}_{[2,2]}}_{[1]} =A^2D_{[1]}{H}^{{\rm tor}_{[2,2]}}_{[1]}\nn\\
\lock_{[1]}^\emptyset + D_{\rm adj}
\cdot\lock_{[1]}^{\rm adj}\cdot A^{2k}
= D_{[1]}{H}^{{\rm twist}_k}_{[1]}
= D_{[1]}\Big(1 + \{Aq\}\{A/q\}\frac{1-A^{2k}}{1-A^{-2}}\Big)
\ee
Since
\be
D_{\rm adj}=\frac{\{Aq\}\{A/q\}}{\{q\}^2},\ \ \ \ \ \ \ \
D_{[1]}=\frac{\{A\}}{\{q\}}
\ee
we obtain
\be
\lock_{[1]}^{\emptyset} =A\ {A^2 q^2-q^4+q^2-1\over\{q\}q^2},\ \ \ \ \ \ \ \lock_{[1]}^{\rm adj} =-A\ \{q\}
\ee
which is perfectly consistent with (\ref{lockS}). Similarly,
\be
\tilde\lock_{[1]}^\emptyset
={A^2q^4-A^2q^2+A^2-q^2\over A\{q\}q^2},\ \ \ \ \ \tilde\lock_{[1]}^{\rm adj} = -A^{-2}\  \{q\}
\ee
With these values of $\lock_{[1]}$ it is easy to check that, indeed,
\be
\left(\tilde\lock^{\emptyset}_{[1]}\right)^2 + D_{\rm adj}\cdot
\left(\tilde\lock^{\rm adj}_{[1]}\right)^2 =  \overline{{\cal \H}^{{\rm tor}_{[2,4]}}_{[1]}}
= D_{[1]}\cdot {A^2q^8-A^2q^6+A^2q^4-A^2q^2+A^2-q^6+q^4-q^2\over q^4A^5\{q\}}
\ee
and
\be
\left( \lock^{\emptyset}_{[1]}\right)^2 + D_{\rm adj}\cdot
\left( \lock^{\rm adj}_{[1]}\right)^2 =  {\cal H}^{\widetilde{\rm tor}_{[2,4]}}_{[1]}
\ \stackrel{\text{ \cite{evo}} }{=}\
 1+A^4\cdot \frac{ \{Aq\}\{A/q\}}{\{q\}^2}
\ee
where $\widetilde{\rm tor}_{[2,4]}$ is the torus link ${\rm tor}_{[2,4]}$
with inverse  orientation of one component.

All this continues to hold in higher representations.

\subsection{The Whitehead block\label{3W}}

One can use equations (\ref{Wtilde}) for the fundamental representations $R_1=R_2=[1]$ and $S=[1]$ in order to provide
the first Whitehead blocks. Since
\be
{\cal H}_{[1],[1]}^{L_{5a1}}=D_{[1]}^2\Big(1+ {\{Aq\}\{A/q\}\{q\}^2\over A\{A\}}\Big)\,
\ee
one obtains
\be
\tilde W_{[1]\times [\bar 1]}^{\emptyset\times [1]} = D_{[1]}\Big(1+ {\{Aq\}\{A/q\}\{q\}^2\over A\{A\}}\Big)
\ee
and
\be
\nn \\
\tilde W_{[1]\times [\bar 1]}^{{\rm adj}\times [1]} =
-{A^2q^6-q^8-A^2q^4+3q^6+A^2q^2-5q^4+3q^2-1\over Aq^4\{q\}}
\ee
Similarly for $W^Q$ from (\ref{W2})-(\ref{W3}) and taking into account that
\be
\mu_{[2]}={q\over A},\ \ \ \mu_{[1,1]}=-{1\over qA},\ \ \ \ D_{[2]}={\{A\}\{Aq\}\over \{q\}\{q^2\}},\ \ \ \ D_{[1,1]}=
{\{A\}\{A/q\}\over \{q\}\{q^2\}}
\ee
what is again sufficient in the case of $R_1=R_2=[1]$ and $S=[1]$, one has
\be
W_{[1]\times [1]}^{[2]\times [1]}={A^2q^4-q^6-A^2q^2+2q^4+A^2-2q^2\over Aq^2\{A\}},\ \ \ \ W_{[1]\times [1]}^{[1,1]\times [1]}=
{A^2q^6-A^2q^4+A^2q^2-2q^4+2q^2-1\over q^4A\{A\}}
\ee
Now one can directly check that (\ref{W1}) is correct. One can also obtain from (\ref{W4}) that
\be\label{Wc1}
W_{[\bar 1]\times [1]}^{\emptyset\times [1]}={1\over q^2A^3\{q\}} \Big(A^4q^2-A^2q^4+A^2q^2-q^4-A^2+2q^2-1\Big)
\ee
Similarly, from (\ref{W3}) it follows that
\be\label{Wc2}
W_{[\bar 1]\times [1]}^{{\rm adj}\times [1]}={A^2q^2-2q^4+3q^2-2\over Aq^2\{q\}}
\ee

\subsection{The switching block\label{sb}}

The simplest way to calculate the switching block

\begin{picture}(300,105)(-140,-40)

\put(0,0){\line(1,0){40}}
\put(0,0){\line(0,1){20}}
\put(0,20){\line(1,0){40}}
\put(40,0){\line(0,1){20}}

\put(8,-20){\vector(0,1){60}}
\put(32,40){\vector(0,-1){60}}
\put(12,-20){\line(0,1){25}}
\put(28,-20){\line(0,1){25}}
\put(12,40){\line(0,-1){25}}
\put(28,40){\line(0,-1){25}}
\qbezier(12,5)(20,10)(28,15)
\qbezier(28,5)(20,10)(12,15)

\put(8,-18){\vector(0,1){2}}
\put(12,-18){\vector(0,1){2}}
\put(28,-18){\vector(0,-1){2}}

\put(12,38){\vector(0,-1){2}}
\put(28,38){\vector(0,1){2}}
\put(32,38){\vector(0,-1){2}}

\put(6,-32){\mbox{$Q$}}
\put(28,-32){\mbox{$\bar Q$}}

\put(6,45){\mbox{$Y$}}
\put(28,45){\mbox{$\bar Y$}}

\put(-50,7){\mbox{${\cal B}_R^{Y|Q}\ = $}}

\end{picture}

\noindent
is to evaluate the HOMFLY polynomials of the double braid. These HOMFLY polynomials can be considered as a generating function for the switching blocks ${\cal B}$. In the double braid, ${\cal K}_1$ and ${\cal K}_2$ (see the figure in s.\ref{sB}) are respectively the
antiparallel and parallel 2-strand braids of lengths $m$ and $n$. Technically, one has to use the evolution method \cite{DMMSS,evo}:
one associates with ${\cal K}_1$ and ${\cal K}_2$ the ${\cal R}$-matrix eigenvalues
$\lambda_Q$ and $\mu_Y$
for parallel and antiparallel 2-strand  braids, accordingly.
Thus,
\be
H_{R}^{(m|n)} = \sum_{Q\in R^{\otimes 2}} \sum_{Y\in R\otimes\bar R}
B^{Y|Q}_R\, \mu_{_Y}^m\,\lambda_{_Q}^n
\ee

For example, for the fundamental representation $R=[1]$ see (123) from \cite{evo}:
\be\label{B}
B_R^{Y|Q} = \left(
\begin{array}{c|cc}
_Y\backslash ^Q & [2] & [11] \\
&\\
\hline
&\\
\emptyset &\frac{\{Aq\}\{q\}}{\{A\}\{q^2\}}& \frac{\{A/q\}\{q\}}{\{A\}\{q^2\}}  \\
&\\
{\rm adjoint} &-\frac{\{Aq\}\{A/q\}}{\{A\}\{q^2\}} & \frac{\{Aq\}\{A/q\}}{\{A\}\{q^2\}}
\end{array}
\right)
\ee
The switching block is constructed from $B^{Y|Q}_R$ in accordance with
\be\label{cB}
{\cal B}_R^{Y|Q}=\mu_Y\lambda_Q {D_R^2\over D_YD_Q}B_R^{Y|Q}
\ee
and is in complete agreement with (\ref{sb1}) and (\ref{sb2}).
In fact, in \cite[s.5.4.3]{evo} there are answers for $B_R^{Y|Q}$ in any symmetric representation. These answers are quite tedious, and we do not write them down here.

\subsection{The switching functor}

We remind that pictorially the switching functor
realizing alternative transformation from $Q$ to $Y$ variables
with $Q\in R_1\otimes R_2$ and $Y\in R_1\otimes \bar R_2$ is

\begin{picture}(300,125)(-150,-43)

\put(0,0){\line(1,0){20}}
\put(0,0){\line(0,1){40}}
\put(0,40){\line(1,0){20}}
\put(20,0){\line(0,1){40}}

\put(5,0){\line(0,-1){20}}
\put(5,40){\vector(0,1){20}}

\qbezier(15,0)(15,-15)(22,-15)
\qbezier(22,-15)(35,-15)(35,60)
\qbezier(15,40)(15,55)(22,55)
\qbezier(22,55)(30,55)(32,30)
\qbezier(34,5)(35,2)(35,-20)

\put(5,-15){\vector(0,1){2}}
\put(35,-15){\vector(0,1){2}}
\put(35,58){\vector(0,1){2}}
\put(15,42){\vector(0,-1){2}}
\put(15,-4){\vector(0,-1){2}}

\put(8,-20){\mbox{$Y$}}
\put(8,50){\mbox{$Y$}}

\put(18,-35){\mbox{$Q$}}
\put(18,65){\mbox{$Q$}}

\put(5,17){\mbox{${\cal C}^ Y$}}

\put(-45,17){\mbox{${C}^Q\ =$}}

{\footnotesize
\put(-2,-22){\mbox{$b$}}
\put(37,-22){\mbox{$j$}}
\put(-2,57){\mbox{$a$}}
\put(37,57){\mbox{$i$}}
\put(18,-7){\mbox{$k$}}
\put(18,45){\mbox{$l$}}
}

\put(110,17){\mbox{
$C^Q_{R_1\times R_2} = \sum_Y {\cal D}^Q_Y\cdot {\cal C}^Y_{R_1\times\bar R_2}$
}}

\end{picture}

\noindent
In the case of fundamental representations $R_1=R_2=[1]$ and $q=1$,
one can easily make a direct calculation.
Using the fact that the ${\cal R}$-matrix acts by multiplication with
$\pm A^{-1}q^{\pm 1} = \pm q^{-N\pm 1}$ in symmetric and antisymmetric representations,
we obtain from (\ref{F1}):
{\footnotesize
\be
\ C_{bj}^{ai} =  \frac{\delta^a_b\delta^i_j+\delta^i_b\delta^a_j}{2}C_{_+}
+  \frac{\delta^a_b\delta^i_j-\delta^i_b\delta^a_j}{2}C_{_-}
= \sum_{k,l} {\cal C}_{bl}^{ak}\cdot {\cal R}_{kj}^{li} =
\sum_{k,l}\left(\frac{\delta^a_l\delta^k_b}{N}{\cal C}_0 +
\Big(\delta^a_b\delta^k_l - \frac{\delta^a_l\delta^k_b}{N}\Big){\cal C}_{\rm adj}
\right)\left(
\frac{\delta^l_k\delta^i_j + \delta^l_j\delta^i_k}{2} -
\frac{\delta^l_k\delta^i_j - \delta^l_j\delta^i_k}{2}
\right)
\nn
\ee
}
which gives
\be
C_\pm = \frac{\pm {\cal C}_\emptyset + (N\mp 1){\cal C}_{\rm adj}}{N}
\ee
and, after the $q$-deformation,
\be
C_+ = \frac{q}{[N]}\left(\frac{{\cal C}_\emptyset}{A} + [N-1]{\cal C}_{\rm adj}\right) \nn \\
C_- = \frac{1}{q[N]}\left(-\frac{{\cal C}_\emptyset}{A} + [N+1]{\cal C}_{\rm adj}\right)
\ee
This satisfies the obvious relation
\be
D_{[2]}C_{_+} + D_{[1,1]}C_{_-} = \frac{[N][N+1]}{[2]}C_{_+} + \frac{[N][N-1]}{[2]}C_{_-} =
D_\emptyset{\cal C}_\emptyset + D_{\rm adj}{\cal C}_{\rm adj} =
{\cal C}_\emptyset + [N+1][N-1]{\cal C}_{\rm adj}
\ee
reflecting invariance under the first Reidemeister move.

One can straightforwardly extend this calculation from the fundamental to arbitrary symmetric
representations \cite{SW},
where the answer will be expressed through the Racah matrices $S$ and $\bar S$,
which play the central role in arborescent calculus of \cite{arbor}.
Since the Racah matrices are already known in various rectangular and even some non-rectangular
representations \cite{Sdifreps},
the switching functor can be probably calculated in full generality,
but this remains one of the many open problems for the future research.

\subsection{Hopf cable
\label{Hopf}}

Many of our examples in this paper involve the simplest Hopf cable tangle:
the one made by cutting one of the circles in the Hopf link.
It is a particular case of the torus tangles, but one of the most important
in simple applications, thus it deserves studying in more details.
The general answer for the Hopf block for any representations is known in a very elegant form \cite{ML,Marrev} (see also \cite{MMH}) and is
associated with the topological vertex in
\cite{GIKV,AK} :
\be
{\cal H}_{R_1\times R_2}^{\rm Hopf} = 
\underbrace{{\rm Schur}_{R_1}\{p^*\}}_{D_{R_1}}
\cdot{\rm Schur}_{R_1}\{p^{(R_2)}\}
= \underbrace{{\rm Schur}_{R_2}\{p^*\}}_{D_{R_2}}
\cdot{\rm Schur}_{R_2}\{p^{(R_1)}\}
\label{Hopffromtv}
\ee
where $p_k^* = \frac{A^k-A^{-k}}{q^k-q^{-k}}$ and
$p^{(R)}_k = p_k^* + A^k\sum_{j=1}^{l_R} q^{-2j-1}(q^{2jr_j}-1)$.
Unfortunately, this expression involves the Schur functions, which are transcendental
special functions difficult to use in practical calculations.
Thus, more explicit formulas are also needed,
and their relation to (\ref{Hopffromtv}) is a certain problem,
which deserves a separate consideration.
In this section, we generalize an old result of \cite{AENV,arthAENV}
from a pair of symmetric representations to the cases when one of them belongs
to the adjoint tower, and when one of them is a two-row representation.

\subsubsection{The Hopf tangle with two symmetric representations}

As we explained in s.\ref{cabH}, if external lines
carry representations $Y$ and $Y'$, while propagating in the
remaining circle is representation $K$,

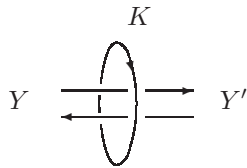
\begin{picture}(300,80)(-150,-30)

\put(0,5){\line(1,0){25}}
\put(32,5){\vector(1,0){18}}
\put(25,-5){\vector(-1,0){25}}
\put(32,-5){\line(1,0){18}}

\qbezier(25,-20)(32,0)(25,20)
\qbezier(25,-20)(18,-30)(15,-8)
\qbezier(25,20)(18,30)(15,8)
\qbezier(14.5,-3)(14.2,0)(14.5,3)

\put(-20,-2){\mbox{$Y$}}
\put(60,-2){\mbox{$Y'$}}
\put(25,30){\mbox{$K$}}
\put(25,20){\vector(1,-4){2}}

\end{picture}

\noindent
then this Hopf tangle is just
\be
{\rm Hopf}^{Y,Y'|K} = \frac{\delta_{Y,Y'}}{D_Y}\cdot {\cal H}^{{\rm tor}_{[2,2]}}_{Y\times K}
\label{Hopftanglevspol}
\ee
and it is independent of $R$.
Thus, it remains to find the double-colored HOMFLY invariant for the Hopf link.
In principle, since the Hopf link is a torus link, the answer is provided
by the Rosso-Jones formula, but it is not very practical.
When both $Y$ and $K$ are symmetric representations, the answer
is known from \cite{arthAENV}:
\be
H^{{\rm tor}_{[2,2]}}_{[r]\times[s]} =
\frac{{\cal H}^{{\rm tor}_{[2,2]}}_{[r]\times[s]}}{D_{[r]}D_{[s]}}=
1 + \sum_{i=1}^{{\rm min}(r,s)} (-A)^{-i} q^{\frac{i(i+3)}{2}-i(r+s)}
\prod_{j=0}^{i-1} \frac{\{q^{r-j}\}\{q^{s-j}\}}{\{Aq^j\}}
\label{Hopfrs}
\ee

\subsubsection{Hopf invariants with one two-line representation}

It has a straightforward generalization to the two-line representation,
\be
H^{\rm Hopf}_{[i,k]\times [s]}=
{\{q\}^{s+1}\over
q^{is-2}\prod_{j=0}^{s-1} \{Aq^j\}}
\sum_{p=0}^{s-1}\left(\prod_{j=1}^pD_{i+j-1} \frac{[s]!}{[p]!}\right)\
{y_{s-p}(k, i)\over A^{s-p}}
\label{Hopfrrs}
\ee
where  $i\ge k$ and
$y_j(k,i)$ are the functions of $q$ only:
\be
y_1(k,i)=-q^{i-k}[k-1]-[i+1] \nn\\
y_2(k,i)=q^k[i+1][i+k-2]+[k-2][k-1]-q^{2i}[k][k-1]\Big(1-{\{q\}\over
q^{2k-3}}\Big)\nn\\
\ldots\\
y_n(k,i)={1\over \{q\}^n q^{ni}q^{2kn}}\sum_{j_1,j_2=0}^n q^{2j_1i+2j_2k }\nu_{j_1,j_2,n}\nn\\
\ee
where
\be
\nu_{i,j,n}:=
\left[{\{q\}^{n-j-1}q^{nj-(j-3)(j+2i)/2}(-1)^{j+n}\over [|n-j/2-i|+j/2]!}\theta(i+j-n)\theta(n-j-1)+\right.\nn\\ +
\left.{[n+i]q^{n(n+1)/2-(n-2)i}\over [|n/2-i|+n/2]!}\theta(j-n)\right]{(-1)^iq^{-2}\over [1+\sum_{l=0}\theta(n-2-i-l)\theta(i+j-n-2-l)]!}
\nn
\ee
and $\theta(x)$ is the Heaviside function. We improve this representation in s.4.2.

\subsubsection{Hopf invariant with one representation from the adjoint tower\label{3Hopfa}}

In this paper we often need $Y$ which belongs to $R\otimes \bar R$,
especially the ones, which belong to
the so called "$E_8$-sector" \cite{MMkrM}, or adjoint tower
beginning from the singlet and adjoint representations. These representations are nothing but the composite representations $(R,R)$, \cite{Koike,MarK}. We consider here only symmetric representations $R=[r]$ and, for the sake of brevity, denote them just $(r)$. In particular, the adjoint representation is $(1)$. Dimensions of these representations $(r) = [2r,r^{N-2}]$ at $A=q^N$
are equal to
\be
D_{(r)} = [N+2r-1][N-1]\cdot \left(\frac{[N+r-1]!}{[N]![r-1]!}\right)^2
\ee
For link polynomials colored with composite representations, in particular, for those in this subsection, we do not omit the $U(1)$-factor $U$, (\ref{U1fac}).

The HOMFLY invariant for the adjoint representation can be deduced from the Rosso-Jones answers
for the family $R=[21^{N-2}]$ at $A=q^N$, by making analytical
continuation in $N$ or, alternatively, from the arborescent calculus
with the known adjoint Racah matrices $S$ and $\bar S$.
However, the latter approach is currently impractical for
more complicated representations, because the Racah matrices are not available,
thus only the tedious former method remains.

In this way, we obtain
\be
H^{\rm Hopf}_{\emptyset\times Y} = 1
\nn \\
H^{\rm Hopf}_{adj\times [s]} = {\{q\}\over \{Aq\}  }
\left(Aq^{2s} + \frac{1}{Aq^{2s}}+ \frac{\{A/q\}}{\{q\}}\right)
\nn\\
H^{\rm Hopf}_{(2)\times [s]}={\{q\}\{q^2\}\over
\{A\}\{Aq^3\} }\left(\Big(Aq^{2s}\Big)^2 + \Big(\frac{1}{Aq^{2s}}\Big)^2+
\frac{\{A/q\}}{\{q\}}\Big(Aq^{2s} + \frac{1}{Aq^{2s}}\Big)
+ \frac{\{A\}\{A/q\}}{\{q\}\{q^2\}}\right)
\label{Hopfadj}
\ee

{\footnotesize
\be
H^{\rm Hopf}_{(3)\times [s]}=  {\{q\}\{q^2\}\{q^3\}\over
\{A \}\{Aq \}\{Aq^5\} }
\left(\Big(Aq^{2s}\Big)^3 + \Big(\frac{1}{Aq^{2s}}\Big)^3+
\frac{\{A/q\}}{\{q\}}\left(\Big(Aq^{2s}\Big)^2 + \Big(\frac{1}{Aq^{2s}}\Big)^2\right)
+\frac{\{A\}\{A/q\}}{\{q\}\{q^2\}}\Big(Aq^{2s} + \frac{1}{Aq^{2s}}\Big)
+ \frac{\{Aq\}\{A\}\{A/q\}}{\{q\}\{q^2\}\{q^3\}}\right)
\nn
\ee
}

\noindent
and in general\footnote{Note that the topological $U(1)$-factor $U$ (\ref{U1fac}), which we preserve for invariants colored with composite representations is equal, in this formula, to $q^{2rs}$.
}
\be\label{Hopf(k)}
H^{\rm Hopf}_{(r)\times [s]}=  \frac{\{q^r\}}{\{Aq^{2r-1}\}}
\prod_{j=1}^{r-1} \frac{\{q^j\}}{\{Aq^{j-1}\}}\cdot
\left(\prod_{j=1}^r \frac{\{Aq^{j-2}\}}{\{q^j\}} + \sum_{i=1}^{r}
 \left(\Big(Aq^{2s}\Big)^i + \Big(\frac{1}{Aq^{2s}}\Big)^i\right)
 \cdot\prod_{j=1}^{r-i} \frac{\{Aq^{j-2}\}}{\{q^j\}}
\right)
= \nn \\
\boxed{
= 1 + \{q\}^2\cdot \frac{[r][s] \{Aq^{r-1}\}\{Aq^s\}}{\{A\}\{Aq^{2r-1}\}}\cdot
\left(1\ + \ \sum_{i=1}^{r-1}\ \{q\}^i\cdot \frac{[r-1]!}{[r-1-i]!}\cdot
\frac{(A^{2i}q^{2is}+q^{-2is})}{\prod_{j=r-i}^{r-1} A\{Aq^j\}}
\right)
}
\ee
i.e. as a function of $s$ it is a sum of powers of $Aq^{2s}$ from $-r$ to $r$ with
coefficients made from the ratios of the differentials,
as usual for the differential expansions.
These link invariants turns into unity at $s=0$ as it should be.

At $A=q^N$, one gets
\be
H^{\rm Hopf}_{(\bar r,r)\times [s]} =  H^{\rm Hopf}_{[2r,1^{N-2}]\times[s]}
\ \ \ {\rm at} \ \ \  A=q^N
\ee
which reduces further at $A=q^2$:
\be
H^{\rm Hopf}_{(\bar r,r)\times [s]} =  H^{\rm Hopf}_{[2r]\times[s]}
\ \ \ {\rm at} \ \ \  A=q^2
\ee
where the r.h.s. is provided by (\ref{Hopfrs}).
Note that for this relation to hold, one has to multiply the invariant in (\ref{Hopfrs}) with the $U(1)$-factor $q^{2rs}$.

\subsection{The trident block}

For $R=[1]$ the product $[1]\otimes \overline{[1]} = \emptyset + {\rm adjoint}$, so that both $[1]\otimes \emptyset = [1]$ and $[1]\otimes {\rm adj}$ contain $[1]$ and the trident block is a $1\times 2$ matrix. Two elements of this matrix $\mathcal{T}^{(k)}=[u_k,v_k]$ can be defined by the set of relations
(omitting unnecessary $R=[1]$ from the indices):
\be
{\cal T}^{(k)}_{  \emptyset }{\cal T}^{(0)}_{  \emptyset }
=\cfrac{\mathcal{\H}^{[2,k]}_{[1]}}{D_{[1]}} =A^k\cdot\frac{q^{-k}\{Aq\} + (-q)^k\{A/q\}}{\{q^2\}}
\ee
and
\be
 \left({\cal T}^{(k)}_{ \emptyset}{\cal T}^{(-k)}_{ \emptyset}
 +{\cal T}^{(k)}_{\rm adj }{\cal T}^{(-k)}_{\rm adj }\right)
 = D_{[1]} = \left(\frac{\{A\}}{\{q\}}\right)
\ee
and  we expect a lot of relations parameterized by arbitrary pairs $k_1,k_2$:
\be
\mathcal{\H}^{[2,k_1+k_2]}_{[1]} =\text{Tr}\ {\cal T}^{(k_1)}\overline{{\cal T}^{(k_2)}}
= D_{[1]}\Big({\cal T}^{(k_1)}_{ \emptyset }{\cal T}^{(k_2)}_{ \emptyset }
 +  {\cal T}^{(k_1)}_{ adj }{\cal T}^{(k_2)}_{ adj }\Big)
\label{2strandfundsumrule}
\ee
For $k=0$ there is only one term, ${\cal T}^{(0)} = u_0\delta_{Y,\emptyset}$, and since the square is just a pair of unknots, we have
\be
({\cal T}^{(0)})^2 = u_0^2 = D_{[1]} \ \
\Longrightarrow \ \ u_0 = \sqrt{D_{[1]}}
\ee
Similarly, for any $k$ we have a pair of relations:
\begin{equation}
\begin{array}{l}
\Tr {\cal T}^{(k)}\overline{{\cal T}^{(0)}} = u_k u_0 =\cfrac{\mathcal{\H}^{[2,k]}_{[1]}}{D_{[1]}}
\\
\Tr {\cal T}^{(k)}\overline{{\cal T}^{(k)}}
= (u_k^2+v_k^2) =\cfrac{\mathcal{\H}^{[2,2k]}_{[1]}}{D_{[1]}}
\end{array}
\end{equation}
For $k=\pm 1$, where $[2,\pm 1]$ and $[2,\pm 2]$ are
the unknot and the Hopf link,
this gives
\begin{equation}
\begin{array}{lcl}
\cfrac{1}{D_{[1]}}{\cal H}_{[1]}^{[2,\pm 1]}=1
&& u_{\pm 1} = \cfrac{\mathcal{H}^{[2,\pm 1]}}{u_0 D_{[1]}} = \cfrac{1}{\sqrt{D_{[1]}}}
\\
&\Longrightarrow \\
\cfrac{1}{D_{[1]}}{\cal \H}_{[1]}^{[2,\pm 2]}=
A^{\pm 2}\cdot \frac{q^{\mp 2}\{Aq\} + q^{\pm 2}\{A/q\}}{\{q^2\}}
&&
v_{\pm 1} = \sqrt{\cfrac{\mathcal{\H}^{[2,\pm 2]} - 1}{D_{[1]}}}
= A^{\pm 1}\sqrt{\frac{\{Aq\}\{A/q\}}{\{A\}\{q\}}}
\end{array}
\end{equation}
so that
\be
\Tr {\cal T}^{(1)}  \overline{{\cal T}^{(-1)} }  = (u_1u_{-1}+v_1v_{-1})\cdot D_{[1]}
=   D_{[1]}^2 = {\cal H}_{[1]}^{[2,0]}
\ee
Likewise for generic $k$
\begin{equation}
u_k = \frac{{\mathcal{\H}}^{[2,k]}_{[1]}}{\sqrt{D^3_{[1]}}} =
\frac{A^k}{\sqrt{D_{[1]}}}\cdot \frac{q^{-k}\{Aq\} + (-q)^k\{A/q\}}{\{q^2\}}
\end{equation}
\begin{equation}
v_k = \sqrt{\cfrac{\mathcal{\H}^{[2,2k]}}{D_{[1]}} - \frac{\left(\mathcal{\H}^{[2,k]}\right)^2}{D^3_{[1]}}}
= A^k\cdot \frac{q^{-k}-(-q)^{k}}{q+q^{-1}}\cdot  \sqrt{\frac{\{Aq\}\{A/q\}}{\{A\}\{q\}}}
\end{equation}
Now we can check (\ref{2strandfundsumrule}):
\begin{equation*}
\boxed{
\begin{array}{l}
\text{Tr}\ \mathcal{T}^{(k_1)}\overline{\mathcal{T}^{(k_2)}}= (u_{k_1}u_{k_2} + v_{k_1}v_{k_2})\cdot D_{[1]}={\cal \H}^{[2,k_1+k_2]}_{[1]}=
\\ \\
= \cfrac{\mathcal{\H}_{[1]}^{[2,k_1]}\mathcal{\H}_{[1]}^{[2,k_2]}}{D^2_{[1]}} + \cfrac{\{Aq\}\{A/q\}}{\{q\}^2}\cdot A^{k_1+k_2}\cdot
\cfrac{(q^{k_1}-(-q)^{-k_1})(q^{k_2}-(-q)^{-k_2})}{(q+q^{-1})^2}
\end{array}
}
\end{equation*}
\vspace{-0.4cm}

\section{Evaluating link invariants}

Now we can use the knowledge of elementary tangle blocks
and perform more sophisticated calculations for links and knots,
which are made from them.

\subsection{Composing links from the lock block}

For instance, one can construct links from a combination of the lock block of s.\ref{lock} and of the cut-and-join blocks of s.\ref{cj} with a few additional twists of double lines added. Below we describe how it works and list a few examples evaluating uncolored link polynomials.

\subsubsection{Composing links: $B(2,-2)$ + lock\label{4.1.1}}

These links belong to the family

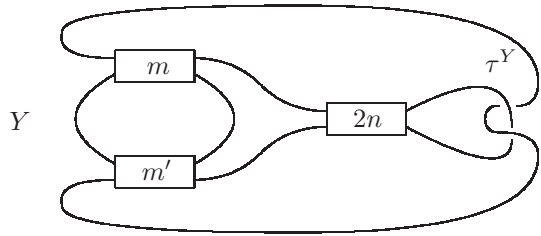
\begin{picture}(300,110)(-150,-65)

\put(0,0){\line(1,0){30}}
\put(0,0){\line(0,1){12}}
\put(30,0){\line(0,1){12}}
\put(0,12){\line(1,0){30}}

\put(0,-40){
\put(0,0){\line(1,0){30}}
\put(0,0){\line(0,1){12}}
\put(30,0){\line(0,1){12}}
\put(0,12){\line(1,0){30}}
}

\qbezier(0,3)(-30,-14)(0,-31)
\qbezier(30,3)(60,-14)(30,-31)

\put(80,-20){
\put(0,0){\line(1,0){30}}
\put(0,0){\line(0,1){12}}
\put(30,0){\line(0,1){12}}
\put(0,12){\line(1,0){30}}
\put(10,3){\mbox{$2n$}}
}

\qbezier(30,9)(45,9)(55,-1)
\qbezier(55,-1)(65,-11)(80,-11)

\qbezier(30,-37)(45,-37)(55,-27)
\qbezier(55,-27)(65,-17)(80,-17)

\put(12,3){\mbox{$m$}}
\put(10,-37){\mbox{$m'$}}

\put(-40,-18){\mbox{$Y$}}


\qbezier(0,9)(-23,9)(-20,19)
\qbezier(-20,19)(-20,29)(70,29)
\qbezier(70,29)(160,29)(160,0)

\qbezier(0,-37)(-23,-37)(-20,-47)
\qbezier(-20,-47)(-20,-57)(70,-57)
\qbezier(70,-57)(160,-57)(160,-28)

\qbezier(140,-14)(140,-9)(145,-9)
\qbezier(140,-14)(140,-19)(145,-19)

\qbezier(152,-9)(160,-9)(160,-0)
\qbezier(145,-19)(160,-19)(160,-28)

\qbezier(110,-11)(150,10)(150,-17)
\qbezier(110,-17)(150,-38)(150,-22)

\put(140,4){\mbox{$\tau^Y$}}
\end{picture}

\noindent
where the box denotes $2n$ possible twists. Here we consider $m=2$, $m'=-2$.
For calculations, we use explicit expressions for the lock block and for the Hopf invariant from ss.\ref{3lock} and \ref{3Hopfa}. The link without additional twists, $n=0$ is the Whitehead, $L_{5a1}$:

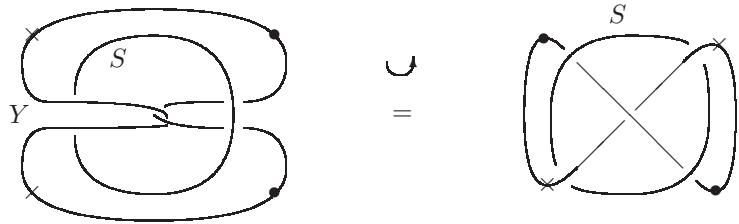
\begin{picture}(400,100)(-150,-45)

\qbezier(-30,8)(-30,30)(0,30)
\qbezier(0,30)(30,30)(30,0)
\qbezier(0,-30)(30,-30)(30,0)
\qbezier(0,-30)(-30,-30)(-30,-8)

\qbezier(-40,5)(5,5)(5,-1)
\qbezier(-40,-5)(0,-5)(5,-4)

\qbezier(26,5)(5,5)(4,3)
\qbezier(26,-5)(5,-5)(0,0)

\qbezier(-40,5)(-50,5)(-50,20)
\qbezier(0,40)(-50,40)(-50,20)
\qbezier(0,40)(50,40)(50,20)
\qbezier(34,5)(50,5)(50,20)

\qbezier(-40,-5)(-50,-5)(-50,-20)
\qbezier(0,-40)(-50,-40)(-50,-20)
\qbezier(0,-40)(50,-40)(50,-20)
\qbezier(34,-5)(50,-5)(50,-20)

\put(-55,-3){\mbox{$Y$}}
\put(-17,18){\mbox{$S$}}

\put(-50,28){\mbox{$\times$}}
\put(43,28){\mbox{$\bullet$}}
\put(-50,-32){\mbox{$\times$}}
\put(43,-32){\mbox{$\bullet$}}

\put(90,-2){\mbox{$=$}}

\put(93,0){
\qbezier(-5,20)(-5,15)(0,15)
\qbezier(5,20)(5,15)(0,15)
\put(5,20){\vector(0,1){2}}
}

\put(180,0){
\qbezier(-30,0)(-30,-14)(-28,-19)
\qbezier(-30,0)(-30,30)(0,30)
\qbezier(0,30)(18,30)(22,27)
\qbezier(0,-30)(30,-30)(30,0)
\qbezier(30,0)(30,14)(28,19)
\qbezier(0,-30)(-18,-30)(-22,-27)

\put(-20,20){\line(1,-1){40}}
\put(-20,-20){\line(1,1){18}}
\put(20,20){\line(-1,-1){18}}

\qbezier(20,20)(40,40)(40,0)
\qbezier(25,-25)(40,-40)(40,0)
\qbezier(-20,-20)(-40,-40)(-40,0)
\qbezier(-25,25)(-40,40)(-40,0)

\put(-8,35){\mbox{$S$}}

\put(30,24){\mbox{$\times$}}
\put(-35,26.5){\mbox{$\bullet$}}
\put(30,-31){\mbox{$\bullet$}}
\put(-35,-29){\mbox{$\times$}}

}

\end{picture}

\be
{\cal H}^{L_{5a1}}_{R\times S}
= \sum_Y D_Y\tau^Y_{R}\left({{\cal H}^{\rm Hopf}_{{\rm Y}\times S}\over D_Y}\right)
\ \ {\Longrightarrow}\ \
{\cal H}^{L_{5a1}}_{[1]\times [1]}=\tau^{[0]}_{[1]}D_{[1]}+
\tau^{{\rm adj}}_{[1]}{\cal H}^{\rm Hopf}_{{\rm adj}\times [1]}=\nn\\
=-{A^4q^6-A^2q^8-2A^4q^4+2A^2q^6+A^4q^2-3A^2q^4+q^6+2A^2q^2-q^4-A^2+q^2\over q^4\{q\}}\ D_{[1]}
\ee
Similarly, the links with two twists, $n=\pm 1$ are $L_{7n2}$ and $L_{8n2}$ accordingly:
\be
{\cal H}^{L_{7n2}}_{R\times S}
= \sum_Y D_Y\tau^Y_{R}\mu_Y^2\left({{\cal H}^{\rm Hopf}_{{\rm Y}\times S}\over D_Y}\right)
\ \ {\Longrightarrow}\ \
{\cal H}^{L_{7n2}}_{[1]\times [1]}=\tau^{[0]}_{[1]}D_{[1]}+
A^2\tau^{{\rm adj}}_{[1]}{\cal H}^{\rm Hopf}_{{\rm adj}\times [1]}=\nn\\
=-{A^4q^6-A^2q^8-A^4q^4+A^2q^6+A^4q^2-3A^2q^4+2q^6+A^2q^2-2q^4-A^2+2q^2\over q^4\{q\}}\ A\ D_{[1]}
\ee

\be
{\cal H}^{L_{8n2}}_{R\times S}
= \sum_Y D_Y\tau^Y_{R}\mu_Y^{-2}\left({{\cal H}^{\rm Hopf}_{{\rm Y}\times S}\over D_Y}\right)
\ \ {\Longrightarrow}\ \
{\cal H}^{L_{8n2}}_{[1]\times [1]}=\tau^{[0]}_{[1]}D_{[1]}+
A^{-2}\tau^{{\rm adj}}_{[1]}{\cal H}^{\rm Hopf}_{{\rm adj}\times [1]}=\nn\\
={A^6q^4-2A^4q^6+A^2q^8+2A^4q^4-A^2q^6-2A^4q^2+2A^2q^4-q^6-A^2q^2+q^4+A^2-q^2\over A^3q^4\{q\}}\ D_{[1]}
\ee

\subsubsection{Composing links: $\tilde W$ + lock}

These links are described by the same figure with the block $B(2,-2)$ replaced with $\tilde W$, and again the box denotes $2n$ possible twists. For calculations, we use explicit expressions for the lock block and for $\tilde W$-block from ss.\ref{3lock} and \ref{3W}. The link without additional twists, $n=0$ is just unknot:

\be
{\cal H}^{{\rm unknot}}_{R\times S}
= \sum_Y D_Y\tau^Y_{R}\tilde W^{Y\times S}_{R\times\bar R}
\ \ {\Longrightarrow}\ \
{\cal H}^{{\rm unknot}}_{[1]\times [1]}=\tau^{[0]}_{[1]}\tilde W^{\emptyset\times S}_{[1]\times [\bar 1]}+D_{{\rm adj}}
\tau^{{\rm adj}}_{[1]}\tilde W^{{\rm adj}\times S}_{[1]\times [\bar 1]}=D_{[1]}^2
\ee
Similarly, the links with two twists, $n=\pm 1$ are $L_{8a4}$ and $L_{8a2}$ accordingly:
\be
{\cal H}^{L_{8a4}}_{R\times S}
= \sum_Y D_Y\tau^Y_{R}\tilde W^{Y\times S}_{R\times\bar R}
\ \ {\Longrightarrow}\ \
{\cal H}^{L_{8a4}}_{[1]\times [1]}=\tau^{[0]}_{[1]}\tilde W^{\emptyset\times S}_{[1]\times [\bar 1]}+A^2D_{{\rm adj}}
\tau^{{\rm adj}}_{[1]}\tilde W^{{\rm adj}\times S}_{[1]\times [\bar 1]}=
\ee
\be
={A^4q^2(2q^8+3q^4+2)-\Big(A^6q^6+4A^4q^6-q^6\{q\}^2-A^2q^2(q^8-2q^6+4q^4-2q^2+1)\Big)(q^2-1+q^{-2})
\over Aq^6\{q\}}\ D_{[1]}\nn
\ee

\be
{\cal H}^{L_{8a2}}_{R\times S}
= \sum_Y D_Y\tau^Y_{R}\tilde W^{Y\times S}_{R\times\bar R}
\ \ {\Longrightarrow}\ \
{\cal H}^{L_{8a2}}_{[1]\times [1]}=\tau^{[0]}_{[1]}\tilde W^{\emptyset\times S}_{[1]\times [\bar 1]}+A^{-2}D_{{\rm adj}}
\tau^{{\rm adj}}_{[1]}\tilde W^{{\rm adj}\times S}_{[1]\times [\bar 1]}=\\
={A^2q^2\Big(A^4q^4+(q^4+1)^2\Big)(q^2-1+q^{-2})-2A^2q^6(1+A^2)(q^2-1+q^{-2})^2-q^{10}+3q^8-5q^6+3q^4-q^2
\over A^3q^6\{q\}}\ D_{[1]}\nn
\ee

\bigskip

One can also consider the lock $\bar\tau$ with the opposite crossings. Then, the link without additional twists, $n=0$ is $L_{7a4}$:
\be
{\cal H}^{L_{7a4}}_{R\times S}
= \sum_Y D_Y\overline{\tau^Y_{R}}\tilde W^{Y\times S}_{R\times\bar R}
\ \ {\Longrightarrow}\ \
{\cal H}^{L_{7a4}}_{[1]\times [1]}=\overline{\tau^{[0]}_{[1]}}\tilde W^{\emptyset\times S}_{[1]\times [\bar 1]}+D_{{\rm adj}}
\overline{\tau^{{\rm adj}}_{[1]}}\tilde W^{{\rm adj}\times S}_{[1]\times [\bar 1]}=\nn
\\
={A^2q^4(A^4-\{q\}^2)(q^2-1+q^{-2})-A^4(q^8-3q^6+5q^4-3q^2+1)+q^4\{q\}^2
\over A^5q^4\{q\}}\ D_{[1]}
\ee
Similarly, the links with two twists, $n=\pm 1$ are $L_{9a8}$ and $L_{7a3}$ accordingly:
\be
{\cal H}^{L_{9a8}}_{R\times S}
= \sum_Y D_Y\overline{\tau^Y_{R}}\tilde W^{Y\times S}_{R\times\bar R}
\ \ {\Longrightarrow}\ \
{\cal H}^{L_{9a8}}_{[1]\times [1]}=\overline{\tau^{[0]}_{[1]}}\tilde W^{\emptyset\times S}_{[1]\times [\bar 1]}+A^2D_{{\rm adj}}
\overline{\tau^{{\rm adj}}_{[1]}}\tilde W^{{\rm adj}\times S}_{[1]\times [\bar 1]}=
\ee
{\footnotesize
\be
={A^4q^2(A^4q^4+5A^2q^4+q^8+3q^4+1)(q^2-1+q^{-2})-A^6q^2(2q^8+3q^4+2)
-A^2q^6(1+3A^2)(q^2-1+q^{-2})^2+q^6\{q\}^2(1-A^2\{q\}^2)
\over A^5q^6\{q\}}\ D_{[1]}\nn
\ee}

\be
{\cal H}^{L_{7a3}}_{R\times S}
= \sum_Y D_Y\overline{\tau^Y_{R}}\tilde W^{Y\times S}_{R\times\bar R}
\ \ {\Longrightarrow}\ \
{\cal H}^{L_{7a3}}_{[1]\times [1]}=\overline{\tau^{[0]}_{[1]}}\tilde W^{\emptyset\times S}_{[1]\times [\bar 1]}+A^{-2}D_{{\rm adj}}\overline{
\tau^{{\rm adj}}_{[1]}}\tilde W^{{\rm adj}\times S}_{[1]\times [\bar 1]}=\nn\\
={A^4q^4(1+q^2)(q^2-1+q^{-2})-A^2(q^{12}-2q^{10}+4q^8-3q^6+4q^4-2q^2+1)+q^6(q^2-1+q^{-2})^2
\over A^5q^6\{q\}}\ D_{[1]}
\ee

\subsection{Link $L_7a3$}

There is another possibility to construct link $L_{7a3}$: to combine the cut-and-join block $B(2,-2)$ (s.\ref{cj}) and the switching block ${\cal B}$ (s.\ref{sb}), one can obtain the series of links $L_{BHn}$ that contains $L_{7a3}$:

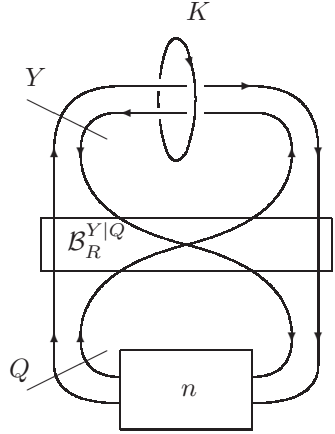
\begin{picture}(300,180)(-150,-135)

\put(0,5){\line(1,0){25}}
\put(32,5){\vector(1,0){18}}
\put(25,-5){\vector(-1,0){25}}
\put(32,-5){\line(1,0){18}}

\qbezier(25,-20)(32,0)(25,20)
\qbezier(25,-20)(18,-30)(15,-8)
\qbezier(25,20)(18,30)(15,8)
\qbezier(14.5,-3)(14.2,0)(14.5,3)

\put(-35,0){\line(2,-1){30}}
\put(-36,5){\mbox{$Y$}}
\put(25,30){\mbox{$K$}}
\put(25,20){\vector(1,-4){2}}
\put(-35,-110){\line(2,1){30}}
\put(-42,-105){\mbox{$Q$}}

\qbezier(0,-5)(-15,-5)(-15,-20)     \qbezier(50,-5)(65,-5)(65,-20)
\qbezier(25,-55)(-15,-45)(-15,-20)  \qbezier(25,-55)(65,-45)(65,-20)
\qbezier(25,-55)(-15,-65)(-15,-90)  \qbezier(25,-55)(65,-65)(65,-90)
\qbezier(0,-105)(-15,-105)(-15,-90) \qbezier(50,-105)(65,-105)(65,-90)

\qbezier(0,5)(-25,5)(-25,-20)            \qbezier(50,5)(75,5)(75,-20)
\qbezier(0,-115)(-25,-115)(-25,-100)     \qbezier(50,-115)(75,-115)(75,-100)
\put(-25,-100){\line(0,1){80}}           \put(75,-100){\line(0,1){80}}

\put(0,-95){\line(1,0){50}}   \put(0,-95){\line(0,-1){30}}
\put(0,-125){\line(1,0){50}}  \put(50,-95){\line(0,-1){30}}
\put(23,-112){\mbox{$n$}}

\put(-15,-90){\vector(0,1){2}}   \put(65,-90){\vector(0,-1){2}}
\put(-25,-90){\vector(0,1){2}}   \put(75,-90){\vector(0,-1){2}}

\put(-15,-20){\vector(0,-1){2}}   \put(65,-20){\vector(0,1){2}}
\put(-25,-20){\vector(0,1){2}}   \put(75,-20){\vector(0,-1){2}}

\put(-30,-45){\line(1,0){110}}   \put(-30,-45){\line(0,-1){20}}
\put(-30,-65){\line(1,0){110}}  \put(80,-45){\line(0,-1){20}}
\put(-20,-57){\mbox{${\cal B}^{Y|Q}_R$}}

\end{picture}

\noindent
Since $B(2,-2)$ is obtained from the Hopf link, we arrive at
\be
{\cal H}^{L_{BHn}}_{R|K} = {1\over D_R}\sum_{Q\in R^{\otimes 2}} \sum_{Y\in R\otimes\bar R} D_YD_Q{\cal B}_R^{Y|Q} \cdot
 \lambda_Q^n\cdot  \frac{{\cal H}^{\rm Hopf}_{Y|K}}{D_Y}
\ee
In particular, $L_{7a3}$ emerges at $n=3$. In the example of one fundamental and one arbitrary symmetric representations,
\be
H^{L_{7a3}}_{[1]|[k]} =\sum_{Q\in \{[2],[11]\}} \sum_{Y\in \{\emptyset,{\rm adj}\}} B_{[1]}^{Y|Q} \cdot
\mu_Y \lambda_Q^4\cdot  H^{\rm Hopf}_{Y|K}=
{A^2q^4+A^2-q^2\over q^2A^4}+{[4]\over [2]}{\{q\}\{q^k\}\{Aq^k\}\{A/q\}\over A^3\{A\}}
\ee
since $\mu_\emptyset=1$, $\mu_{{\rm adj}}=-A$, $\lambda_{[2]}=q/A$, $\lambda_{[11]}=-1/(qA)$ and we used (\ref{B}), (\ref{cB}) and (\ref{Hopfadj}). This answer coincides with the result obtained in \cite{Rama5}. Moreover, using the expressions for
for $B_R^{Y|Q}$ in any symmetric representation $R=[r]$ from \cite[s.5.4.3]{evo} and the explicit formula for the Hopf invariant (\ref{Hopf(k)}), one can evaluate the link polynomial for $L_{7a3}$ colored with the pair of two arbitrary symmetric representations. The results coincides with the answer obtained in \cite{Rama5}.

\subsection{Open and closed chains}

The simplest links to deal with are chains made entirely from the $\lock$-blocks
introduced in sec.\ref{lock}:

\begin{picture}(300,100)(-150,-50)

\put(-3,0){\mbox{$\ldots$}}
\put(-58,0){\circle{6}}
\put(63,0){\circle{6}}
\put(-40,0){\circle*{6}}
\put(-20,0){\circle*{6}}
\put(30,0){\circle*{6}}
\put(50,0){\circle*{6}}

{\footnotesize
\put(-65,10){\mbox{$R_1\otimes \bar R_1$}}
\put(-45,-15){\mbox{$R_2\otimes \bar R_2$}}
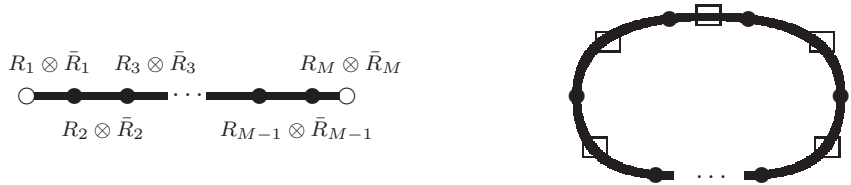
\put(-25,10){\mbox{$R_3\otimes \bar R_3$}}
\put(15,-15){\mbox{$R_{M-1}\otimes \bar R_{M-1}$}}
\put(45,10){\mbox{$R_M\otimes \bar R_M$}}
}

\put(195,-31){\mbox{$\ldots$}}
\put(150,0){\circle*{6}}
\put(250,0){\circle*{6}}
\put(185,29){\circle*{6}}
\put(215,29){\circle*{6}}
\put(180,-30){\circle*{6}}
\put(220,-30){\circle*{6}}

\put(195,30){\mbox{$\boxed{ \phantom. }$}}
\put(157,20){\mbox{$\boxed{ \phantom. }$}}
\put(152,-20){\mbox{$\boxed{ \phantom. }$}}
\put(238,20){\mbox{$\boxed{ \phantom. }$}}
\put(240,-20){\mbox{$\boxed{ \phantom. }$}}

\linethickness{1mm}

\put(-55,0){\line(1,0){50}}
\put(10,0){\line(1,0){50}}

\qbezier(150,0)(150,30)(200,30)
\qbezier(150,0)(150,-30)(185,-30)
\qbezier(250,0)(250,30)(200,30)
\qbezier(250,0)(250,-30)(215,-30)

\end{picture}

Here the boxes denote an additional possible $k_i$-fold twist of the double lines (of the fat lines in the figure), the dependence on this twist is a particular feature of the closed chain.

Open chains $C_M$ are composite links of the Hopf links, i.e. their reduced HOMFLY invariants are just the product of the reduced Hopf invariants.
Technically their HOMFLY polynomials are made from
$\lock^\emptyset_{R_1\times R_2}$ only,
which are just the Hopf polynomials.
Thus the answers
\be
{\cal \H}^{C_M}={\cal V}_{R_1}\cdot\left(\prod_{i=1}^{M-1} \lock^\emptyset_{R_i\times R_{i+1}}\right)
\cdot{\cal V}_{R_M}
= \frac{\prod_{i=1}^{M-1} {\cal \H}^{\rm Hopf}_{R_i\times R_{i+1}} }{\prod_{i=2}^{M-1} D_{R_i}}
\label{OCh}
\ee
are just products of those,
one should only care about orientations.

For the necklaces (closed chains), we need more: $\lock^Y_{R_1\times R_2}$ in various
representations $Y$, the singlet $Y=\emptyset$
is insufficient:
\be
\sum_Y  \mu_Y^{\sum_{i=1}^M k_i}\cdot \prod_{i=1}^{M-1} \lock^Y_{R_i\times R_{i+1}}
\label{CCh}
\ee
Moreover, the closed chain of a given length can describe several topologically different
links, depending on the number and cyclic ordering of $\lock$ and $\bar \lock$ blocks and on the number of twists $\sum_{i=1}^M k_i$.
For a given topology, one can still change $\bar \tau$ for $\tilde\tau$,
this affects the orientation.

\subsection{Open triple chains (\ref{OCh})}

For $\lock^\emptyset_{R_1\times R_2}$, which are just the Hopf invariants,
we already know the answers,
thus one can immediately handle the open chains only taking care about the orientations.

\subsubsection{Chain colored with $[r]\times[s]\times\overline{[r]}$}

Let us consider the open chain $C_3$. Then, it is decomposed into the sum of adjoint Hopf invariants:
\be\label{parHopf0}
\boxed{
{\cal \H}_{R_1\times S\times R_2}^{\rm Hopf}=D_{S}\sum_{Q\in R_1\otimes R_2 } D_Q\cdot   \H^{\rm Hopf}_{Q\times S}
= {\cal V}_{R_1} {\cal V}_{R_2}   \lock^\emptyset_{R_1\times S} \lock^\emptyset_{R_2\times S}}\\
= \frac{1}{D_{S}}\cdot\ {\cal \H}^{\rm Hopf}_{R_1\times S}
 {\cal \H}^{\rm Hopf}_{R_2\times S}
= D_{S}D_{R_1}D_{R_2}\cdot\ {\H}^{\rm Hopf}_{R_1\times S}
 {\H}^{\rm Hopf}_{R_2\times S}\nn
\ee
In the particular choice of coloring the components  $[r]\times[s]\times\overline{[r]}$, one obtains
\be
{\cal \H}^{C_3}_{[r]\times[s]\times\overline{[r]}} =
D_{[s]}\cdot \sum_{p=0}^r D_{(p)}\cdot \H^{\rm Hopf}_{(p)\times [s]}
=
{\cal V}_{[r]}^2 \tau^\emptyset_{[r]\times[s]}\tilde\tau^\emptyset_{[r]\times [s]}
\ee
In order to simplify formulas, we use below for the link invariants the standard framing instead of the differential one.

As we explained, since this link is composite, its invariant is a product of two Hopf invariants, $H^{\rm Hopf}_{[r]\times [s]}(A,q)$  and $H^{\rm Hopf}_{[r]\times [s]}(A^{-1},q^{-1})$: the r.h.s. changes if we substitute $\overline{[r]}$ by $[r]$,
this is equivalent to mirror-reflecting one component of the composite link. Thus, the answer finally reads
\be
{\cal H}^{C_3}_{[r]\times[s]\times\overline{[r]}} =\sum_{p=0}^r {\cal H}^{\rm Hopf}_{(p)\times [s]}=
D_{[s]}\cdot \sum_{p=0}^r D_{(p)}\cdot H^{\rm Hopf}_{(p)\times [s]}= \frac{1}{D_{[s]}}\cdot {\cal H}^{\rm Hopf}_{[r]\times [s]}\cdot \overline{{\cal H}^{\rm Hopf}_{[r]\times [s]}}
\ee
where
\be
D_{(p)}={\{Aq^{2p-1}\}\{A/q\}\over\{q\}^2}\left(\prod_{j=0}^{p-2}{\{Aq^j\}\over \{q^{j+2}\}}
\right)^2
\ee
As a trivial check, for $r=1$ one gets
\be
1 + \frac{\{Aq\}\{A/q\}}{\{q\}^2}\cdot {\{q\}\over \{Aq\}  }
\left(Aq^{2s} + \frac{1}{Aq^{2s}}+ \frac{\{A/q\}}{\{q\}}\right)
= D_{[1]}^2 \left(1 -q^{1-s}A^{-1}\frac{\{q\}\{q^s\}}{\{A\}}\right)
\left(1 +q^{-1+s}A^{1}\frac{\{q\}\{q^s\}}{\{A\}}\right)
\ee

\subsubsection{Chain colored with $[r]\times[s]\times [r]$}

Similarly, one can reproduce ${\cal H}^{C_3}_{[r]\times[s]\times [r]}$,
this is equivalent to mirror-reflecting one component of the composite link.
Now it is reproduced by the sum over the Hopf invariants with  parallel lines.

In this case, one needs the decomposition of the product of two symmetric representations
\be
[r]\otimes [r]=\oplus_{m=0}^r [r+m,r-m]
\ee
their quantum dimensions being
\be
D_{[r+m,r-m]}={[2m+1]\over [r+m+1]![r-m]!}\prod_{j=0}^{2r-1}{\{Aq^j\}\over\{q\}}\prod_{i=0}^{r-m-1}{{\{Aq^{i-1}}\}\over
\{Aq^{r+m+i}\}}
\ee
Then one obtains from (\ref{parHopf0})
\be\label{2line}
{\cal H}^{C_3}_{[r]\times[s]\times [r]} =\sum_{p=0}^r {\cal H}^{\rm Hopf}_{[r+p,r-p]\times [s]}=
D_{[s]}\cdot \sum_{p=0}^r D_{[r+p,r-p]}\cdot H^{\rm Hopf}_{[r+p,r-p]\times [s]}= \frac{1}{D_{[s]}}\cdot \Big({\cal H}^{\rm Hopf}_{[r]\times [s]}\Big)^2
\ee
In particular, for the fundamental representations $r=1$
and arbitrary $s$
\be
D_{[2]}H^{\rm Hopf}_{[2]\times[s]} + D_{[11]}H^{\rm Hopf}_{[11]\times[s]} =
D_{[1]}^2 \cdot\left(1 -q^{1-s}A^{-1}\frac{\{q\}\{q^s\}}{\{A\}}\right)^2
\ee
Note that at the l.h.s. one now needs $H^{\rm Hopf}_{Q\times [s]}$ with
not only with symmetric, but also with two-row representations $Q$.
From this relation, we can {\it deduce}:
\be
H^{\rm Hopf}_{[11]\times[s]} = \frac{(A^2-q^4)
+ (q^4-1)\cdot q^{-2s}}{A^2-1}
= 1 - \frac{[2][s](q^2-1)^2}{q^{s}A\{A\}}
\ee
For $A=q^2$ this gives $q^{-2s}$ which reduces to $1$ with account of the $U(1)$-factor, which is exactly $U=q^{4s/N}=q^{2s}$ in this case (or, equivalently, in the differential framing, which would give rise to the same factor).

In other words, we could actually {\it deduce} Hopf link expressions
not from the direct calculations of sec.\ref{Hopf}, but
from study of the open-chain composite link.
This is a simple but typical illustration of the possibilities
opened by the tangle blocks method:
one can extract tangle blocks from many different sources
and use them in many other circumstances.

In the next section, we extend this example a little further to demonstrate how
expressions from s.\ref{Hopf} can be either re-deduced or checked,
whatever one prefers by the open-chain study. In fact, we will obtain a more elegant form for them.

\subsection{Closed chains}

As already mentioned, for necklaces we need to know $\lock^Y_{R_1\times R_2}$ in all
representations $Y$. In particular, for $R_1=R_2=[r]$, in all representations $(r)$ from the adjoint-tower.
As we already know from sec.\ref{lock}, they are expressed through the HOMFLY invariants for
links that are more complicated than Hopf: generically these are the 3-strand
links from the following series:

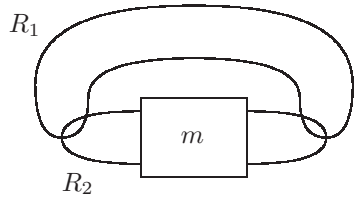
\begin{picture}(300,100)(-175,-50)

\qbezier(-20,0)(-20,30)(40,30) \qbezier(40,30)(100,30)(100,0)
\qbezier(-20,0)(-20,-20)(-10,-20)\qbezier(-10,-20)(0,-20)(0,-5)\qbezier(0,-5)(0,10)(40,10)
\qbezier(100,0)(100,-20)(90,-20)\qbezier(90,-20)(80,-20)(80,-5)\qbezier(80,-5)(80,10)(40,10)
\qbezier(-10,-20)(-10,-10)(20,-10)\qbezier(-10,-20)(-10,-30)(20,-30)
\qbezier(90,-20)(90,-10)(60,-10)\qbezier(90,-20)(90,-30)(60,-30)
\put(20,-35){\line(1,0){40}}
\put(20,-35){\line(0,1){30}}
\put(20,-5){\line(1,0){40}}
\put(60,-35){\line(0,1){30}}
\put(35,-22){\mbox{$m$}}
\put(-30,20){\mbox{$R_1$}}
\put(-10,-40){\mbox{$R_2$}}
\end{picture}

\noindent
\be
{\cal H}_{R_1\times R_2} = \sum_Y \Tr \Big(\big(\tau^Y_{R_1\times R_2}\big)^2\cdot \mu_Y^m\Big)
\ee

Once the lock blocks are known, one can calculate closed chains.
We give just two examples of a 4-component link.
The first example is $L_{8n8}$ with three different orientations:

\begin{picture}(300,275)(-220,-220)

\put(-190,0){
\qbezier(-40,0)(-40,20)(0,20) \qbezier(40,0)(40,20)(0,20)
\qbezier(-40,-10)(-40,-30)(0,-30) \qbezier(40,-10)(40,-30)(0,-30)

\put(0,-80){
\qbezier(-40,0)(-40,20)(0,20) \qbezier(40,0)(40,20)(0,20)
\qbezier(-40,-10)(-40,-30)(0,-30) \qbezier(40,-10)(40,-30)(0,-30)
}

\qbezier(-40,-5)(-60,-5)(-60,-45)\qbezier(-40,-85)(-60,-85)(-60,-45)
\qbezier(-40,-5)(-20,-5)(-20,-25)\qbezier(-40,-85)(-20,-85)(-20,-65)
\qbezier(-19.5,-32)(-19,-45)(-19.5,-58)

\qbezier(40,-5)(60,-5)(60,-45)\qbezier(40,-85)(60,-85)(60,-45)
\qbezier(40,-5)(20,-5)(20,-25)\qbezier(40,-85)(20,-85)(20,-65)
\qbezier(19.5,-32)(19,-45)(19.5,-58)

\put(-60,-45){\vector(0,1){2}}
\put(-19,-45){\vector(0,-1){2}}
\put(0,20){\vector(1,0){2}}
\put(0,-30){\vector(-1,0){2}}

\put(60,-45){\vector(0,-1){2}}
\put(19,-45){\vector(0,1){2}}
\put(0,-110){\vector(-1,0){2}}
\put(0,-60){\vector(1,0){2}}

\put(-75,-20){\mbox{$R_1$}}
\put(-20,25){\mbox{$R_2$}}
\put(65,-20){\mbox{$R_3$}}
\put(-20,-120){\mbox{$R_4$}}

{\footnotesize
\put(-50,-150){\mbox{$\Tr \lock_{12} \bar\lock_{23}\lock_{34} \bar\lock_{41} =$}}
\put(-90,-170){\mbox{$
= \sum_Y D_Y\cdot \lock^Y_{R_1\times R_2} \overline{\lock^Y_{R_2\times R_3}}
\lock^Y_{R_3\times R_4} \overline{\lock^Y_{R_4\times R_1}} $}}
\put(-50,-200){\mbox{$= \sum_{\stackrel{Y\in R_1\times \bar R_3}{Y'\in R_2\otimes \bar R_4}}
{\cal H}^{\rm Hopf}_{Y,Y'}$}}
}
}

\put(0,0){

\qbezier(-40,0)(-40,20)(0,20) \qbezier(40,0)(40,20)(0,20)
\qbezier(-40,-10)(-40,-30)(0,-30) \qbezier(40,-10)(40,-30)(0,-30)

\put(0,-80){
\qbezier(-40,0)(-40,20)(0,20) \qbezier(40,0)(40,20)(0,20)
\qbezier(-40,-10)(-40,-30)(0,-30) \qbezier(40,-10)(40,-30)(0,-30)
}

\qbezier(-40,-5)(-60,-5)(-60,-45)\qbezier(-40,-85)(-60,-85)(-60,-45)
\qbezier(-40,-5)(-20,-5)(-20,-25)\qbezier(-40,-85)(-20,-85)(-20,-65)
\qbezier(-19.5,-32)(-19,-45)(-19.5,-58)

\qbezier(40,-5)(60,-5)(60,-45)\qbezier(40,-85)(60,-85)(60,-45)
\qbezier(40,-5)(20,-5)(20,-25)\qbezier(40,-85)(20,-85)(20,-65)
\qbezier(19.5,-32)(19,-45)(19.5,-58)

\put(-60,-45){\vector(0,1){2}}
\put(-19,-45){\vector(0,-1){2}}
\put(0,20){\vector(1,0){2}}
\put(0,-30){\vector(-1,0){2}}

\put(60,-45){\vector(0,-1){2}}
\put(19,-45){\vector(0,1){2}}
\put(0,-110){\vector(1,0){2}}
\put(0,-60){\vector(-1,0){2}}

\put(-75,-20){\mbox{$R_1$}}
\put(-20,25){\mbox{$R_2$}}
\put(65,-20){\mbox{$R_3$}}
\put(-20,-120){\mbox{$R_4$}}

{\footnotesize
\put(-50,-140){\mbox{$\Tr \lock_{12} \bar\lock_{23}\tilde\lock_{34} \bar{\tilde\lock}_{41} =$}}
\put(-90,-160){\mbox{$
= \sum_Y D_Y\cdot \lock^Y_{R_1\times R_2} \overline{\lock^Y_{R_2\times R_3}}
\tilde\lock^Y_{R_3\times R_4} \overline{\tilde\lock^Y_{R_4\times R_1}} $}}
\put(-50,-190){\mbox{$= \sum_{\stackrel{Y\in R_1\times \bar R_3}{Q\in R_2\otimes  R_4}}
{\cal H}^{\rm Hopf}_{Y,Q}$}}
}
}

\put(190,0){

\qbezier(-40,0)(-40,20)(0,20) \qbezier(40,0)(40,20)(0,20)
\qbezier(-40,-10)(-40,-30)(0,-30) \qbezier(40,-10)(40,-30)(0,-30)

\put(0,-80){
\qbezier(-40,0)(-40,20)(0,20) \qbezier(40,0)(40,20)(0,20)
\qbezier(-40,-10)(-40,-30)(0,-30) \qbezier(40,-10)(40,-30)(0,-30)
}

\qbezier(-40,-5)(-60,-5)(-60,-45)\qbezier(-40,-85)(-60,-85)(-60,-45)
\qbezier(-40,-5)(-20,-5)(-20,-25)\qbezier(-40,-85)(-20,-85)(-20,-65)
\qbezier(-19.5,-32)(-19,-45)(-19.5,-58)

\qbezier(40,-5)(60,-5)(60,-45)\qbezier(40,-85)(60,-85)(60,-45)
\qbezier(40,-5)(20,-5)(20,-25)\qbezier(40,-85)(20,-85)(20,-65)
\qbezier(19.5,-32)(19,-45)(19.5,-58)

\put(-60,-45){\vector(0,1){2}}
\put(-19,-45){\vector(0,-1){2}}
\put(0,20){\vector(1,0){2}}
\put(0,-30){\vector(-1,0){2}}

\put(60,-45){\vector(0,1){2}}
\put(19,-45){\vector(0,-1){2}}
\put(0,-110){\vector(1,0){2}}
\put(0,-60){\vector(-1,0){2}}

\put(-75,-20){\mbox{$R_1$}}
\put(-20,25){\mbox{$R_2$}}
\put(65,-20){\mbox{$R_3$}}
\put(-20,-120){\mbox{$R_4$}}

{\footnotesize
\put(-50,-150){\mbox{$\Tr \lock_{12} \bar{\tilde\lock}_{23}\lock_{34} \bar{\tilde\lock}_{41} =$}}
\put(-90,-170){\mbox{$
= \sum_Y D_Y\cdot \lock^Y_{R_1\times R_2} \overline{\tilde\lock^Y_{R_2\times R_3}}
\lock^Y_{R_3\times R_4} \overline{\tilde\lock^Y_{R_4\times R_1}} $}}
\put(-50,-200){\mbox{$= \sum_{\stackrel{Q\in R_1\times   R_3}{Q'\in R_2\otimes   R_4}}
{\cal H}^{\rm Hopf}_{Q,Q'}$}}
}
}

\end{picture}

\noindent
For illustration, when all the four representations are fundamental,
$R_1 = R_2=R_3=R_4=[1]$, explicit formulas in the three cases are (all formulas for the corresponding Hopf polynomials can be found in s.\ref{summary}):
\be
{\cal \H}^{L_{8n8}}_{[1]\times[1]\times[1]\times[1]}  =
\Big(\lock_\emptyset \overline{\lock_\emptyset}\Big)^2
+ D_{\rm adj}\cdot\Big(\tau_{\rm adj} \overline{\tau_{\rm adj}}\Big)^2 =
\nn \\
= 1 + D_{\rm adj}\cdot\Big(1+ (q^2-1+q^{-2})^2 D_{[1]}^2\Big) =
\underbrace{{\cal H}^{\rm Hopf}_{\emptyset\times\emptyset}
+2{\cal H}^{\rm Hopf}_{{\rm adj}\times\emptyset }}_{1+2D_{\rm adj}}
+{\cal H}^{\rm Hopf}_{{\rm adj}\times{\rm adj}}
\ee
\be
{\cal \H}^{L_{8n8}}_{[1]\times[1]\times[1]\times\overline{[1]}} =
\lock_\emptyset\overline{\lock_\emptyset}\tilde\lock_\emptyset\overline{\tilde\lock_\emptyset}
+ D_{\rm adj}\cdot\lock_{\rm adj}\overline{\lock_{\rm adj}}\tilde\lock_{\rm adj}
\overline{\tilde\lock_{\rm adj}}
={\cal H}^{\rm Hopf}_{[2],\emptyset}+{\cal H}^{\rm Hopf}_{[2],{\rm adj}}
+{\cal H}^{\rm Hopf}_{[1,1],\emptyset}+{\cal H}^{\rm Hopf}_{[1,1],{\rm adj}}
\ee
which is actually the same as
$\Big(\lock_\emptyset \overline{\lock_\emptyset}\Big)^2
+ D_{\rm adj}\cdot\Big(\tau_{\rm adj} \overline{\tau_{\rm adj}}\Big)^2$
in the previous case, because $\tilde\lock = \overline{\mu_Y\lock}$ and $\mu_Y\bar \mu_Y=1$.
Note that this expression is simultaneously equal to
\be
{\H}^{L_{8n8}}_{[1]\times[1]\times[1]\times\overline{[1]}} ={1\over D_{[1]}^2}\cdot\left(1+
D_{\rm adj}\cdot \Big({H}^{\rm Hopf}_{[1],{\rm adj}}\Big)^2\right)
\ee
in complete agreement with (\ref{decof}).

The third orientation gives
\be
{\cal \H}^{L_{8n8}}_{[1]\times[1]\times\overline{[1]}\times\overline{[1]}} =
\lock_\emptyset^4 + A^2\cdot D_{\rm adj}\cdot\lock_{\rm adj}^4 =
A^8\Big(
{\cal H}^{\rm Hopf}_{[2],[2]}+2{\cal H}^{\rm Hopf}_{[2],[1,1]}+{\cal H}^{\rm Hopf}_{[1,1],[1,1]}
\Big)
\ee
where we substituted
$\lock_{\rm adj}^2 \overline{\tilde\lock_{\rm adj}^2} = A^{2}\lock_{\rm adj}^4$.

There is also the fourth case, ${\cal H}^{L_{8n8}}_{[1]\times\overline{[1]}\times[1]\times\overline{[1]}}$, however, it reduces to the first one. Thus, there are totally two basically distinct answers depending on the orientations.

\bigskip

In the sense, which is clear from these examples, $L_{8n8}$ is a closed
chain of the type $\lock\bar\lock\lock\bar\lock$, and it turns to be alternatively expressible through the Hopf block.
To demonstrate what happens if we change the type,
we consider the second example, $L_{8a21}$, which is of the type $\lock^4$,
it is similarly expressible through the Whitehead block:

\begin{picture}(300,275)(-220,-220)

\put(-190,0){
\qbezier(-40,0)(-40,20)(0,20) \qbezier(40,-5)(40,20)(0,20)
\qbezier(-40,-10)(-40,-30)(15,-27) \qbezier(40,-5)(40,-25)(25,-26)

\put(0,-80){
\qbezier(-40,-5)(-40,15)(-25,16) \qbezier(40,0)(40,20)(-15,17)
\qbezier(-40,-5)(-40,-30)(0,-30) \qbezier(40,-10)(40,-30)(0,-30)
}

\qbezier(-40,-5)(-60,-5)(-60,-45)\qbezier(-45,-85)(-60,-85)(-60,-45)
\qbezier(-40,-5)(-20,-5)(-20,-22)\qbezier(-35,-85)(-20,-85)(-20,-65)
\qbezier(-19.5,-32)(-19,-45)(-20,-65)

\qbezier(45,-5)(60,-5)(60,-45)\qbezier(40,-85)(60,-85)(60,-45)
\qbezier(35,-5)(20,-5)(20,-25)\qbezier(40,-85)(20,-85)(20,-68)
\qbezier(20,-25)(19,-45)(19.5,-58)

\put(-60,-45){\vector(0,1){2}}
\put(-19,-45){\vector(0,-1){2}}
\put(0,20){\vector(1,0){2}}
\put(0,-27){\vector(-1,0){2}}

\put(60,-45){\vector(0,-1){2}}
\put(19,-45){\vector(0,1){2}}
\put(0,-110){\vector(-1,0){2}}
\put(0,-63){\vector(1,0){2}}

\put(-75,-20){\mbox{$R_1$}}
\put(-20,25){\mbox{$R_2$}}
\put(65,-20){\mbox{$R_3$}}
\put(-20,-120){\mbox{$R_4$}}

{\footnotesize
\put(-50,-150){\mbox{$\Tr \lock_{12}  \lock_{23}\lock_{34}  \lock_{41} =$}}
\put(-90,-170){\mbox{$
= \sum_Y D_Y\cdot \lock^Y_{R_1\times R_2}  {\lock^Y_{R_2\times R_3}}
\lock^Y_{R_3\times R_4}  {\lock^Y_{R_4\times R_1}} $}}
\put(-50,-200){\mbox{$=
\sum_{\stackrel{Y\in R_1\times \bar R_3}{Y'\in R_2\otimes \bar R_4}}
D_YW_{\bar R_1\times R_3}^{Y\times R_2}W_{\bar R_1\times R_3}^{Y\times R_4}$}}
}
}

\put(0,0){
\qbezier(-40,0)(-40,20)(0,20) \qbezier(40,-5)(40,20)(0,20)
\qbezier(-40,-10)(-40,-30)(15,-27) \qbezier(40,-5)(40,-25)(25,-26)

\put(0,-80){
\qbezier(-40,-5)(-40,15)(-25,16) \qbezier(40,0)(40,20)(-15,17)
\qbezier(-40,-5)(-40,-30)(0,-30) \qbezier(40,-10)(40,-30)(0,-30)
}

\qbezier(-40,-5)(-60,-5)(-60,-45)\qbezier(-45,-85)(-60,-85)(-60,-45)
\qbezier(-40,-5)(-20,-5)(-20,-22)\qbezier(-35,-85)(-20,-85)(-20,-65)
\qbezier(-19.5,-32)(-19,-45)(-20,-65)

\qbezier(45,-5)(60,-5)(60,-45)\qbezier(40,-85)(60,-85)(60,-45)
\qbezier(35,-5)(20,-5)(20,-25)\qbezier(40,-85)(20,-85)(20,-68)
\qbezier(20,-25)(19,-45)(19.5,-58)

\put(-60,-45){\vector(0,1){2}}
\put(-19,-45){\vector(0,-1){2}}
\put(0,20){\vector(1,0){2}}
\put(0,-27){\vector(-1,0){2}}

\put(60,-45){\vector(0,-1){2}}
\put(19,-45){\vector(0,1){2}}
\put(0,-110){\vector(1,0){2}}
\put(0,-63){\vector(-1,0){2}}

\put(-75,-20){\mbox{$R_1$}}
\put(-20,25){\mbox{$R_2$}}
\put(65,-20){\mbox{$R_3$}}
\put(-20,-120){\mbox{$R_4$}}

{\footnotesize
\put(-50,-150){\mbox{$\Tr \lock_{12}  \lock_{23}\tilde\lock_{34}  \tilde\lock_{41} =$}}
\put(-90,-170){\mbox{$
= \sum_Y D_Y\cdot \lock^Y_{R_1\times R_2}  {\lock^Y_{R_2\times R_3}}
\tilde\lock^Y_{R_3\times R_4}  {\tilde\lock^Y_{R_4\times R_1}} $}}
\put(-50,-200){\mbox{$=
\sum_{\stackrel{Y\in R_1\times \bar R_3}{Y'\in R_2\otimes \bar R_4}}
D_YW_{\bar R_1\times R_3}^{Y\times R_2}W_{\bar R_1\times R_3}^{Y\times \bar R_4}$}}
}
}

\put(190,0){
\qbezier(-40,0)(-40,20)(0,20) \qbezier(40,-5)(40,20)(0,20)
\qbezier(-40,-10)(-40,-30)(15,-27) \qbezier(40,-5)(40,-25)(25,-26)

\put(0,-80){
\qbezier(-40,-5)(-40,15)(-25,16) \qbezier(40,0)(40,20)(-15,17)
\qbezier(-40,-5)(-40,-30)(0,-30) \qbezier(40,-10)(40,-30)(0,-30)
}

\qbezier(-40,-5)(-60,-5)(-60,-45)\qbezier(-45,-85)(-60,-85)(-60,-45)
\qbezier(-40,-5)(-20,-5)(-20,-22)\qbezier(-35,-85)(-20,-85)(-20,-65)
\qbezier(-19.5,-32)(-19,-45)(-20,-65)

\qbezier(45,-5)(60,-5)(60,-45)\qbezier(40,-85)(60,-85)(60,-45)
\qbezier(35,-5)(20,-5)(20,-25)\qbezier(40,-85)(20,-85)(20,-68)
\qbezier(20,-25)(19,-45)(19.5,-58)

\put(-60,-45){\vector(0,1){2}}
\put(-19,-45){\vector(0,-1){2}}
\put(0,20){\vector(1,0){2}}
\put(0,-27){\vector(-1,0){2}}

\put(60,-45){\vector(0,1){2}}
\put(19,-45){\vector(0,-1){2}}
\put(0,-110){\vector(1,0){2}}
\put(0,-63){\vector(-1,0){2}}

\put(-75,-20){\mbox{$R_1$}}
\put(-20,25){\mbox{$R_2$}}
\put(65,-20){\mbox{$R_3$}}
\put(-20,-120){\mbox{$R_4$}}

{\footnotesize
\put(-50,-150){\mbox{$\Tr \lock_{12}  \tilde\lock_{23}\lock_{34}  \tilde\lock_{41} =$}}
\put(-90,-170){\mbox{$
= \sum_Y D_Y\cdot  \lock^Y_{R_1\times R_2}  \tilde\lock^Y_{R_2\times R_3}
\lock^Y_{R_3\times R_4}  \tilde\lock^Y_{R_4\times R_1} $}}
\put(-50,-200){\mbox{$=
\sum_{\stackrel{Y\in R_1\times \bar R_3}{Y'\in R_2\otimes \bar R_4}}
D_YW_{\bar R_1\times \bar R_3}^{Y\times R_2}W_{\bar R_1\times \bar R_3}^{Y\times \bar R_4}$}}
}
}

\end{picture}

\noindent
To compare with the previous example, we draw the same three orientations.
For all fundamental representations,
$R_1 = R_2=R_3=R_4=[1]$, we obtain this time:
\be
{\cal \H}^{L_{8a21}}_{[1]\times[1]\times[1]\times[1]}  =
\lock_\emptyset^4
+ D_{\rm adj}\cdot\tau_{\rm adj} ^4 =A^{-8}\left(\Big[W_{[\bar 1]\times [1]}^{\emptyset\times [1]}\Big]^2+D_{\rm adj}\cdot
\Big[W_{[\bar 1]\times [1]}^{{\rm adj}\times [1]}\Big]^2\right)
\ee
where we used (\ref{Wc1}) and (\ref{Wc2}).

\section{Hopf invariants revisited}

The main tool for our new attack on the Hopf invariants will be formula (\ref{parHopf0})
which we rewrite in a more concise form:
\be
\boxed{
\frac{1}{D_S}\cdot{\cal H}^{\rm Hopf}_{R_1,S}\cdot{\cal H}^{\rm Hopf}_{R_2,S}
\ \ =
\sum_{R\in R_1\otimes R_2}N^R_{R_1R_2}\cdot
{\cal H}^{\rm Hopf}_{R,S}
 }
\label{parHopf}
\label{HvsHH}
\ee
where $N^R_{R_1R_2}$ are the integer-valued
Littlewood-Richardson coefficients,
those appearing in decomposition of the product of the Schur functions (characters),
\be
\chi_{R_1}\{p\}\cdot\chi_{R_2}\{p\} \ \ = \sum_{R\in R_1\otimes R_2}N^R_{R_1R_2}\cdot \chi_{R}\{p\}
\ee
This opens a possibility of treating colored Hopf invariants as values of
the character at a particular color-dependent point:
\be
\frac{1}{D_S}\cdot {\cal H}^{\rm Hopf}_{R,S} \ \stackrel{?}{=}\ \chi_R\{p^S\}
\label{Hopfchar}
\ee
This is indeed true, with
\be
p^S_k = \frac{A^k-A^{-k}}{q^k-q^{-k}} =\sum_{j=1}^{l_S} A^kq^{(1-2j)k}(q^{2ks_j}-1)
\ee
this fact is behind the relation between the Hopf invariants
and the topological vertex \cite{GIKV,AK,Nag}.
Below in this section, we give some examples of {\it direct} application of
(\ref{HvsHH}) to big classes of representations.
In this paper, we restrict ourselves to the cases of specific $S$:
symmetric representations $S=[s]$  and the adjoint tower $S=(s)$.
An exhaustive solution to the Hopf polynomial problem will be presented elsewhere.

\subsection{Symmetric $S=[s]$ and arbitrary $R$}

If all the three representations in (\ref{parHopf}) are symmetric,
then $Q\in [r_1]\otimes[r_2]$ are the two-row Young diagrams
$[r_1+r_2-i,i]$ with $i = 0,\ldots,r_2$.
This allows one to find ${\cal H}_{[a,b]\times [s]}$ from (\ref{parHopf}) successively
one after another. Say, choose $(r_1,r_2) = (n-1,1)$, then only $Q=[n]$ and $Q=[n-1,1]$ contribute. As soon as
${\cal H}^{\rm Hopf}_{[n]\times S}$
is already known for symmetric $S=[s]$ from  (\ref{Hopfrs}), one calculates ${\cal H}^{\rm Hopf}_{[n-1,1]\times [s]}$.  In this way, using (\ref{Hopfrs}),
\be
H^{{\rm Hopf} }_{[r]\times[s]} =
1 + \sum_{i=1}^{{\rm min}(r,s)} (-)^i q^{\frac{i(i-1)}{2}-i(r+s)}
\frac{[r]![s]!}{[r-i]![s-i]!}(q^2-1)^{2i}
\prod_{j=0}^{i-1} \frac{1}{A\{Aq^j\}}
= \nn \\
\boxed{=
1 - \frac{q^{-1}}{[r+1]}\cdot
\frac{ [s](q^2-1)^2}{q^{s }\prod_{j=0}^{r-1} A\{Aq^j\}}\cdot
\sum_{k=0}^{r-1} \frac{ [r+1]!(q^2-1)^k \prod_{j=0}^{r-k-2} A\{Aq^j\}}
{[r-k-1]!\cdot q^{2ks}}
}
\label{Hopfrs1}
\ee
we easily obtain an alternative form of (\ref{Hopfrs}):
\be
H^{\rm Hopf}_{[r,1]\times[s]} =
1 - \frac{1}{[r]}\cdot
\frac{ [s](q^2-1)^2}{q^{s }\prod_{j=0}^{r-1} A\{Aq^j\}}\cdot
\sum_{k=0}^{r-1} \frac{ [r+1]!(q^2-1)^k \prod_{j=0}^{r-k-2} A\{Aq^j\}}
{[r-k-1]!\cdot q^{2ks}}
\ee
Likewise, choosing further
$(r_1,r_2)=(n-2,2)$, one obtains ${\cal H}^{\rm Hopf}_{[n-2,2]\times [s]}$ the only new contribution to (\ref{parHopf}), since
${\cal H}^{\rm Hopf}_{[n]\times [s]}$ and ${\cal H}^{\rm Hopf}_{[n-1,1]\times [s]}$ are already known:
\be
H^{\rm Hopf}_{[r,2]\times[s]} =
 1 +\frac{[2][s](q^2-1)^2}{q^{s-r}[r-1](A^2-1)}
\ -\  \frac{q}{[r-1]}\cdot\frac{q\cdot [s](q^2-1)^2}{q^{s }\prod_{j=0}^{r-1} A\{Aq^j\}}\cdot
\sum_{k=0}^{r-1} \frac{ [r+1]!(q^2-1)^k \prod_{j=0}^{r-k-2} A\{Aq^j\}   }
{[r-k-1]!\cdot q^{2ks}}
\ee
Repeating the recursive procedure, we finally obtain for an arbitrary two-row $R=[r_1,r_2]$ and symmetric $S=[s]$
{
\be
H^{\rm Hopf}_{[r_1,r_2]\times[s]} =
1 \ + \
\frac{q^{r_1}}{[r_1-r_2+1]}\cdot\frac{[r_2][r_2-1] [s](q^2-1)^2}{q^{s }
\prod_{j=0}^{r_2-2} A\{Aq^j\}}\cdot
\sum_{k=0}^{r_2-2} \frac{ [r_2-2]!(q^2-1)^k \prod_{j=0}^{r-k-3} A\{Aq^j\}   }
{[r_2-k-2]!\cdot q^{2ks}}
-
\nn\\ 
\ -\  \frac{q^{r_2-1}}{[r_1-r_2+1]}\cdot\frac{ [s](q^2-1)^2}{q^{s}
\prod_{j=0}^{r_1-1} A\{Aq^j\}}\cdot
\sum_{k=0}^{r_1-1} \frac{ [r_1+1]!(q^2-1)^k \prod_{j=0}^{r_1-k-2} A\{Aq^j\}   }
{[r_1-k-1]!\cdot q^{2ks}}
\nn
\ee
}
or\footnote{The sums in these formulas
are understood as MAPLE's operation {\it add}, i.e. $\sum_{k=0}^{r-1} f(k) =0$ for $r\leq 0$.
\label{sumadd}
}
\be
H^{\rm Hopf}_{[r_1,r_2]\times[s]} =
\label{Hopfr1r2s}
\ee
\vspace{-0.5cm}
\be
\nn
\boxed{
=
1 \ + \
\frac{q^{r_1+r_2-s+2}\{q\}^2[s]}{[r_1-r_2+1]}
\left(
\sum_{k=0}^{r_2-2} \frac{ q^{-2ks-r_2}[r_2 ]!(q^2-1)^k    }
{[r_2-2-k]!\cdot \prod_{j=r_2-2-k}^{r_2-2} A\{Aq^j\}}
\ -\
\sum_{k=0}^{r_1-1} \frac{q^{-2ks-r_1-1} [r_1+1]!(q^2-1)^k    }
{[r_1-1-k]!\cdot \prod_{j=r_1-1-k}^{r_1-1} A\{Aq^j\}}
\right)
}
\ee
At $A=q^2$, we have a reduction relation:
\be
q^{(r_1+r_2)s}H_{[r_1,r_2]\times[s]} (A=q^2,q)=   q^{(r_1-r_2)s}H_{[r_1-r_2]\times [s]}(A=q^2,q)
\ee
where we have manifestly included the $U(1)$-factors: $q^{2(r_1+r_2)s/N}=q^{(r_1+r_2)s}$ and $q^{2(r_1-r_2)s/N}=q^{(r_1-r_2)s}$ at $N=2$ accordingly. In fact, a more general formula is correct
\be
\H_{[r_1,r_2]\times[s]} - \H_{[r_1-r_2]\times [s]} \sim \{A/q^2\}
\ee
and the factors $q^{(r_1+r_2)s}$ and $q^{(r_1-r_2)s}$ come, in this case, from the differential framing \cite{China1}.

Formula (\ref{Hopfr1r2s}) is definitely equivalent to (\ref{Hopfrrs}), however, is simpler and can be easily generalized. Indeed,
using the answers for two-row representations, we can proceed
to triple- and quadruple-line ones, and so on:
(\ref{parHopf}) is a very simple  {\bf non-linear recursion in representation space}. In this way,
eq.(\ref{Hopfr1r2s}) can be easily generalized to the multi-row Young diagram $R$,
for example,
\be
H^{\rm Hopf}_{[r_1,r_2,1]\times[s]} =
\label{Hopfr1r21s}
\ee
\vspace{-0.5cm}
\be
= 1+
\frac{q^{r_1+r_2+3-s}\{q\}^2[s]}{[r_1-r_2+1]}
\left(
\sum_{k=0}^{r_2-2} \frac{  q^{-2ks-r_2}[r_2+1][r_2-1 ]!(q^2-1)^k    }
{[r_2-2-k]!\cdot \prod_{j=r_2-2-k}^{r_2-2} A\{Aq^j\}}
\ -\
\sum_{k=0}^{r_1-1} \frac{q^{-2ks-r_1-1} [r_1+2] [r_1]!(q^2-1)^k    }
{[r_1-1-k]!\cdot \prod_{j=r_1-1-k}^{r_1-1} A\{Aq^j\}}
\right)
\nn
\ee
and, for the arbitrary Young diagram $R=\{r_1\geq r_2\geq \ldots \geq r_l>0\}$,
\be
\boxed{\boxed{
H^{\rm Hopf}_{[r_1\ldots r_l]\times[s]} =
1 - q^{r_1+\ldots+r_l} \sum_{i=1}^l \frac{q^{-r_i+i-2}[l+r_i-i]!}{\prod_{j\neq i}^l
[r_i-r_j-i+j]}\cdot\sigma_s(r_i-i+1)
}}
\label{HopfRs}
\ee
where
\be
\sigma_s(r) := {[s]}\cdot\sum_{k=0}^{r-1}\frac{q^{-(2k+1)s}\cdot (q^2-1)^{k+2}}{[r-k-1]!
\prod_{j=r-1-k}^{r-1}A\{Aq^j\}}
\ee
The sum here is understood as MAPLE's operation {\it add}, i.e. $\sigma_s(r)=0$ for $r\leq 0$,
see footnote \ref{sumadd}.

Since all the expressions for the Hopf invariants are basically given by the Rosso-Jones formula, i.e. are sums of quantum dimensions with coefficients expressed through characters of the permutation group, it should not come as a surprise that particular quantities in (\ref{HopfRs}) are typical for the permutation group calculus. For instance, the standard permutation group formula
\be
{1\over d_R \cdot |R|!} =\frac{ \prod_i[l+r_i-i]!}{\prod_{j\neq i}^l[r_i-r_j-i+j]}
\ee
for the dimension $d_R|R|!$ of representation $R$ of the symmetric group $S_{|R|}$ contains the same typical quantity as appeared in (\ref{HopfRs}). Note that this ratio is in fact independent of $l$: if one adds $r_{l+1}=0$, this provides
an extra factor $\frac{[l+1+r_i-i]}{[r_i-i-0+l+1]}=1$, and $\sigma_s(r)=0$ for $r\leq 0$.

One could definitely obtain (\ref{HopfRs}) directly from the Rosso-Jones formula, because the coefficients $c_{Q,W}$ in (\ref{RJ}) are, in the case of the Hopf link, just the Littlewood-Richardson coefficients which are especially simple in the case of one of the representations symmetric: there are no multiplicities in this case.

Note that similarly one can construct Hopf invariants for various other representations. For instance, one can extract
\be
{H}^{\rm Hopf}_{\overline{[1]}\times [s]} =
{1\over qD_{\overline{[1]}}} \cdot \left(A  q^{2s}+\frac{\{A/q\}}{\{q\}}\right)
\ee
from
\be
{\cal H}^{\rm Hopf}_{\overline{[1]}\times [s]}
\cdot {\cal H}^{\rm Hopf}_{{[1]}\times [s]} = D_{[s]}\cdot
\Big( {\cal H}^{\rm Hopf}_{\emptyset\times [s]} +
{\cal H}^{\rm Hopf}_{{\rm adj}\times [s]} \Big)
\ \stackrel{(\ref{Hopfadj})}{=}\
D_{[s]}^2\left(1+   D_{adj}\cdot \frac{\{q\}}{\{Aq\}}\Big(Aq^{2s} + \frac{1}{Aq^{2s}}
+ \frac{\{A/q\}}{\{q\}}\Big)            \right)
\ee
The calculation involves the adjoint Hopf polynomial, hence, we restore everywhere the $U(1)$ factor in order to match (\ref{Hopfadj}).

However, the most interesting case we consider in the next subsection.

\subsection{Involving adjoint representations}

Let us consider now the cases when, at least, one of the representations $R$, $S$ is adjoint.

For the adjoint representation $R={\rm adj}=[21^{N-2}]$, $l_{\rm adj}=N-1$
and $|{\rm adj}|=N$, and only the term with $i=1$ contributes
to the sum (\ref{HopfRs}), in all other terms $\sigma_s$ have negative arguments
and vanish.
Then, since $q^N[N] = q\cdot \frac{A^2-1}{q^2-1} $, we obtain
\be
H_{[21^{N-2}]\times [s]}^{\rm Hopf} =
1 - q^N\cdot \frac{q^{-3}[N]!}{[N-1]!}\cdot \sigma_s(2)
= 1 - \frac{q^{-2}[s] (A^2-1)(q^2-1)}{q^s}
\left(\frac{1}{A\{Aq\}} + \frac{(q^2-1)}{q^{2s}A^2\{Aq\}\{A\}}\right)
= \nn \\
= 1 - \frac{1-q^{-2s}}{qA\{Aq\}}   \Big(A^2-1 + \frac{q^2-1}{q^{2s}}\Big)
= q^{-2s} \cdot\frac{\{q\}}{\{Aq\}}\left(Aq^{2s}+\frac{1}{Aq^{2s}}+\frac{\{A/q\}}{\{q\}}\right)
\ee
what coincides with the second formula in  (\ref{Hopfadj})
up to the $U(1)$-factor.
In the same way, one can derive the general formula  (\ref{Hopfadj}) with arbitrary $r$
from (\ref{HopfRs}) with $R=[2r,r^{N-2}]$.

\bigskip

One can also apply (\ref{parHopf}) to the case of adjoint representation $S={\rm adj}$
and obtain, for example:
\be
\frac{1}{D_{\rm adj}}\Big({\cal H}^{\rm Hopf}_{[1]\times{\rm adj}}\Big)^2
=   {\cal H}^{\rm Hopf}_{[2]\times{\rm adj}} + {\cal H}^{\rm Hopf}_{[1,1]\times{\rm adj}}
\ \ \Longrightarrow \ \
{\cal H}^{\rm Hopf}_{[1,1]\times{\rm adj}}(A,q) =
{\cal H}^{\rm Hopf}_{[2]\times{\rm adj}}(A,q^{-1})
\ee
which can be further generalized to
\be
{\cal H}^{\rm Hopf}_{[1^r]\times{\rm adj}}(A,q) =
{\cal H}^{\rm Hopf}_{[r]\times{\rm adj}}(A,q^{-1})
\ee
At the next step, one can consider
\be
\frac{1}{D_{\rm adj}} \cdot {\cal H}^{\rm Hopf}_{[1]\times{\rm adj}}
\cdot {\cal H}^{\rm Hopf}_{\overline{[1]}\times{\rm adj}}
= \frac{1}{D_{\rm adj}} \cdot {\cal H}^{\rm Hopf}_{[1]\times{\rm adj}}
\cdot {\cal H}^{\rm Hopf}_{ {[1^{N-1}]}\times{\rm adj}}
=   {\cal H}^{\rm Hopf}_{\emptyset\times{\rm adj}}
+ {\cal H}^{\rm Hopf}_{{\rm adj}\times{\rm adj}}
\ee
at $A=q^N$ and deduce that
\be
{\cal H}^{\rm Hopf}_{{\rm adj}\times{\rm adj}} = D_{\rm adj}\Big(-1+
(q^2-1+q^{-2})^2\cdot D_{[1]}^2\Big)
\ee
As usual for the invariants involving adjoint representations, we do not omit the $U(1)$-factor. This is necessary in order to
respect relations like $\overline{[1]}=[1^{N-1}]$ for $sl_N$.
For example, there is a coefficient $q^4$ in the relation
\be
{\cal H}^{\rm Hopf}_{{\rm adj}\times{\rm adj}} \ \stackrel{A=q^2}{=}
q^4\cdot {\cal H}^{\rm Hopf}_{[2]\times [2]}
\ee
unless one either manifestly adds the $U(1)$-factor to ${\cal H}^{\rm Hopf}_{[2]\times [2]}$ or use the differential framing.

\subsection{Composite representations}

This result can now be used in the new recursion:
\be
{\cal H}^{\rm Hopf}_{\overline{[1]}\times [s]}
\cdot {\cal H}^{\rm Hopf}_{R\times [s]} = D_{[s]}\cdot\!\!\!\!\!
\sum_{Y\in R\otimes \overline{[1]}}{\cal H}^{\rm Hopf}_{Y\times [s]}
\ee
Consider the composite (or rational, \cite{Koike,Kanno}; or coupled, \cite{Vafa}) representations $(R,[p])$ \cite{Koike,GW,Vafa,Kanno,MarK}, which are associated with the Young diagram obtained by putting $R$ atop of  $p$ lines of the length $N-1$, i.e. $(R,[p])= [r_1+p,\ldots,r_{l_R}+p,p^{N-1-l_R}]$:

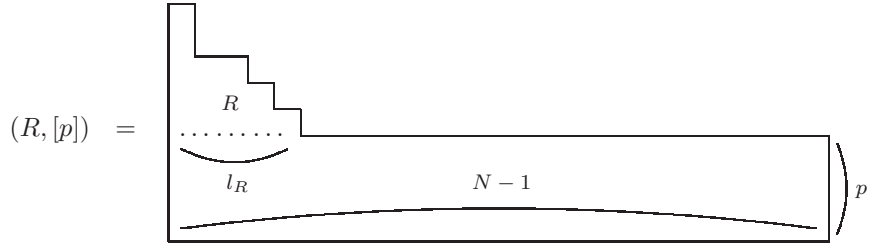
\begin{picture}(300,120)(-110,-20)

\put(0,0){\line(0,1){90}}
\put(0,0){\line(1,0){250}}
\put(50,40){\line(1,0){200}}
\put(250,0){\line(0,1){40}}

\put(0,90){\line(1,0){10}}
\put(10,90){\line(0,-1){20}}
\put(10,70){\line(1,0){20}}
\put(30,70){\line(0,-1){10}}
\put(30,60){\line(1,0){10}}
\put(40,60){\line(0,-1){10}}
\put(40,50){\line(1,0){10}}
\put(50,50){\line(0,-1){10}}

\put(-60,40){\mbox{$(R,[p]) \ \ =$}}

{\footnotesize
\put(20,50){\mbox{$R$}}
\qbezier(253,3)(260,20)(253,37)
\put(260,18){\mbox{$p$}}
\qbezier(5,5)(125,20)(245,5)
\put(115,20){\mbox{$N-1$}}
\qbezier(5,35)(25,25)(45,35)
\put(22,20){\mbox{$l_R$}}
}

\put(4,40){\mbox{$\ldots$}}
\put(18,40){\mbox{$\ldots$}}
\put(32,40){\mbox{$\ldots$}}

\end{picture}

\noindent
so that $\overline{[1]}=[1^{N-1}]=(\emptyset,[1])$
and the adjoint itself is
${\rm adj} = ([1],[1])$, we get in this way:
\be
\phantom.[2]\times \overline{[1]} = ([2],[1]) + [1] & \Longrightarrow &
{\cal H}_{([2],[1])\times [1]}^{{\rm Hopf}} = \frac{\{Aq^2\}\{A\}\{A/q\}}{\{q^2\}\{q\}^2}\cdot
\left({A\over q}+{1\over Aq^5}+{\{A/q^2\}\over q^2\{q\}}\right)
\nn \\
\phantom.[1,1]\times \overline{[1]} = ([1,1],[1]) + [1] & \Longrightarrow &
{\cal H}_{([1,1],[1])\times [1]}^{\rm Hopf} = \frac{\{Aq\}\{A\}\{A/q^2\}}{\{q^2\}\{q\}^2}\cdot
\left({A\over q}+{[2]\over Aq^2}+{\{A/q^3\}\over q\{q\}}\right)
\nn\\
\ldots
\ee
where we have omitted the $U(1)$-factor.
We can immediately obtain similar Hopf polynomials in a slightly more general case: ${\cal H}_{([2],[1])\times [s]}^{\rm Hopf}$ and ${\cal H}_{([1,1],[1])\times [s]}^{\rm Hopf}$. For explicit formulas, we refer the reader to ss.\ref{summary}.

We can  generalize further, by allowing arbitrary composite representations, which are described by "subtracting"
an arbitrary diagram $P$, not just $[p]$:

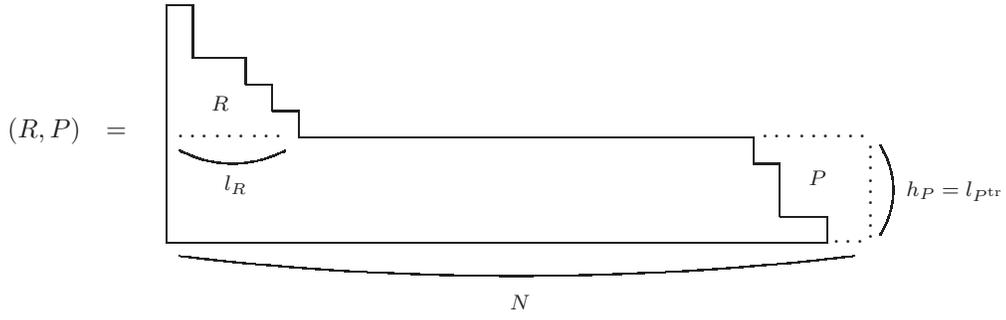
\begin{picture}(300,130)(-110,-30)

\put(0,0){\line(0,1){90}}
\put(0,0){\line(1,0){250}}
\put(50,40){\line(1,0){172}}

\put(0,90){\line(1,0){10}}
\put(10,90){\line(0,-1){20}}
\put(10,70){\line(1,0){20}}
\put(30,70){\line(0,-1){10}}
\put(30,60){\line(1,0){10}}
\put(40,60){\line(0,-1){10}}
\put(40,50){\line(1,0){10}}
\put(50,50){\line(0,-1){10}}

\put(265,2){\mbox{$\vdots$}}
\put(265,15){\mbox{$\vdots$}}
\put(265,28){\mbox{$\vdots$}}

\put(252,0){\mbox{$\ldots$}}
\put(253,40){\mbox{$\ldots$}}
\put(239,40){\mbox{$\ldots$}}
\put(225,40){\mbox{$\ldots$}}

\put(222,40){\line(0,-1){10}}
\put(222,30){\line(1,0){10}}
\put(232,30){\line(0,-1){20}}
\put(232,10){\line(1,0){18}}
\put(250,0){\line(0,1){10}}

\put(0,90){\line(1,0){10}}
\put(10,90){\line(0,-1){20}}
\put(10,70){\line(1,0){20}}
\put(30,70){\line(0,-1){10}}
\put(30,60){\line(1,0){10}}
\put(40,60){\line(0,-1){10}}
\put(40,50){\line(1,0){10}}
\put(50,50){\line(0,-1){10}}

\put(-60,40){\mbox{$(R,P) \ \ =$}}

{\footnotesize
\put(17,50){\mbox{$R$}}
\put(243,22){\mbox{$ P$}}
\qbezier(270,3)(280,20)(270,37)
\put(280,18){\mbox{$h_P = l_{P^{\rm tr}}$}}
\qbezier(5,-5)(132,-20)(260,-5)
\put(130,-25){\mbox{$N $}}
\qbezier(5,35)(25,25)(45,35)
\put(22,20){\mbox{$l_R$}}
}

\put(4,40){\mbox{$\ldots$}}
\put(18,40){\mbox{$\ldots$}}
\put(32,40){\mbox{$\ldots$}}

\end{picture}

\noindent
will be the first ("maximal"), contributing to the product $R\otimes \bar P$. It can be manifestly obtained from the tensor products (i.e. as a projector from  $R\otimes \bar P$) by formula \cite{Koike}
\be
(R,P)=\sum_{Y,Y_1,Y_2}(-1)^{l(Y)}N^R_{YY_1}N^{P}_{Y^tY_2}\ Y_1\otimes\overline{Y_2}
\ee
where the superscript "t" denotes transposition.

One can now calculate, for instance, ${\cal H}_{\overline{[1,1]}\times [s]}^{\rm Hopf} =
{\cal H}_{(\emptyset,[1,1])\times [s]}^{\rm Hopf}$ in this notation by a direct calculation
and then obtain ${\cal H}_{[21^{N-3}]\times [s]}^{\rm Hopf} =
{\cal H}_{([1],[1,1])\times [s]}^{\rm Hopf}$ and ${\cal H}_{[31^{N-3}]\times [s]}^{\rm Hopf} =
{\cal H}_{([2],[1,1])\times [s]}^{\rm Hopf}$
as implications of (\ref{parHopf}), see ss.\ref{summary}.

An additional important fact about the Hopf link,
\be
{\cal H}^{\rm Hopf}_{R_1\times\bar R_2}(A,q) = {\cal H}^{\rm Hopf}_{R_1\times R_2}(A^{-1},q^{-1})
\label{conjHopf}
\ee
is obvious from the picture

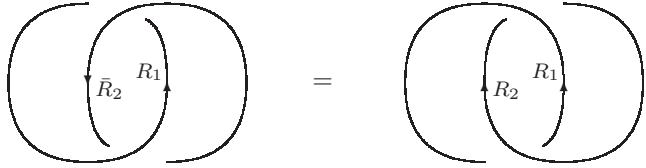
\begin{picture}(300,100)(-120,-50)

\qbezier(-30,0)(-30,30)(0,30)
\qbezier(-30,0)(-30,-30)(0,-30)
\qbezier(30,0)(30,-30)(0,-30)

\qbezier(30,0)(30,20)(22,24)

\put(30,0){
\qbezier(-30,0)(-30,30)(0,30)
\qbezier(30,0)(30,30)(0,30)
\qbezier(30,0)(30,-30)(0,-30)

\qbezier(-30,0)(-30,-20)(-22,-24)
}

\put(0,1){\vector(0,-1){2}}
\put(30,-1){\vector(0,1){2}}
{\footnotesize
\put(3,-5){\mbox{$\bar R_2$}}
\put(18,2){\mbox{$R_1$}}
}

\put(85,-2){\mbox{$=$}}

\put(150,0){
\qbezier(-30,0)(-30,30)(0,30)
\qbezier(-30,0)(-30,-30)(0,-30)
\qbezier(30,0)(30,30)(0,30)

\qbezier(30,0)(30,-20)(22,-24)

\put(30,0){
\qbezier(-30,0)(-30,-30)(0,-30)
\qbezier(30,0)(30,30)(0,30)
\qbezier(30,0)(30,-30)(0,-30)

\qbezier(-30,0)(-30,20)(-22,24)
}

\put(0,-1){\vector(0,1){2}}
\put(30,-1){\vector(0,1){2}}

{\footnotesize
\put(3,-5){\mbox{$R_2$}}
\put(18,2){\mbox{$R_1$}}
}

}

\end{picture}

\noindent
supplemented by the property of ${\cal R}$-matrix, that its inversion is equivalent to
inversion of $A$ and $q$.
Note that the
equality (\ref{conjHopf}) specifically holds for the Hopf link despite ${\cal R}_{R_1\times\bar R_2}(A,q)\neq {\cal R}_{R_1\times R_2}(A^{-1},q^{-1})$.
The simplest example of (\ref{conjHopf}) is
\be
{\cal H}^{\rm Hopf}_{[1]\times\overline{[1]}} =q^{-1}D_{[1]}\Big(Aq^2+{\{A/q\}\over\{q\}}\Big)=A^2+ {\{Aq\}\{A/q\}\over
\{q\}^2}
= \overline{{\cal H}^{\rm Hopf}_{[1]\times {[1]}}}
\ee
For illustration purposes, we present several more explicit calculations
with use of the above properties referring again to ss.\ref{summary}. First of all,
since $[1]\otimes{\rm adj} = ([2],[1])+([1,1],[1])+[1]$,
\be
{\cal H}^{\rm Hopf}_{([2],[1])\times {\rm adj}}
+ {\cal H}^{\rm Hopf}_{[1,1]\times{\rm adj}}
+ {\cal H}^{\rm Hopf}_{[1]\times{\rm adj}}
\ \stackrel{(\ref{HvsHH})}{=}\
\frac{1}{D_{\rm adj}} \cdot {\cal H}^{\rm Hopf}_{{\rm adj}\times {\rm adj}}
\cdot {\cal H}^{\rm Hopf}_{ {[1]}\times {\rm adj}}
\ee
and we obtain ${\cal H}^{\rm Hopf}_{([2],[1])\times {\rm adj}}$. The same result can be obtained from $[2]\otimes\overline{[1]}=([2],[1])+[1]$,
i.e. from
\be
{\cal H}^{\rm Hopf}_{([2],[1])\times {\rm adj}}
+ {\cal H}^{\rm Hopf}_{[1]\times{\rm adj}}
\ \stackrel{(\ref{HvsHH})}{=}\
\frac{1}{D_{\rm adj}} \cdot {\cal H}^{\rm Hopf}_{\overline{[1]}\times {\rm adj}}
\cdot {\cal H}^{\rm Hopf}_{ {[2]}\times {\rm adj}}
\ee
From
\be
{\cal H}^{\rm Hopf}_{([2],[2])\times {\rm adj}}
+ {\cal H}^{\rm Hopf}_{{\rm adj} \times {\rm adj}}
+ {\cal H}^{\rm Hopf}_{\emptyset \times {\rm adj}}
\ \stackrel{(\ref{HvsHH})}{=}\
\frac{1}{D_{\rm adj}} {\cal H}^{\rm Hopf}_{[2]\times {\rm adj}}
 {\cal H}^{\rm Hopf}_{\overline{[2]}\times {\rm adj}}
\ \stackrel{(\ref{conjHopf})}{ =} \
\frac{1}{D_{\rm adj}}\cdot {\cal H}^{\rm Hopf}_{[2]\times {\rm adj}}(A,q)
 \cdot {\cal H}^{\rm Hopf}_{ {[2]}\times {\rm adj}}(A^{-1},q^{-1})
\ee
we  deduce the first item at the l.h.s., ${\cal H}^{\rm Hopf}_{([2],[2])\times {\rm adj}}$ and, in general,
\be
{\cal H}^{\rm Hopf}_{([m],[m])\times {\rm adj}}  =
\frac{1}{D_{\rm adj}}\cdot\Big( {\cal H}^{\rm Hopf}_{[m]\times {\rm adj}}(A,q)
 \cdot {\cal H}^{\rm Hopf}_{ {[m]}\times {\rm adj}}(A^{-1},q^{-1})
 - {\cal H}^{\rm Hopf}_{[m-1]\times {\rm adj}}(A,q)
 \cdot {\cal H}^{\rm Hopf}_{ {[m-1]}\times {\rm adj}}(A^{-1},q^{-1})\Big)
 = \nn \\
 = D_{([m],[m])}\cdot\left\{-1 +
\left(\frac{Aq^{2m}}{q} + \frac{q}{Aq^{2m}} + \frac{\{A/q^2\}}{\{q\}}\right)^2
 \right\} \ \ \ \ \ \ \ \ \ \
 \ee
Similarly,
\be
{\cal H}^{\rm Hopf}_{([1^m],[1^m])\times {\rm adj}}  =
\frac{1}{D_{\rm adj}}\cdot\Big( {\cal H}^{\rm Hopf}_{[1^m]\times {\rm adj}}(A,q)
 \cdot {\cal H}^{\rm Hopf}_{ {[1^m]}\times {\rm adj}}(A^{-1},q^{-1})
 - {\cal H}^{\rm Hopf}_{[1^{m-1}]\times {\rm adj}}(A,q)
 \cdot {\cal H}^{\rm Hopf}_{ {[1^{m-1}]}\times {\rm adj}}(A^{-1},q^{-1})\Big)
 = \nn \\
 =\frac{1}{D_{\rm adj}}\cdot\Big( {\cal H}^{\rm Hopf}_{[m]\times {\rm adj}}(A,q^{-1})
 \cdot {\cal H}^{\rm Hopf}_{ {[m]}\times {\rm adj}}(A^{-1},q)
 - {\cal H}^{\rm Hopf}_{[1^{m-1}]\times {\rm adj}}(A,q^{-1})
 \cdot {\cal H}^{\rm Hopf}_{ {[ {m-1}]}\times {\rm adj}}(A^{-1},q )\Big)
 = \nn \\
 = {\cal H}^{\rm Hopf}_{([m],[m])\times {\rm adj}}(A,q^{-1})
 = D_{([1^m],[1^m])}\cdot\left\{-1 +
\left(\frac{Aq}{q^{2m}} + \frac{q^{2m}}{Aq} - \frac{\{Aq^2\}}{\{q\}}\right)^2
 \right\} \ \ \ \ \ \ \ \ \ \
 \ee
These both latter examples involve the adjoint representations, hence, they both are given {\bf with} the $U(1)$-factor included. In fact, it is necessary to take into account the $U(1)$-factor, since (\ref{conjHopf}) is correct only with this factor.

\subsection{Table of Hopf polynomials\label{summary}}

Here we give a short summary of our results for the Hopf polynomials,
which are obtained from the Hopf tangles by division over dimensions,
see (\ref{Hopftanglevspol}).\footnote{Here is a simple illustration of what is the decomposition of products of representations necessary for dealing with formula (\ref{HvsHH}) and how it is realized in terms of dimensions:
\be
\!\!\!\!\!\!\!\!\!\!\!\!\!\!\!\!
\phantom.[1]\otimes (\emptyset,[1^p])
= ([1],[1^p])   +(\emptyset,[1^{p-1}])
& \longrightarrow &[N] \cdot \frac{[N]!}{[p]![N-p]!} = \frac{[N+1]!}{ [p]![N-p-1]![N-p+1]}
 + \frac{[N ]!}{[p-1]![N-p+1]! }
\nn
\\
\phantom.[2]\otimes (\emptyset,[1^p])
= ([2],[1^p])   +([1],[1^{p-1}])
&\longrightarrow  & \frac{[N][N-1]}{[2]}\cdot \frac{[N]!}{[p]![N-p]!} = \frac{[N+2]!}{[2][p]![N-p-1]![N-p+2]}
 + \frac{[N+1]!}{[p-1]![N-p]![N-p+2]}
\nn
\ee
\be
\phantom.[1]\otimes ([1],[1^p])&
=& ([2],[1^p])  + ([1,1],[1^p])+([1],[1^{p-1}])\ \ \ \ \ \longrightarrow \nn\\
\phantom.[N]\cdot \frac{[N+1]!}{[p]![N-p-1]![N-p+1]} &=& \frac{[N+2]!}{[2][p]![N-p-1]![N-p+2]}
+ \frac{[N+1]![N] }{[2][p]![N-p-2]![N-p][N-p+1]} + \frac{[N+1]!}{[p-1]![N-p]![N-p+2]}
\nn
\ee
In particular,
\be
[1]\otimes ([1],[1]) = ([2],[1])  + ([1,1],[1])+[1]\ :
& [N]\cdot  {[N+1] [N-1]}  = \frac{[N+2][N+1][N-1][N-2]}{[2]^2}
+ \frac{[N+1][N]^2[N-3]}{[2]^2} + [N]
\nn \\
\phantom.[1]\otimes ([1],[1,1])
= ([2],[1,1])  + ([1,1],[1,1])+([1],[1])\ :
& [N]\cdot \frac{[N+1]N[N-2]}{[2]} = \frac{[N+2][N+1][N-1][N-2]}{[2]^2}
+ \frac{[N+1][N]^2[N-3]}{[2]^2} + [N+1][N-1]
\nn
\ee
}

\subsubsection{$H_{R\times [s]}$:}
\be
\left\{\begin{array}{rl}
{H}_{[r]\times [s]} = &
1 + \sum_{i=1}^{{\rm min}(r,s)} (-A)^{-i} q^{\frac{i(i+3)}{2}-i(r+s)}
\prod_{j=0}^{i-1} \frac{\{q^{r-j}\}\{q^{s-j}\}}{\{Aq^j\}}
\\ & \\
{H}_{[1^r]\times [s]} =& 1 - \frac{\{q^r\}\{q^s\}}{q^{s-r}A\{A\}}
\\ &\\
H_{R\times[s]} =&
1 - q^{r_1+\ldots+r_l-s}[s](q^2-1)^2
\sum_{i=1}^l \frac{q^{-r_i+i-2}[l+r_i-i]!}{\prod_{j\neq i}^l
[r_i-r_j-i+j]}\cdot\sum_{k=0}^{r_i-i}\frac{q^{-2ks}(q^2-1)^k}{[r_i-i-k ]!\,
\prod_{j=r_i-i-k}^{r_i-i}A\{Aq^j\}}
\end{array}\right.
\ee
These answers are given in the standard framing without the $U(1)$-factor.

\subsubsection{$H_{(R,P)\times [s]}$:}
\be
\left\{\begin{array}{rcl}
{H}_{\overline{[r]}\times [s]}&=&{H}_{(\emptyset,[r])\times [s]}
\ \stackrel{(\ref{conjHopf})}{ =}\overline{
{H}_{{[r]}\times [s]}}\\ & &\\
{H}_{\overline{[1^r]}\times [s]}&=&{H}_{(\emptyset,[1^r])\times [s]}
\ \stackrel{(\ref{conjHopf})}{ =}\overline{
{H}_{{[1^r]}\times [s]}}\\ && \\
{H}_{([r],[r])\times [s]}&
=& 1 + \{q\}^2\cdot \frac{[r][s] \{Aq^{r-1}\}\{Aq^s\}}{\{A\}\{Aq^{2r-1}\}}\cdot
\left(1\ + \ \sum_{i=1}^{r-1}\ \{q\}^i\cdot \frac{[r-1]!}{[r-1-i]!}\cdot
\frac{(A^{2i}q^{2is}+q^{-2is})}{\prod_{j=r-i}^{r-1} A\{Aq^j\}}
\right)\\ && \\
{H}_{([2],[1])\times [s]}
&=& {q^{2s/N+1}\over \{A\}\{Aq^2\}}\cdot
\left(Aq^{2s}\{A\}\{q\} +\{A\}\{A/q\} + \frac{\{q^2\}\{A/q\}}{Aq^{2s}}
+\frac{\{q^2\}\{q\}}{A^2q^{4s}}
\right) \\ && \\
{H}_{([1,1],[1])\times [s]}
&= &{q^{2s/N+1}\over \{A\}\{Aq\}}\cdot
\left( Aq^{2s}\{q\}\{A/q\} +  \{A/q\}^2
+ \frac{\{A\}\{q^2\}}{Aq^{2s}}
\right) \\ && \\
{H}_{([1],[1,1])\times [s]}
& \stackrel{(\ref{conjHopf})}{=}&\overline{
{H}_{([1,1],[1])\times [s]}}
\\ && \\
H_{([1],[2])\times [s]} & \stackrel{(\ref{conjHopf})}{=}&\overline{
{H}_{([2],[1])\times [s]}}
\\ && \\
H_{([1,1],[2])\times [s]}&=&
{q^{4s/N+2s}\over \{Aq\}\{Aq^2\}} \cdot
\left({A^2}{\{q^2\}\{q\}}+
\frac{\{q^2\}\{A^2q^{2s}\}}{q^{4s+1}}-{{\{q^2\}} \{q^{2s}\}\over q^{4s-1}}
  +q^{-4s} \{A\}\{A/q\}
\right)
\\ && \\
{H}_{([2],[1,1])\times [s]}
& \stackrel{(\ref{conjHopf})}{=}&\overline{
H_{([1,1],[2])\times [s]}}
\\ && \\
{H}_{([1,1],[1,1])\times [s]} &=& {q^{4s/N}\over \{A\}\{Aq\}}\cdot
\Big(\{q^2\}\{A^2q^{2s-1}\} + \{A/q\}\{A/q^2\} - (q^{s+1}+q^{-s-1})\{q^2\}\{q^s\}\Big)
\end{array}\right.
\ee
These answers are given in the standard framing with the $U(1)$-factor.

\subsubsection{$H_{R\times {\rm adj}}$:}
\be
\left\{\begin{array}{rl}
{H}_{[s]\times {\rm adj}}
=& {H}_{([1],[1])\times [s]}
= {\{q\}\over \{Aq\}}\cdot
\left(A q^{2s}  +{ \{A/q\}\over \{q\}} + {1\over Aq^{2s}}
\right)
\\ & \\
{H}_{[1^r]\times{\rm adj}}(A,q) =&
{H}_{[r]\times{\rm adj}}(A ,q^{-1})
\\ & \\
{H}_{\overline{[s]}\times {\rm adj}} =&
{H}_{(\emptyset,[s])\times {\rm adj}}
\ \stackrel{(\ref{conjHopf})}{ =} \overline{
{\cal H}_{{[s]}\times {\rm adj}}}
\end{array}\right.
\ee
These answers are given in the standard framing with the $U(1)$-factor.

\subsubsection{$H_{(R,P)\times {\rm adj}}$:}

\be
\left\{\begin{array}{rcl}
{H}_{{\rm adj}\times{\rm adj}}& =&
{H}_{([1],[1])\times {\rm adj}} =
{1\over D_{\rm adj}}\cdot\left\{-1+ (q^2-1+q^{-2})^2 \cdot D_{[1]}^2\right\}\\
& &\\
{H}_{([2],[1])\times {\rm adj}}& =&
{\{q\}\over \{A/q\}}(q^2-1+q^{-2})\left(Aq^2+{1\over Aq^2}+{\{A/q^3\}\over\{q\}}\right) \\
&& \\
{H}_{([1],[2])\times {\rm adj}} & \stackrel{(\ref{conjHopf})}{=} &
\overline{{H}_{([2],[1])\times {\rm adj}}}  \\
&& \\
{H}_{([r],[r])\times {\rm adj}} & =&
{1\over D_{{\rm adj}}}\cdot\left\{ -1+
\left( Aq^{2r-1}+ \frac{1}{Aq^{2r-1}} + \frac{\{A/q^2\}}{\{q\}}\right)^{\!2}
 \right\}\\
&&\\
{H}_{([1^r],[r])\times {\rm adj}}  &=&
{1\over D_{\rm adj}}\cdot\left\{-1+ \left(q^{s+1}-q^{s-1}+{1\over q^{s+1}}\right)\left(q^{s+1}-{1\over q^{s-1}}+{1\over q^{s+1}}\right) \cdot D_{[1]}^2\right\}
\\
&&\\
{H}_{([ r],[1^r])\times {\rm adj}} & =&\overline{
{H}_{([1^r],[r])\times {\rm adj}}}
\\
&&\\
{H}_{([1^r],[1^r])\times {\rm adj}}&
= &{H}_{([r],[r])\times {\rm adj}}(A,q^{-1})
\end{array}\right.
\ee
These answers are given in the standard framing with the $U(1)$-factor.

\section{Conclusion}

In this paper, we addressed the main implication of the Reshetikhin-Turaev
formulation (RT) \cite{RT} of knot theory
as a lattice theory on arbitrary graphs:
it respects cutting and gluing, i.e. the {\bf multiplicative structures}
underlying knot theory, which are obscured in the consideration of knot polynomial
{\it per se}.
Normally this is cured by lifting to the categorified knot theory,
but it is very difficult to develop it in a form useable in practical calculations.
The tangle calculus provides a tool unifying advantages of the both approaches:
explicit multiplicative structure on one side and calculability on the other.

The RT theory associates with each tangle a tensor of rank equal to the number
of external legs, and indices are just convoluted (with quantum weights)
when the ends merge.
However, in this form, this is hardly an efficient way of thinking and,
especially, calculating.
The breakthrough comes, as in all applications of RT theory, when one
takes into account the special property of ${\cal R}$-matrices: that they
are essentially unit matrices in irreducible representations,
thus the only things which matter are the eigenvalues (expressed through the
quadratic Casimirs) and the Racah (rather than Clebsh-Gordan) coefficients.
This modern version of RT approach \cite{modRT1}-\cite{modRT2}
leads to incredible simplifications and to very efficient calculational methods.
The cut-and-join technique within this context already led to a powerful
arborescent calculus \cite{arbor} applicable to a special class
of tangles,  and time is coming to study the method in full generality.
This paper is the first step in this direction:
we demonstrate that the arborescent calculus can be extended, not only conceptually,
but practically, and this provides us with new insights and new results.

The main points that we discussed are nearly obvious:
\begin{itemize}
\item When the link diagram is cut into tangles, they are glued back
with the "small" sum over representations only (representation spaces
are taken into account by quantum dimension factors), or, better to say, over intertwining operators.
This is the main point of the {\it modern} RT formalism of \cite{modRT1}-\cite{modRT2},
and it is nicely applicable to arbitrary tangles.

\item The same tangles enter the construction of different knots and links,
which provides a lot of  relations, predominantly non-linear. The most important are the gluing of tangles, which is actually a multiplication of tangle blocks, and switching or reshuffling operations
which change arrow directions and representations in different channels.
\end{itemize}

\noindent
Starting from these two main statements, one can choose different roads to investigate:

\begin{itemize}
\item Can the net of non-linear relations be reduced to quadratic
and to reveal some hidden integrability?

\item Knots and links with a given number of intersections can be made from
a finite number of tangles. Of course, everything can be build from the ${\cal R}$-matrix, which is an elementary tangle in RT theory, but clearly this is not the most economic and effective option for the tangle calculus. What is the optimal choice of tangles which minimizes the number of building blocks at each level?

\item What is an efficient way to evaluate tangles?
\end{itemize}

In this paper, we demonstrate that one can really move along these directions,
obtain non-trivial relations between already known knot polynomials
and deduce new answers from the previously known.

\section*{Acknowledgements}

This work was performed at the Institute for Information Transmission Problems with the financial
support of the Russian Science Foundation (Grant No.14-50-00150).

\end{document}